\documentclass[aps,prx,superscriptaddress,twocolumn,nofootinbib,nobibnotes,floatfix,showpacs,reprint,longbibliography]{revtex4-2}
\usepackage{siunitx}
\usepackage[utf8]{inputenc}
\usepackage{amsmath,amsfonts,amssymb}
\usepackage{color}
\usepackage{soul} 
\usepackage{xcolor}         
\usepackage{subfigure}
\usepackage{multirow}
\usepackage{dcolumn} 
\usepackage{bm} 
\usepackage{graphicx} 
\usepackage{hyperref} 
\hypersetup{
    colorlinks={true},
    linkcolor={blue},
    citecolor={blue},
    urlcolor={blue}
}
\usepackage[sectionbib]{bibunits}
\defaultbibliographystyle{revtex4-2}

\makeatletter
\def\maketitle{
\@author@finish
\title@column\titleblock@produce
\suppressfloats[t]}
\makeatother

\newcommand{\nocontentsline}[3]{}
\let\origcontentsline\addcontentsline
\newcommand\stoptoc{\let\addcontentsline\nocontentsline}
\newcommand\resumetoc{\let\addcontentsline\origcontentsline}

\begin{document}

\title{Intrinsic Berry Curvature Driven Anomalous Hall and Nernst Effect in Co$_2$MnSn}

\author{Bishal Das}
\thanks{These authors contributed equally to this work.}
\affiliation{Department of Physics, Indian Institute of Technology Bombay, Mumbai 400076, India}

\author{Arnab Bhattacharya}
\thanks{These authors contributed equally to this work.}
\affiliation{CMP Division, Saha Institute of Nuclear Physics, A CI of Homi Bhabha National Institute, Kolkata-700064, India}

\author{Amit Chanda}
\affiliation{Department of Physics, University of South Florida, Tampa FL 33620, USA}
\affiliation{Department of Energy Conversion and Storage, Technical University of Denmark, 2800 Kgs. Lyngby, Denmark}

\author{Chanchal K. Barman}
\affiliation{Dipartimento di Fisica, Universit\`a di Cagliari, Cittadella Universitaria, Monserrato (CA) 09042, Italy}

\author{Jadupati Nag}
\affiliation{Department of Materials Science and Engineering, \& Materials Research Institute, Pennsylvania State University, Millennium Sciences Complex Building University Park, PA 16802, USA}

\author{Hariharan Srikanth}
\affiliation{Department of Physics, University of South Florida, Tampa FL 33620, USA}

\author{Aftab Alam}
\email{aftab@iitb.ac.in}
\affiliation{Department of Physics, Indian Institute of Technology Bombay, Mumbai 400076, India}

\author{ I. Das}
\email{indranil.das@saha.ac.in}
\affiliation{CMP Division, Saha Institute of Nuclear Physics, A CI of Homi Bhabha National Institute, Kolkata-700064, India}

\begin{abstract}
    Magnetic topological semimetals often exhibit unusual electronic and thermal transport due to nontrivial bulk band crossings, enabling simultaneous realization of large anomalous Hall and Nernst conductivities ($\sigma_{xy}$ and $\alpha_{xy}$).  Here,  a comprehensive experimental and theoretical study of the anomalous transport properties of ferromagnetic Co$_2$MnSn is reported. First-principles calculations reveal topological Weyl points producing significant Berry curvature, driving dominant intrinsic anomalous Hall/Nernst effects. Electronic and thermal transport measurements demonstrate robust anomalous transport with substantial conductivity values that persist at room temperature ($\sigma_{xy}\sim$~500 S/cm, $\alpha_{xy}\sim$~1.3 A/m/K).  We also show how the chemical substitution (via tuning Fermi level) can boost these effects (up to $\sigma_{xy}\sim$~1376 S/cm, $\alpha_{xy}\sim$~1.49 A/m/K at 150 K). These findings position Co$_2$MnSn as a compelling platform for exploring topological transport phenomena and advancing next-generation thermoelectric and spintronic technologies.
\end{abstract}

\maketitle

\stoptoc
\section{Introduction}

Research based on thermal gradient-driven electronic transport in magnetic materials have attracted tremendous attention to understand the mechanism of anomalous Nernst effect (ANE), a phenomenon with profound implications for thermoelectric devices \cite{Hirohata2014}. The ANE induces a transverse voltage perpendicular to both the thermal gradient ($\Delta T$) and magnetization ($M$) direction, namely \textbf{E}$_{\mathrm{NE}} = Q_\mathrm{S}(\mu_0 \textbf{\textit{M}} \times \bf \Delta \textit{T})$, where $Q_\mathrm{S}$ and $\mu_0$ are the anomalous Nernst coefficient and vacuum permeability \cite{Ikhlas2017, Pu2008}. In conventional magnets, the size of ANE remains negligibly small, unsuitable for practical applications. Theoretical band structure calculations revealed certain magnetic materials, hosting linearly dispersed band (anti) crossings near the Fermi energy (\textit{E}$_\mathrm{F}$). Such band features governed by Weyl Hamiltonian,  facilitate promising platform for discovering new materials with pronounced ANE \cite{Armitage2018, Breidenbach2022, Sakuraba2020, Hu2018, Hu2020, Mende2021, De2021} and anomalous Hall effect (AHE) \cite{Xiao2006, p3, Li2020, p7, p10, p13, p15}. This is driven by the hotspot of Berry curvature concentrated around the Weyl nodes in the vicinity of \textit{E}$_\mathrm{F}$. The intrinsic contributions to AHE and ANE from non-zero Berry curvature is given by the expressions \cite{Wang2006, Xiao2006}:

\begin{equation}
    \sigma_{xy}^A = -\frac{e^2}{\hbar}\int\,[dk]\,\Theta(E-E_k)\,\Omega^z(k)
\end{equation}
\begin{equation}
    \alpha_{xy}^A = -\frac{1}{e}\int\,dE\,\frac{\partial f}{\partial \mu}\,\sigma_{xy}^A(E)\,\frac{E-\mu}{T}
\end{equation}
where, $\sigma_{xy}^A$ and $\alpha_{xy}^A$ are the anomalous Hall and anomalous Nernst conductivity (AHC, ANC), $e$, $\hbar$, $\mu$ and $\Omega^z(k)$ are the electronic charge, reduced Planck's constant, chemical potential and $z$-component of Berry curvature, respectively. $\Theta$ is the step function and $f$ is the Fermi-Dirac distribution at temperature $T$. These expressions highlight that, while AHC arises from the integration of Berry curvature below \textit{E}$_\mathrm{F}$, ANC is governed by effectively sampling the Berry curvature around the Fermi surface \cite{Xiao2006}. This provides a reliable backdrop for simultaneously realizing substantial electrical and thermoelectrical response by tuning $\mu$ in novel functional materials. 

In this scenario, Weyl semimetallic full Heusler alloys (fHA) have garnered immense interest due to the presence of multiple Weyl points near \textit{E}$_\mathrm{F}$ \cite{Sakai2018, Guin2019, Cox2019, Breidenbach2022, Sakuraba2020, Hu2018, Hu2020, Mende2021, De2021}. Notably, within this family, Co$_2$MnGa \cite{Sakai2018} has exhibited largest intrinsic AHC (= 870 S cm$^{-1}$) as well as high anomalous Nernst coefficient in its bulk single crystalline form ($\sim 6$ $\mu\mathrm{V \,K^{-1}}$) \cite{Sakai2018, Guin2019} and epitaxial thin film ($\sim 3$ $\mu\mathrm{V \,K^{-1}}$) \cite{Park2020}, polycrystalline thin film ($\sim 5.4$ $\mu\mathrm{V \,K^{-1}}$) \cite{Uesugi2023} as well as in bulk polycrystalline form ($\sim 6.8$ $\mu\mathrm{V \,K^{-1}}$) \cite{Zhou2023}. Intriguingly, amorphous materials \cite{p20} exhibit comparable Berry-curvature-driven AHC and ANC to their single-crystalline counterparts \cite{p21,p22}. This pose a fundamental inquiry into whether the enhanced anomalous transport is inherently dictated by the proximity of nodal lines or Weyl points to \textit{E}$_\mathrm{F}$. More specifically, the extent to which anomalous transport varies when these nodal lines are positioned farther from \textit{E}$_\mathrm{F}$ remains an important aspect to explore. Co$_2$MnX (X = Si, Ge, Sn, Al, Ga) fHA provides an excellent platform to systematically tune nodal positions relative to \textit{E}$_\mathrm{F}$ using controlled modification of the valence electron count ($N_\mathrm{V}$) via different \textit{X}-element substitution \cite{p26, p27, p28}. At this juncture, Co$_2$MnSn stands out with its unique electronic structure, giving an excellent stage to probe these critical questions.

In this work, we present a comprehensive experimental and theoretical investigation of the anomalous transport properties of ferromagnetic (FM) Co$_2$MnSn, driven by finite Berry curvature. First-principles calculations establish Co$_2$MnSn to be a topological semimetal, characterized by the presence of nodal lines and Weyl points in the bulk band structure. These topological degeneracies generate substantial Berry curvature, giving rise to pronounced intrinsic anomalous Hall and Nernst effects. Notably, the anomalous transport in Co$_2$MnSn is primarily governed by the Weyl nodes near \textit{E}$_\mathrm{F}$, which serve as the dominant source of Berry curvature rather than gapped nodal lines. Our experimental synthesis confirms a prototype L2$_1$ structure for Co$_2$MnSn, with saturation magnetization of 5 $\mu_B$/f.u., agreeing well with the Slater-Pauling rule \cite{p30}.  Transport measurements validate the theoretical predictions, revealing significant AHC and ANC persisting all the way up to room temperature.

\section{Methods}

\subsection{Sample Synthesis}
Bulk polycrystalline samples were synthesized using the stoichiometric ratio of high-purity constituent elements and melting them under partial argon pressure in an arc furnace. The samples were melted several times to maintain uniformity and subsequently annealed at 800$^{\circ}$C followed by quenching in an ice bath.  The phase purity of the samples was verified by the Rietveld refinement of X-ray diffraction data obtained at room temperature (RT) using a Rigaku TTRX-III diffractometer. The compositional uniformity was checked using a transmission electron microscope (TEM) equipped with an energy-dispersive X-ray (EDX) spectrometer. 

\subsection{Magnetometry measurements}
The magnetic measurements were done using SQUID-VSM (QD, USA) in the temperature range of 2 K - 380 K using several samples of rectangular geometry with dimensions ranging between 1.88-2 mm $\times$ 3.9-4.3 mm $\times$ 0.49-0.52 mm. The effective magnetic field applied on the sample, $H_e$, is determined by $H_e = H_{app} - N_dM$, where $N_d = 1$ is the demagnetisation constant for plate-geometry samples and $M$ is the magnetisation of the system. 

\subsection{Electrical Transport}
The same samples were used for magnetotransport measurements, which were performed to minimise the demagnetisation effect, using the standard four-probe method with current and voltage contacts made of 50 $\ mu$m gold wires and conducting silver epoxy. The measurements were conducted using a PPMS (QD, USA) equipped with a 9 T superconducting magnet. Before each isothermal magnetization and transverse resistivity measurement, the sample was zero field cooled (ZFC) from high temperature to the target temperature and subsequently the field was applied. The transverse resistivity was symmetrized using the relation of $\rho_{xy} = [\,\rho_{xy}(+H)-\rho_{xy}(-H)\,]/2$. 

\subsection{Thermal Transport}
Anomalous Nernst measurements on the Co$_2$MnSn sample were performed by employing a custom-designed spin-caloritronic setup integrated with the PPMS. The sample was sandwiched between two copper plates in such a way that a temperature gradient is applied along the thickness of the sample ($y$-direction), and the direction of the temperature gradient is transverse to the direction of the external magnetic field ($z$-direction) provided by the PPMS. The bottom copper plate (hot) was thermally insulated from the base of the PPMS puck by a 4 mm thick Teflon block, whereas the top copper plate (cold) was thermally attached to the base of the PPMS puck using a pair of molybdenum screws. The induced anomalous Nernst voltage in the sample in presence of the applied temperature gradient and the external magnetic field was recorded by a Keithley 2182A nanovoltmeter along the $x$-direction. A detailed description of our spin-caloritronic setup is reported in our previous studies \cite{Chanda2022_1, Chanda2022_2, Chanda2023_1,Chanda2023_2, Chanda2024}.

\subsection{First-principles calculations}
Density functional theory (DFT) \cite{Hohenberg_Kohn1964, Kohn_Sham1965} calculations were carried out using the Vienna Ab-initio Simulation Package (VASP) \cite{Kresse1993, Kresse1996_1, Kresse1996_2} which is based on projector augmented wave (PAW) formalism \cite{Kresse1999, Blochl1994_paw}. Although the generalized gradient approximation (GGA) describes the exchange and correlation effects fairly well in many systems, one needs to go beyond GGA to account for strong electronic correlations. We have used the Strongly Constrained and Appropriately Normed (SCAN) \cite{Sun2015} meta-GGA functional which is known to satisfy all 17 exact constraints that a semilocal functional should satisfy, hence being mathematically robust. SCAN includes the local kinetic energy density within the exchange-correlation potential as correction to the non-interacting Kohn-Sham kinetic energy density, hence better describing strong electronic correlations. Brillouin zone (BZ) integration was performed on a 12$\times$12$\times$12 $\Gamma$-centered $k$-mesh using the Blöchl corrected linear tetrahedron method \cite{Blochl1994_tet} with the total energy convergence criteria set to $10^{-6}$ eV. A plane wave energy cutoff of 500 eV was used for all the calculations.

\subsection{Berry curvature and anomalous transport calculations}
A tight-binding Hamiltonian was constructed from pre-converged DFT results using Wannier90 \cite{Marzari1997, Souza2001, Marzari2012, wannier90v1, wannier90v2, wannier90v3}. A total of 160 bands were wannierized with projections on atomic sites including Co($s, p, d$), Mn($d$), and Sn($s, p$) orbitals as the basis set. From the constructed tight-binding Hamiltonian in the basis of atomic Wannier functions, the anomalous Hall conductivity $\sigma_{ab}$ and anomalous Nernst conductivity $\alpha_{ab}$ were calculated by our in-house developed numerical code on a $\Gamma$-centered $200\times200\times200$ $k$-mesh using the following relations \cite{Wang2006, Xiao2006}:
\begin{eqnarray}
    \sigma_{ab}(\varepsilon) &=& -\frac{e^2}{\hbar}\int\frac{d^3k}{(2\pi)^3}\,\Omega_{c}(\vec{k}) \\
    \alpha_{ab}(\mu) &=& -\frac{1}{e}\int d\varepsilon \left(-\frac{\partial f}{\partial\varepsilon}\right)\,\frac{\varepsilon-\mu}{T}\,\sigma_{ab}(\varepsilon)
\end{eqnarray}
where, $\Omega_{c}(\vec{k})=\sum_n f_n(\vec{k},\varepsilon)\,\Omega_{c}^n(\vec{k})$ is the total Berry curvature calculated by summing over the band-resolved Berry curvature $\Omega_{c}^n(\vec{k})$ with Fermi weights given by $f_n(\vec{k},\varepsilon)=\left[1+\text{exp}\left(\frac{E_{n}(\vec{k})-\varepsilon}{k_{B}T}\right)\right]^{-1}$. The band-resolved Berry curvature for the $n$-th band is defined as \cite{Wang2006}:
\begin{equation}
    \Omega_{c}^n(\vec{k})=-2\,\text{Im}\sum_{m\neq n} \frac{\langle n|\frac{\partial H}{\partial k_{a}}|m\rangle\langle m|\frac{\partial H}{\partial k_{b}}|n\rangle}{\left(E_m(\vec{k})-E_n(\vec{k})\right)^2}
\end{equation}
where, $|m\rangle$ and $|n\rangle$ are eigenstates of Hamiltonian $H$ with eigenenergies $E_m$ and $E_n$ respectively. In all expressions above, the indices $a,b,c \in \left\lbrace x,y,z \right\rbrace $ refer to the cartesian directions.

\subsection{Carrier concentration calculations}
Additionally, the carrier concentration was evaluated by integrating the total density of states obtained from DFT calculations on an energy grid with 8001 points, using the relations:
\begin{eqnarray}
    N_e(\mu)&=&\int_{E_C}^{E_{max}} g(E)\,f(E,\mu)\,dE \\
    N_h(\mu)&=&\int_{E_{min}}^{E_V} g(E)\,[1-f(E,\mu)]\,dE
\end{eqnarray}
where, $N_e$ and $N_h$ are n-type and p-type carrier concentrations respectively, $E_{min}$ and $E_{max}$ are the minimum and maximum of the chosen energy grid respectively and $E_{V}$ and $E_{C}$ are the valence and conduction band edges respectively. Since Co$_2$MnSn is metallic as discussed earlier, $E_{V}$ and $E_{C}$ have been chosen to be the Fermi level. Here, the Fermi function at energy $E$ for a given chemical potential $\mu$ is denoted by $f(E,\mu)$ and $g(E)$ is the density of states.

\section{Results and Discussions}

\begin{figure*}[!t]
  \centering
  \includegraphics[width=\linewidth]{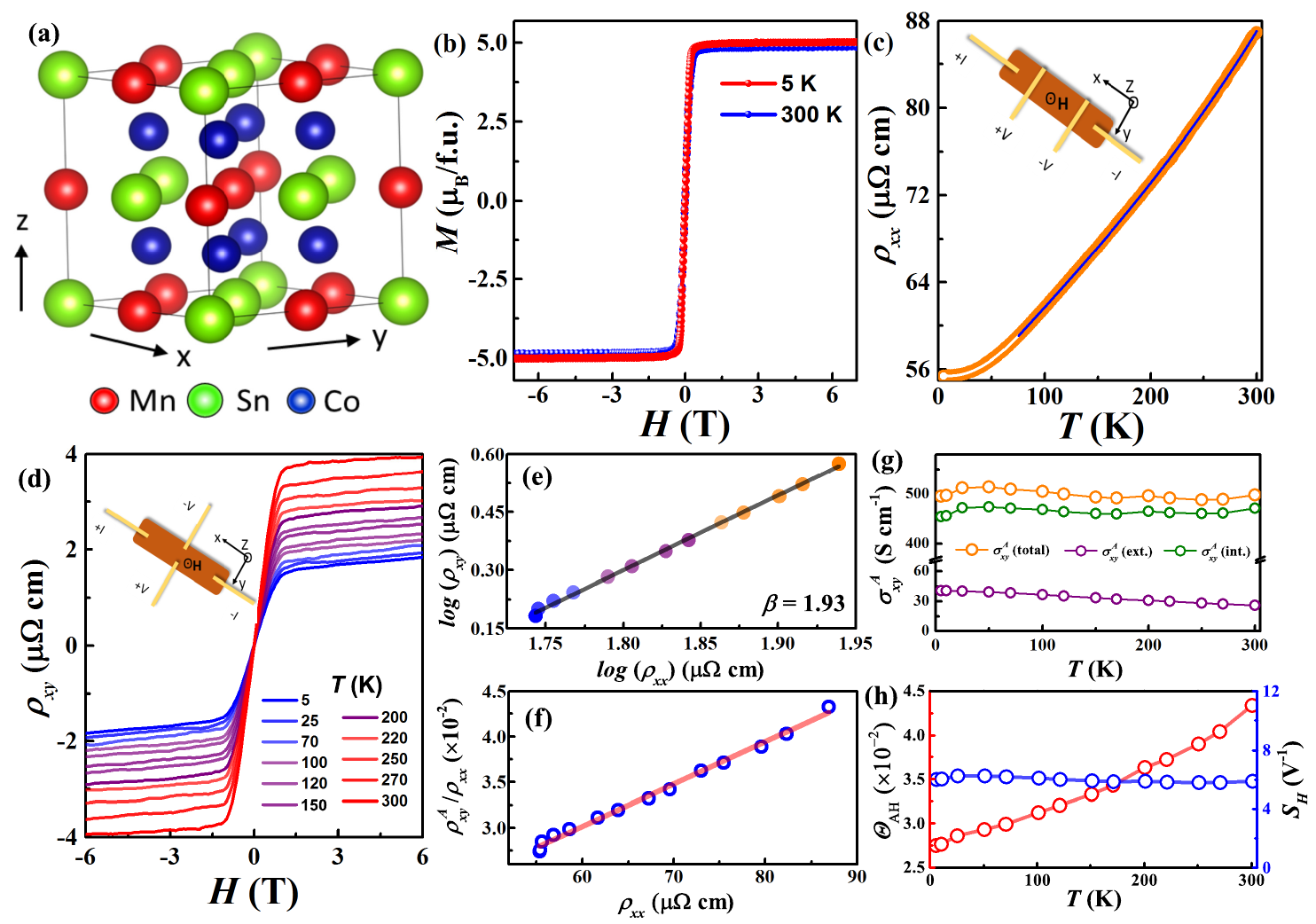}
  \caption{(a) Crystal structure of Co$_2$MnSn. (b) Isothermal field-dependent DC magnetization, $M(H)$ at $T$ = 5, 300 K. (c) $T$-dependent zero-field longitudinal resistivity($\rho_{xx}$). (d) Isothermal $H$-dependent transverse resistivity ($\rho_{xy}$)  from 5 to 300 K. (e) log($\rho_{xy}$) vs. log($\rho_{xx}$) curve shows quadratic dependence. (f) TYJ scaling relation. (g) $T$-dependence of total, intrinsic, and extrinsic AHC. (h) $T$-dependence of anomalous Hall angle and Hall factor. }
   \label{F1}
\end{figure*}

\subsection{Crystal Structure}
Rietveld refinement of the room temperature XRD pattern of Co$_2$MnSn  (see Table~\ref{XRDt} and Fig.~\ref{XRD} of Supplementary Material SM \cite{supp}) confirms that it crystallizes in the Cu$_2$MnAl-type L2$_1$ structure with a lattice parameter of 5.99(6)~\AA. The crystal structure is shown in Fig.~\ref{F1}(a).

\subsection{Magnetization}
Isothermal magnetization $M(H)$ at 5 K and 300 K indicates soft ferromagnetism with a small hysteresis loop. The saturation magnetic moment ($M_S$) of 5 $\mu_\mathrm{B}$/f.u. aligns with the Slater-Pauling (SP) rule, $M_S$ = N$_V$ – 24 $\mu_\mathrm{B}$/f.u., where N$_V$ = 29 is the system’s total valence electrons in a fully ordered FM state. Figure~\ref{F1}(c) shows the $T$-dependence of longitudinal resistivity ($\rho_{xx}$), with a residual resistivity ratio ($\rho_{\mathrm{300 K}}/\rho_{\mathrm{5 K}}$) of 1.56, and residual resistivity of 55.4 $\mu\Omega$cm.

\subsection{Anomalous Hall Transport}
Figure~\ref{F1}(d) shows transverse resistivity ($\rho_{xy}$) isotherms from 5–300 K, with a sharp rise for $H<$ 1 T followed by weak field dependence up to 6 T. In ferro/ferrimagnetic systems, $\rho_{xy}$ comprises normal Hall resistivity (NHE) and anomalous Hall effect (AHE): $\rho_{xy} = \rho_{xy}^N + \rho_{xy}^A = R_0H + R_S M$, where $ R_0$ and $R_S$ are the normal and anomalous Hall coefficients. Considering a single-band model,  $ R_0 = -1/ n_0|e|$, where $n_0$ is the carrier concentration. From the positive high-field (4–7 T) slope, $n_0 = 1.17 \times 10^{21}$ cm$^{-3}$, confirming holes as majority carriers. Since AHE is a low-field effect, $\rho_{xy}^A$ is extracted by extrapolating high-field data to zero. Power-law dependence, $\rho_{xy}^A \propto \rho_{xx}^\beta$ (Fig.~\ref{F1}(c)), helps to distinguish the intrinsic/extrinsic mechanism to AHE \cite{Nagaosa2010}. A log($\rho_{xy}^A$) vs. log($\rho_{xx}$) fiting yields $\beta \approx 1.93$, implying a Berry curvature-dominant intrinsic or side-jump (sj) mechanism \cite{p3,p8,p7}. We further use TYJ scaling, $\rho_{xy}^A/\rho_{xx} = a  + b \rho_{xx}$ (Fig.~\ref{F1}(c)) \cite{p18}, where $a$ and $b$ indicate skew-scattering (sk) and intrinsic/sj contributions, respectively. The linear fit gives $a$ = 2.26$\times10^{-3}$ and $b$  $(=|\sigma_{xy,int}^A|)\approx$ 464 S cm$^{-1}$. Figure~\ref{F1}(g) depicts the $T$-dependence of total AHC ($\sigma_{xy}^A$), intrinsic AHC ($\sigma_{xy,int}^A$) and extrinsic contribution $\sigma_{xy,ext}^A (\approx a\rho_{xx}/\rho_{xx}^2$), respectively \cite{p19}. The total AHC is obtained by the tensorial relation, $\sigma_{xy}^A  \approx \rho_{xy}^A/\rho_{xx}^2$.
The nearly $T$-invariant $\sigma_{xy,int}^A$ contributing 93$\%$ of $\sigma_{xy}^A \approx 494$ S cm$^{-1}$ at $T =$ 5 K, confirms Berry curvature-dominant AHE. 
While the contribution of $\sigma_{xy, sj}^A$ and $\sigma_{xy, int}^A$ are difficult to decouple due to reduced phonon scattering, a rough estimate of $\sigma_{xy, sj}^A$ is found  to be two orders of magnitude lower than $\vert\sigma_{xy, int}^A\vert$, reinforcing the key role of intrinsic mechanism in the anomalous transport of Co$_2$MnSn \cite{p12,p13}.

We have also compared the AHE of Co$_2$MnSn with other FM materials using two characteristic parameters, the anomalous Hall angle, $\Theta_{AH} =\sigma_{xy}^A/\sigma_{xx} $ and anomalous Hall factor, $S_H =\sigma_{xy}^A/M$. The $\Theta_{AH}$ defines the fraction of longitudinal current that gets converted to anomalous Hall current mediated by the Berry-curvature-driven Karplus-Luttinger contribution. While $S_H = \sigma_{xy}^A/M$ estimates the magnitude of anomalous Hall current to magnetization, quantifying the strength of the AHE \cite{p10}. Figure~\ref{F1}(h) demonstrates the $T$ variation of $\Theta_{AH}$ and $S_H$. While $\Theta_{AH}$ shows a weak $T$-variation, $S_H$ remains constant over the entire range, a characteristic feature of intrinsic Berry curvature-dominated AHE \cite{p10,p14,p13}. A maximum value of $\Theta_{AH} \approx 4.32\%$ and $S_H \approx 0.065$ V$^{-1}$ was observed at 295 K, comparable with other known Weyl semimetallic systems \cite{p10,p13}. It has always been a challenge to realize materials with simultaneous high value for both $\Theta_{AH}$ and $S_H$, that too at room temperature with its origin lying solely in the involved scattering mechanisms. For large $\Theta_{AH}$, the value of $\sigma_{xy}^A$ needs to be substantially large and remain unaffected for small changes in $\sigma_{xx}$. But in the hopping transport regime, where $\sigma_{xx}$ becomes small, the $\sigma_{xy}^A$ decreases at a substantially faster rate than $\sigma_{xx}$ due to quasiparticle damping \cite{p15} and hence manifest a small $\Theta_{AH}$. Interestingly, $\sigma_{xy}^A$ remains well within the estimated intrinsic-mechanism-regime and nearly unchanged with $\sigma_{xx}$ over the measured $T$-range (see Fig.~\ref{master} of SM \cite{supp}), indicating a scattering independent mechanism driven AHE in Co$_2$MnSn.

\begin{figure*}[!t]
  \centering
  \includegraphics[width=\linewidth]{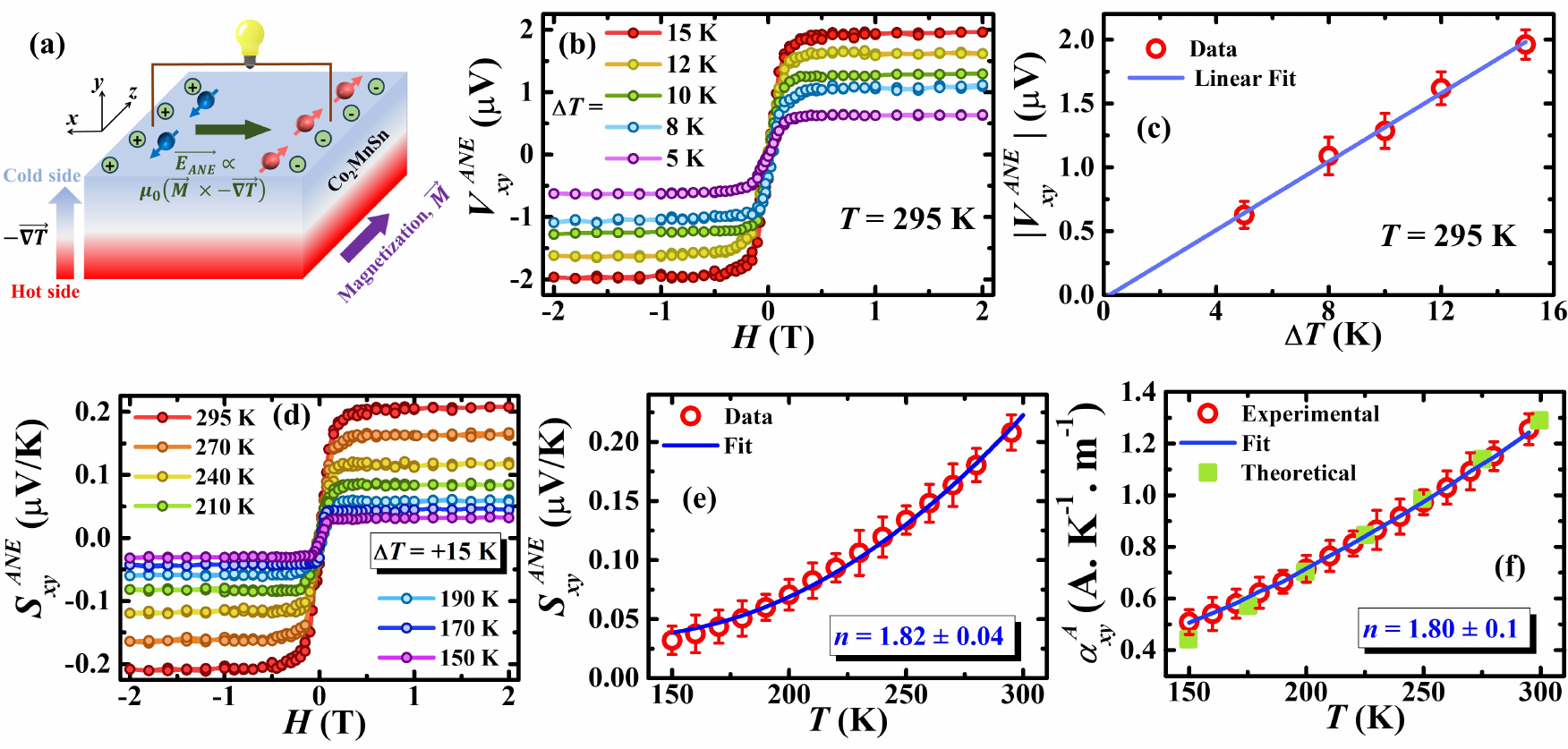}
  \caption{(a) Schematic of the experimental setup for ANE measurements using the spincaloritronic system. (b) Anomalous Nernst voltage ($V_{xy}^{ANE}$) vs. $H$  for various $\Delta T$ from +5 to +15 K at $T$ = 295 K. (c) Linear fit of background-corrected $V_{xy}^{ANE}$ (see Eq.~\eqref{eq3}) vs. $\Delta T$ (d) $H$-dependence of $S_{xy}^{ANE}$ at selected $T$ (150 $\leq T \leq$ 295 K) with $\Delta T$ = +15 K. (e) $S_{xy}^{ANE}(T)$ and (f) $\alpha_{xy}^{A}(T)$ data along with the fitted curves (see Eq.~\eqref{eq7} and Eq.~\eqref{eq8}). Theoretical $\alpha_{xy}^{A}(T)$ values align closely with experimental results (see Fig.~\ref{F3} and related discussions).}
   \label{F2}
\end{figure*}

\subsection{Transverse Magneto-Thermoelectric Transport}
The measurement geometry of the transverse magneto-thermoelectric transport is illustrated Fig.~\ref{F2}(a). Figure~\ref{F2}(b) shows the magnetic field dependence of the anomalous Nernst voltage, $V_{xy}^{ANE}(\mu_0H)$ for different values of $\Delta T = (T_{Hot} - T_{Cold})$ in the range +5 K $\leq \Delta T \leq$ +15 K, for a fixed sample temperature, $T$ = 295 K, where $T_{Hot}$ ($T_{Cold}$) being the temperature of the hot (cold) plate. Evidently, the $\left|V_{xy}^{ANE}(H)\right|$ signal strength increases with increasing $\Delta T$, consistent with the proportionality relation of ANE-induced voltage with applied $\Delta T$ across the sample \cite{Ikhlas2017, Pu2008}. The anomalous Nernst voltage $V_{xy}^{ANE}$ was anti-symmetrized at the saturation magnetic field employing the relation: 
\begin{equation}
\small{V_{xy}^{ANE}(\mu_0H_{sat}) = \frac{1}{2}\left[V_{xy}^{ANE}(+\mu_0H_{sat})\;-\;V_{xy}^{ANE}(-\mu_0H_{sat})\right]}
\label{eq3}
\end{equation}
Figure~\ref{F2}(c) displays the $\Delta T$-dependence of the background corrected anti-symmetrized $V_{xy}^{ANE}(\mu_0H_{sat})$ at $T$ = 295 K, where $\mu_0H_{sat}$ is the saturation magnetic field. A linear relation confirms the intrinsic contribution of the thermally induced ANE for Co$_2$MnSn \cite{Ramos2014, Ghosh2021}. To investigate the $T$-dependence of the ANE signal, we demonstrate the $\mu_0H$ dependence of transverse Seebeck effect $S_{xy}^{ANE}$ (in Fig.~\ref{F2}(d)) at selected average sample temperatures, $T = \left(\frac{T_{Hot} - T_{Cold}}{2} \right)$ in the range 150 K $\leq T \leq $ 295 K for a fixed value of $\Delta T$ = +15 K. The transverse Seebeck effect is defined as:
\begin{widetext}
    \begin{equation}
        S_{xy}^{ANE}(\mu_0H) = \left( \frac{V_{xy}^{ANE}(\mu_0H)}{\Delta T_{eff}}\right) \times \left(\frac{L_y}{L_x}\right) = \frac{1}{2}\left[\frac{V_{xy}^{ANE}(+\mu_0H)\,-\,V_{xy}^{ANE}(-\mu_0H)}{\Delta T_{eff}}\right] \times \left(\frac{L_y}{L_x}\right)
        \label{eq4}
    \end{equation}
\end{widetext}
where, $V_{xy}^{ANE}(\mu_0H)$ is as defined in Eq.~\eqref{eq3} but measured at a fixed temperature, $L_y$ ($\approx$ 0.5 mm) is the thickness of the sample, $L_x$ ($\approx$ 2 mm) is the distance between the contact probes for the ANE voltage measurements and $\Delta T_{eff}$ = ($T_1 - T_2$) is the effective temperature difference between the two opposite isothermal surfaces of the sample, one of which is in contact with the hot copper plate ($T_{Hot}$) and the other surface is in contact with the cold copper plate ($T_{Cold}$). Note that $T_1$ ($T_2$) is the temperature of the surface of the sample which is in contact with the hot (cold) copper plate. Since the applied temperature gradient not only drops across the sample but also at the N-grease layers at the interface between the sample surfaces and the hot/cold plates, $\Delta T \neq \Delta T_{eff}$. In our previous studies \cite{Chanda2023_1, Chanda2024}, we have shown that $\Delta T_{eff}$ can be estimated using the expression: 
\begin{equation}
\Delta T_{eff} = \frac{\Delta T}{\left[ 1 + \left(\frac{2L_{\text{N-Grease}}}{\kappa_{\text{N-Grease}}}\right)\cdot\left(\frac{\kappa}{L_S}\right) \right]}
\label{eq5}
\end{equation}
where $L_S (= L_y)$ and $\kappa$ are the thickness and the thermal conductivity of Co$_2$MnSn sample, and $L_{\text{N-Grease}}$ and $\kappa_{\text{N-Grease}}$ are the thickness and thermal conductivity of the N-grease layers \cite{De2020}. Using the reported values of the temperature dependence of $\kappa_{\text{N-Grease}}$ \cite{Ashworth} and considering  $L_{\text{N-Grease}} \approx$ 10 $\mu$m \cite{Chanda2024}, $L_{S} = L_{y} \approx 0.5$mm, we determined the temperature dependence of $\Delta T_{eff}$ for the Co$_2$MnSn sample, as shown in Fig.~\ref{figS5} of SM \cite{supp}. The temperature dependence of $\kappa(T)$ is shown in Fig.~\ref{figS4}(b) of SM \cite{supp}.The estimated value of $S_{xy}^{ANE}$ at $T$ = 295 K is $(0.208\pm0.02)$ $\mu$V~K$^{-1}$, while its $T$-variation at $\mu_0H$ = 2 T is shown in Fig.~\ref{F2}(e). It is evident from Fig.~\ref{F2}(e) that $S_{xy}^{ANE}$ decreases nonlinearly with decreasing $T$. Now let us understand the origin of the observed behavior of $S_{xy}^{ANE}(T)$ in Co$_2$MnSn. 

According to the Cartesian coordinate system defined for our ANE measurements (see Fig. \ref{F2}(a)), the anomalous Nernst coefficient, $S_{xy}^{ANE}$ can be further expressed as \cite{Breidenbach2022, Ramos2014, Wang2024, Ghosh2019}: 
\begin{equation}
S_{xy}^{ANE} = \left( \rho_{xx}\,\alpha_{xy}^A + \rho_{xy}^{A}\,\alpha_{xx} \right) = \left[ \rho_{xx}\,\alpha_{xy}^A + \left(\frac{\rho_{xy}^{A}}{\rho_{xx}}\right)\,S_{xx} \right]
\label{eq6}
\end{equation}
where $\alpha_{xy}^A$, $S_{xx}= \rho_{xx}\alpha_{xx}$ and $\alpha_{xx}$ are the anomalous off-diagonal transverse thermoelectric conductivity (Peltier conductivity), longitudinal Seebeck coefficient and thermoelectric conductivity, respectively. Now, the resistivity tensor in three dimensions can be expressed as \cite{Li2005}:
\begin{equation*}
\tilde{\rho} =
\left[\begin{smallmatrix}
    \rho_{xx} & \rho_{xy} & \rho_{xz} \\
    \rho_{yx} & \rho_{yy} & \rho_{yz} \\
    \rho_{zx} & \rho_{zy} & \rho_{zz}
\end{smallmatrix}\right]
\end{equation*}
For an isotropic polycrystalline material, we can assume that $\rho_{xx}= \rho_{yy}= \rho_{zz}=\rho=$~longitudinal resistivity and $\rho_{xy}= \rho_{yz}= \rho_{zx}=$~Hall resistivity. Furthermore, in the case of an isotropic polycrystalline material, we can consider that $S_{xx}=S_{yy}= S_{zz}=S=$~longitudinal Seebeck coefficient. Considering these approximations, the anomalous Nernst coefficient for polycrystalline Co$_2$MnSn sample can be written as \cite{Ramos2014}: 
\begin{equation*}
S_{xy}^{ANE} = \left[\rho\,\alpha_{xy}^A + S\,\left(\frac{\rho_{xy}^A}{\rho}\right)\right]
\end{equation*}
where $\rho_{xy}^A$ is the isotropic anomalous Hall resistivity.
Now, considering the tensorial relation of $\sigma_{xy}^A$ and $\rho_{xy}^A$ as mentioned previously \cite{Pu2008, Guin2019, Xu2019}, $S_{xy}^{ANE}$ transforms to:
\begin{widetext}
    \begin{equation*}
    S_{xy}^{ANE} = \left[\rho\,\alpha_{xy}^A + S\,\left(\frac{\rho_{xy}^{A}}{\rho}\right)\right] = \left[\rho\,\alpha_{xy}^A - S\,\rho\,\left(\frac{-\rho_{xy}^A}{\rho^2}\right)\right] = \rho\,\left[\alpha_{xy}^A - S\,\sigma_{xy}^{A}\right]
    \end{equation*}
\end{widetext}
Now, following the Mott’s relations, $\alpha_{xy}^A$ and $S$ can be written as \cite{Pu2008, Mott}:
\begin{eqnarray*}
\alpha_{xy}^A &=& -\frac{\pi^2 k_B^2 T}{3e}\left(\frac{\partial\sigma_{xy}^{A}}{\partial E}\right)_{E=E_F} \\
S &=& \frac{\pi^2 k_B^2 T}{3e\sigma}\left(\frac{\partial\sigma}{\partial E}\right)_{E=E_F}
\end{eqnarray*}
Note that, $\rho_{xy}^{A}$ depends on $\rho$ through the power law as, $\rho_{xy}^{A} = \lambda\rho^n$, where $\lambda$ is the spin-orbit coupling constant and $n$ being equivalent to $\beta$, distinguishing the dominant extrinsic to intrinsic Berry-curvature-driven mechanism contribution to the AHE and ANE \cite{Nagaosa2010}. Considering the power law for AHE, $\alpha_{xy}^A$ and $S$ can be expressed as:
\begin{eqnarray*}
\alpha_{xy}^A &=& -\frac{\pi^2 k_B^2 T}{3e}\left[\frac{\partial}{\partial E}\{-\lambda\rho^{n-2}\}\right]_{E=E_F} \\
S &=& -\frac{\pi^2 k_B^2 T}{3e\rho}\left(\frac{\partial\rho}{\partial E}\right)_{E=E_F}
\end{eqnarray*}
Therefore, combining and substituting the aforementioned expressions in Eq.~\eqref{eq6}, $S_{xy}^{ANE}(T)$ can be written as \cite{Pu2008, Ramos2014}:
\begin{equation}
    \small{{S}_{xy}^{ANE}(T) = \rho^{n-1} \left[\frac{\pi^2 k_B^2 T}{3e}\left(\frac{\partial\lambda}{\partial E}\right)_{E=E_F}\!\!\!\!\! - (n-1)\,\lambda\,S\,\right]}
    \label{eq7}
\end{equation}
where $\rho$, $S$ and $\lambda$ are the longitudinal resistivity, longitudinal Seebeck coefficient and spin-orbit coupling constant respectively and $n$ (similar to $\beta$) dictates the type of mechanism(s) dominating the AHE and ANE \cite{Nagaosa2010}. The $T$-dependence of  $S(T)$ for Co$_2$MnSn is shown in Fig.~\ref{figS4}(a) of SM \cite{supp}. We fitted the $S_{xy}^{ANE}(T)$ data for Co$_2$MnSn using Eq.~\eqref{eq7}, considering $\lambda$, $\left(\frac{\partial\lambda}{\partial E}\right)_{E=E_F}$ and $n$ as the fitting parameters.  The best fit was obtained for $n = (1.82 \pm 0.04)$, indicating the dominance of intrinsic Berry curvature mechanism \cite{Chanda2022_1, Nagaosa2010} to ANE. 
To further verify the origin of ANE, we have estimated and analyzed the $\alpha_{xy}^A(T)$ data, as shown in Fig.~\ref{F2}(f). Since $S_{xy}^{ANE}= \rho\,[\alpha_{xy}^A - S\,\sigma_{xy}^{A}]$, we can express $\alpha_{xy}^A$ as \cite{Wang2024}, $\alpha_{xy}^A=(\sigma\,S_{xy}^{ANE} + S\,\sigma_{xy}^{A})$. Using this expression, we have estimated $\alpha_{xy}^A(T)$ for Co$_2$MnSn. The estimated value of $\alpha_{xy}^A$ at $T$ = 295 K is $(1.25\pm0.06)$ A~m$^{-1}$~K$^{-1}$.
Following the Mott’s relations, $\alpha_{xy}^A$ can further be expressed as \cite{Pu2008, Ramos2014, Ding2019},
\begin{equation}
   \small{{\alpha}_{xy}^A(T) = \rho^{n-2} \left[\frac{\pi^2 k_B^2 T}{3e}\left(\frac{\partial\lambda}{\partial E}\right)_{E=E_F}\!\!\!\!\! - (n-2)\,\lambda\,S\,\right]}
    \label{eq8}
\end{equation}
Figure~\ref{F2}(f) shows the fitting of $\alpha_{xy}^A(T)$ vs. $T$ data  using Eq.~\eqref{eq8}, which yields $n=(1.80 \pm 0.1)$, further confirming the dominance of intrinsic Berry curvature \cite{Nagaosa2010} to ANE.

\subsection{Electronic Structure}
\begin{figure*}[!t]
    \centering
    \includegraphics[width=\linewidth]{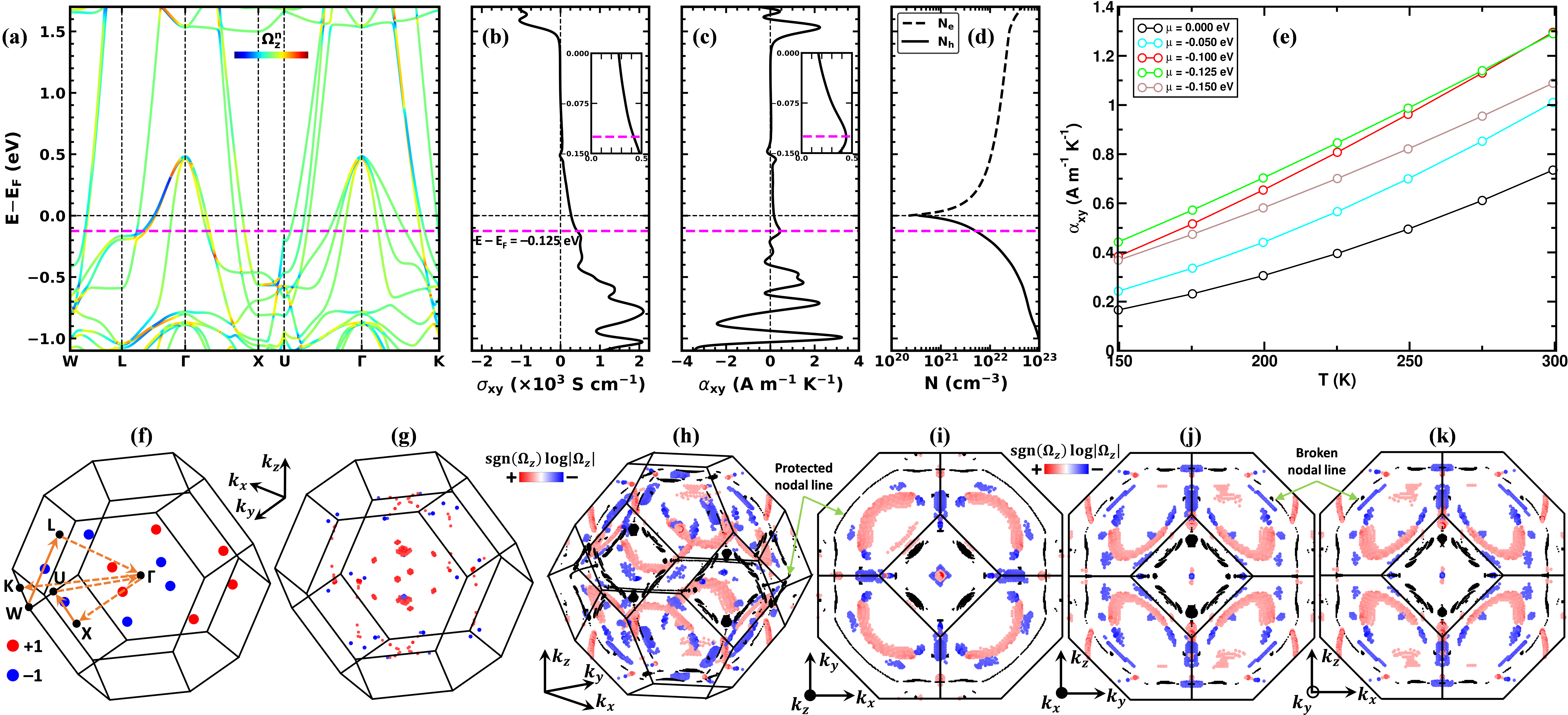}
    \caption{Electronic band structure of Co$_2$MnSn with SOC, overlaid with band-resolved z-component of Berry curvature, $\Omega_z^n(\vec{k})$; red/blue color indicates large positive/negative values, with n being the band index. (b) Anomalous Hall conductivity ($\sigma_{xy}$) and (c) anomalous Nernst conductivity ($\alpha_{xy}$) vs. $\mu$ at $T = 150$ K; insets show closer view in the energy range [-0.15,0] eV. (d) Carrier concentration ($N$) vs. $\mu$ at 150 K, with solid (dashed) lines showing p-type (n-type) carriers. (e) $T$-dependence of $\alpha_{xy}$ for various $\mu$. (f) Weyl node distribution in the bulk Brillouin zone (BZ), marked by high symmetry paths; red (blue) nodes indicate chirality $+1$ ($-1$), all within the $k_z = 0$ plane. Distribution of $\Omega_z(\mu)$  in the BZ at (g) $\mu$ = -0.125 eV and (h) $\mu$ lying in $[-0.65,-0.55]$ eV, overlaid with nodal lines at 150 K. Projection of same data along (i) $k_z$, (j) $k_x$ and (k) $k_y$ axes. In (g-k), only points with $\Omega_z \geq 150$\AA$^2$ are displayed for clarity. The colour scale is set to $\log_{10}|\Omega_z|$ for enhanced contrast.}
    \label{F3}
\end{figure*}

To investigate the origin of anomalous transport, Co$_2$MnSn was simulated in the FM phase using a theoretically optimized lattice parameter $a = 5.98$~\AA, just 0.17\% smaller than the measured value. The spin-polarized band structure is shown in Fig.~\ref{figS6} of SM \cite{supp}, while the band structure with spin-orbit coupling (SOC) is shown in Fig.\ref{F3}(a). With SOC, the simulated magnetic moments at Co, Mn, and Sn are 1.0, 3.45, and -0.13 $\mu_B$, respectively, totalling 5.32 $\mu_B$. The band structure exhibits several SOC-induced band splittings, especially near \textit{E}$_\mathrm{F}$ and down to -0.65 eV, leading to avoided crossings yielding finite Berry curvature in momentum space. The $z$-component of band-projected Berry curvature $\Omega_z^n(\vec{k})$ is overlaid on the band structure in Fig.~\ref{F3}(a). Notably, only a handful number of avoided crossings appear near \textit{E}$_\mathrm{F}$ in an energy range $\sim[-0.15,0]$ eV, resulting in relatively less contribution of $\Omega_z^n(\vec{k})$ in this energy range (except along $\overline{\text{L}\Gamma}$ direction which is a consequence of magnetic order breaking the rotational symmetry about the 3-fold axis). It is also observed that unlike Co$_2$MnGa, which hosts nodal line near \textit{E}$_\mathrm{F}$ \cite{Sakai2018, Guin2019, Sumida2020}, the nodal line in Co$_2$MnSn appears far below \textit{E}$_\mathrm{F}$ in the energy range $E-E_F\sim[-0.65,-0.55]$ eV (along the  $\overline{\text{WL}}\,,\overline{\text{XU}}$ and $\overline{\Gamma\text{X}}$ directions in Fig.~\ref{F3}(a)) and consequently are not expected to play a dominant role in anomalous transport near \textit{E}$_\mathrm{F}$. Despite this, Co$_2$MnSn shows a sizable AHC, $\sigma_{xy}^A=\,272.28$ S cm$^{-1}$ at \textit{E}$_\mathrm{F}$, arising from non-zero intrinsic $\Omega_z$. While this value is moderate compared to Co$_2$MnGa, it remains substantial. Additionally, we have computed anomalous Nernst conductivity (ANC) at $T=\,150$ K using calculated AHC data \cite{Xiao2006}. As expected, ANC ($\alpha_{xy}$) shows a modest value of $0.166$ A m$^{-1}$ K$^{-1}$ at \textit{E}$_\mathrm{F}$ ($\mu=0$ eV), consistent with the presence of a nonzero $\Omega_z$. To better understand this, the variation of $\sigma_{xy}$ and $\alpha_{xy}$ with chemical potentials ($\mu\,=\,E-E_F$) are shown in Figs.~\ref{F3}(b) and (c). Notably, at $\mu=\,-0.125$, $\sigma_{xy}$ attains significantly high value of 423.24 S cm$^{-1}$ (see inset of Fig.~\ref{F3}(b)). Moreover, $\alpha_{xy}$ attains a pronounced peak at the same $\mu$, reaching a maximum value of 0.44 A m$^{-1}$ K$^{-1}$, as shown in Fig.\,\ref{F3}(c) inset. The calculated values of $\sigma_{xy}$ and $\alpha_{xy}$ for $\mu=\,-0.125$ eV are in excellent agreement with our experimental observations at $T=150$ K, validating the robustness of our calculations. At lower $\mu$, the AHC continues to increase significantly, and a similar trend is observed in ANC, as illustrated in Fig.~\ref{F3}(b) and \ref{F3}(c). To gain further insights, we calculated the carrier concentration $N$ as a function of $\mu$ as shown in Fig.~\ref{F3}(d), where $N_e$ ($N_h$) indicates the electron (hole) concentration. Our calculations reveal an intrinsic carrier concentration of $N_0=3.67\times10^{20}$ cm$^{-3}$ at $\mu=0$ eV. Remarkably, at $\mu=-0.125$ eV, where both $\sigma_{xy}$ and $\alpha_{xy}$ reaches a high value, the hole concentration is $N_h\sim5.12\times10^{21}$ cm$^{-3}$, close to the measured value. This downward shift of $\mu$ relative to E$_F$ corroborates the experimental evidence indicating that the hole-type carriers are predominant in Co$_2$MnSn, consistent with the sign of the Hall resistivity (Fig.~\ref{F1}(d)).

While the preceding analysis focuses on a chosen temperature $T=150$ K, exploring the $T$-dependence of anomalous transport is essential for potential spintronics application. Figure~\ref{F3}(e) illustrates the $T$-dependence of $\alpha_{xy}$ in the range $150-300$ K for different $\mu$. The ANC increases monotonically with $T$, closely following the experimental trend, as in Fig.~\ref{F2}(f). The best match and the highest ANC value obtained is $\alpha_{xy}=1.29$ A m$^{-1}$ K$^{-1}$  at $\mu=-0.125$ eV and $T=300$ K, in excellent agreement with our measured value at 295 K. In contrast, the AHC values do not vary much with $T$ (see Fig.~\ref{figS7} of SM \cite{supp}), which is also observed experimentally (see Fig.~\ref{F1}(g)).

The fact that there is non-zero $\Omega_z$ near \textit{E}$_\mathrm{F}$ leading to high AHC and ANC prompted us to search for topologically non-trivial nodal points in the vicinity of \textit{E}$_\mathrm{F}$. Given that Co$_2$MnSn is centrosymmetric but lacks time reversal ($\mathcal{T}$) symmetry, we anticipate presence of Weyl nodes. A distribution of Weyl nodes in the first Brillouin zone (BZ) within the energy range $E-E_\mathrm{F}\sim[-0.15,0]$ eV is shown in Fig.~\ref{F3}(f). We find 12 Weyl nodes, all located on the $k_z=0$ plane, with their respective chiralities marked in red ($+1$) and blue ($-1$). Since both the AHC and ANC reach a high value at $\mu =-0.125$ eV, we further examine the distribution of total Berry curvature at the same energy range in Fig.~\ref{F3}(g). The Berry curvature hotspots in Fig.~\ref{F3}(g) appear due to the avoided crossings and Weyl nodes, reinforcing their role in the anomalous Hall transport. Additionally, the nodal line in Co$_2$MnSn appears within the energy window $E-E_\mathrm{F}\sim[-0.65,-0.55]$ eV. Figures~\ref{F3}(h-k) illustrates nodal line structure and distribution of $z$-component of Berry curvature within the BZ for the energy window $E-E_F\sim[-0.65,-0.55]$ eV. In Fig.~\ref{F3}(h), a perspective view of the BZ reveals nodal line features (black continuous dots) formed from points on the 3D Fermi surface. To enhance visualization, all points with an energy gap $\leq 5$ meV are displayed. Similar to Co$_2$MnGa \cite{Sakai2018, Guin2019, Sumida2020}, Co$_2$MnSn hosts three nodal lines, one on each mirror plane ($k_x$-$k_y$, $k_y$-$k_z$ and $k_z$-$k_x$). For magnetization along $[001]$, the nodal line on the $k_x$-$k_y$ plane remains protected (see Fig.~\ref{F3}(i)), while those on the $k_y$-$k_z$ and $k_x$-$k_z$ planes are broken (see Figs.~\ref{F3}(j) and \ref{F3}(k)). Such breaking of nodal line leads to the substantial accumulation of Berry curvature, further enhancing AHC and ANC within the considered energy window. Consequently, we observe remarkably large magnitude of AHC and ANC, with $\sigma_{xy}^{nodal}=1376.56$ S cm$^{-1}$ and $\alpha_{xy}^{nodal}=1.49$ A m$^{-1}$ K$^{-1}$, respectively, in the chemical potential range $\mu=[-0.65,-0.55]$ eV (see Figs. \ref{F3}(b) and \ref{F3}(c)). These values are comparable to those found in other topological magnets, including Co$_2$MnGa \cite{Sakai2018,Guin2019,Sumida2020}, further underscoring the crucial role of Berry curvature in shaping the anomalous transport properties in Co$_2$MnSn.

\begin{figure*}[!t]
  \centering
  \includegraphics[width=0.98\linewidth]{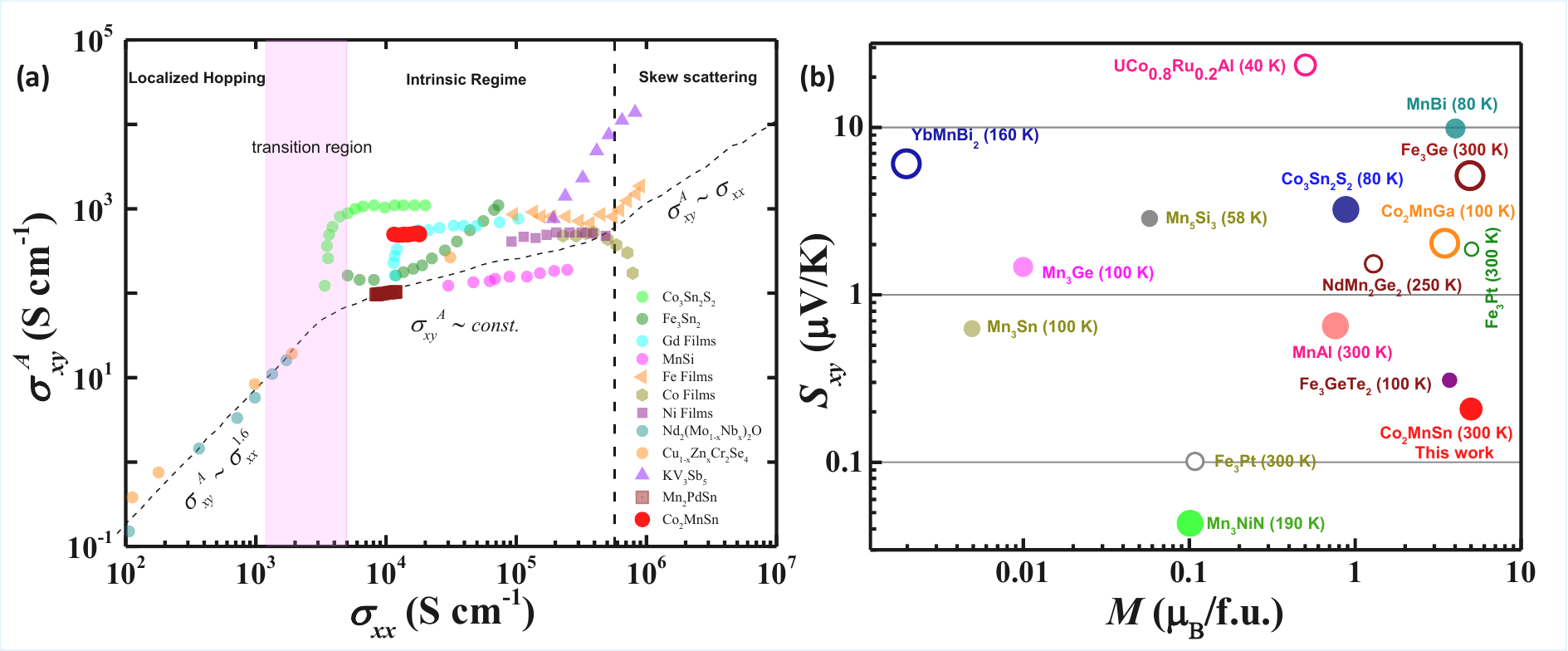}
  \caption{(a) Universal plot of anomalous Hall conductivity ($\sigma_{xy}^A$) as a function of longitudinal conductivity ($\sigma_{xx}$). The red curve depicts our experimental result for Co$_2$MnSn, along with other materials such as MnSi \cite{p23,p24,p21,p13,p25}. For more details, see Fig.~\ref{master} and related discussions in SM \cite{supp}. (b) The highest magnitude of ANE, $S_{xy}$, for different materials is represented as a function of their magnetization \cite{Sakai2018, sakai2020iron, asaba2021colossal, badura2025observation}. }
   \label{F4}
\end{figure*}

\section{Conclusion}
In summary, we present a combined experimental and theoretical analysis of anomalous transport in the ferromagnetic full Heusler alloy Co$_2$MnSn. Figures \ref{F4}(a) and \ref{F4}(b) compare the magnitude of anomalous Hall conductivity and anomalous Nernst effect to longitudinal conductivity and magnetisation per formula unit, respectively. The AHC lies well within the intrinsic limit for Co$_2$MnSn, while the ANE coefficient showcases a considerable value despite being constituted of light elements. {\it Ab-initio} calculations show Co$_2$MnSn hosts topologically non-trivial features like nodal lines and Weyl points, generating significant Berry curvature near the Fermi level. This underpins the intrinsic contributions to the anomalous Hall and Nernst effects. Experimental data align with theory, revealing sizable anomalous conductivities that persist all the way up to room temperature, underscoring their robustness. These results establish Co$_2$MnSn as a promising candidate for unconventional transport needed for room-temperature applications. Furthermore, we emphasize the potential of tuning electronic topology—via chemical substitution or $E_F$ adjustment to enhance anomalous electrical and thermal responses. The rich magnetism-topology interplay in Co$_2$MnSn makes it a promising platform for both fundamental research as well as future spin-caloritronic technologies.

\begin{acknowledgements}
    BD thanks IIT Bombay and Ministry of Education, Govt. of India, for financial support. AB acknowledges SINP, Kolkata, and the Department of Atomic Energy, Govt. of India, for financial support. HS and AC acknowledge support from the U.S. Department of Energy, Office of Basic Energy Sciences, Division of Materials Science and Engineering (Award No. DE-FG02-07ER46438) for magnetotransport, ANE measurements and analysis performed at USF. HS also acknowledges IIT Bombay for a short-term visiting professorship.
\end{acknowledgements}

\bibliography{main}

\begin{thebibliography}{10}%
\makeatletter
\providecommand \@ifxundefined [1]{%
 \@ifx{#1\undefined}
}%
\providecommand \@ifnum [1]{%
 \ifnum #1\expandafter \@firstoftwo
 \else \expandafter \@secondoftwo
 \fi
}%
\providecommand \@ifx [1]{%
 \ifx #1\expandafter \@firstoftwo
 \else \expandafter \@secondoftwo
 \fi
}%
\providecommand \natexlab [1]{#1}%
\providecommand \enquote  [1]{``#1''}%
\providecommand \bibnamefont  [1]{#1}%
\providecommand \bibfnamefont [1]{#1}%
\providecommand \citenamefont [1]{#1}%
\providecommand \href@noop [0]{\@secondoftwo}%
\providecommand \href [0]{\begingroup \@sanitize@url \@href}%
\providecommand \@href[1]{\@@startlink{#1}\@@href}%
\providecommand \@@href[1]{\endgroup#1\@@endlink}%
\providecommand \@sanitize@url [0]{\catcode `\\12\catcode `\$12\catcode `\&12\catcode `\#12\catcode `\^12\catcode `\_12\catcode `\%12\relax}%
\providecommand \@@startlink[1]{}%
\providecommand \@@endlink[0]{}%
\providecommand \url  [0]{\begingroup\@sanitize@url \@url }%
\providecommand \@url [1]{\endgroup\@href {#1}{\urlprefix }}%
\providecommand \urlprefix  [0]{URL }%
\providecommand \Eprint [0]{\href }%
\providecommand \doibase [0]{https://doi.org/}%
\providecommand \selectlanguage [0]{\@gobble}%
\providecommand \bibinfo  [0]{\@secondoftwo}%
\providecommand \bibfield  [0]{\@secondoftwo}%
\providecommand \translation [1]{[#1]}%
\providecommand \BibitemOpen [0]{}%
\providecommand \bibitemStop [0]{}%
\providecommand \bibitemNoStop [0]{.\EOS\space}%
\providecommand \EOS [0]{\spacefactor3000\relax}%
\providecommand \BibitemShut  [1]{\csname bibitem#1\endcsname}%
\let\auto@bib@innerbib\@empty
\bibitem [{\citenamefont {Lee}\ \emph {et~al.}(2007)\citenamefont {Lee}, \citenamefont {Onose}, \citenamefont {Tokura},\ and\ \citenamefont {Ong}}]{Lee2007_sm}%
  \BibitemOpen
  \bibfield  {author} {\bibinfo {author} {\bibfnamefont {M.}~\bibnamefont {Lee}}, \bibinfo {author} {\bibfnamefont {Y.}~\bibnamefont {Onose}}, \bibinfo {author} {\bibfnamefont {Y.}~\bibnamefont {Tokura}},\ and\ \bibinfo {author} {\bibfnamefont {N.~P.}\ \bibnamefont {Ong}},\ }\href {https://doi.org/10.1103/PhysRevB.75.172403} {\bibfield  {journal} {\bibinfo  {journal} {Phys. Rev. B}\ }\textbf {\bibinfo {volume} {75}},\ \bibinfo {pages} {172403} (\bibinfo {year} {2007})}\BibitemShut {NoStop}%
\bibitem [{\citenamefont {Bhattacharya}\ \emph {et~al.}(2024)\citenamefont {Bhattacharya}, \citenamefont {Habib}, \citenamefont {Ahmed}, \citenamefont {Satpati}, \citenamefont {DuttaGupta}, \citenamefont {Dasgupta},\ and\ \citenamefont {Das}}]{Bhattacharya2024_sm}%
  \BibitemOpen
  \bibfield  {author} {\bibinfo {author} {\bibfnamefont {A.}~\bibnamefont {Bhattacharya}}, \bibinfo {author} {\bibfnamefont {M.~R.}\ \bibnamefont {Habib}}, \bibinfo {author} {\bibfnamefont {A.}~\bibnamefont {Ahmed}}, \bibinfo {author} {\bibfnamefont {B.}~\bibnamefont {Satpati}}, \bibinfo {author} {\bibfnamefont {S.}~\bibnamefont {DuttaGupta}}, \bibinfo {author} {\bibfnamefont {I.}~\bibnamefont {Dasgupta}},\ and\ \bibinfo {author} {\bibfnamefont {I.}~\bibnamefont {Das}},\ }\href {https://doi.org/10.1103/PhysRevB.110.014417} {\bibfield  {journal} {\bibinfo  {journal} {Phys. Rev. B}\ }\textbf {\bibinfo {volume} {110}},\ \bibinfo {pages} {014417} (\bibinfo {year} {2024})}\BibitemShut {NoStop}%
\bibitem [{\citenamefont {Ye}\ \emph {et~al.}(2018)\citenamefont {Ye}, \citenamefont {Kang}, \citenamefont {Liu}, \citenamefont {Von~Cube}, \citenamefont {Wicker}, \citenamefont {Suzuki}, \citenamefont {Jozwiak}, \citenamefont {Bostwick}, \citenamefont {Rotenberg}, \citenamefont {Bell}, \citenamefont {Fu}, \citenamefont {Comin},\ and\ \citenamefont {Checkelsky}}]{Ye2018_sm}%
  \BibitemOpen
  \bibfield  {author} {\bibinfo {author} {\bibfnamefont {L.}~\bibnamefont {Ye}}, \bibinfo {author} {\bibfnamefont {M.}~\bibnamefont {Kang}}, \bibinfo {author} {\bibfnamefont {J.}~\bibnamefont {Liu}}, \bibinfo {author} {\bibfnamefont {F.}~\bibnamefont {Von~Cube}}, \bibinfo {author} {\bibfnamefont {C.~R.}\ \bibnamefont {Wicker}}, \bibinfo {author} {\bibfnamefont {T.}~\bibnamefont {Suzuki}}, \bibinfo {author} {\bibfnamefont {C.}~\bibnamefont {Jozwiak}}, \bibinfo {author} {\bibfnamefont {A.}~\bibnamefont {Bostwick}}, \bibinfo {author} {\bibfnamefont {E.}~\bibnamefont {Rotenberg}}, \bibinfo {author} {\bibfnamefont {D.~C.}\ \bibnamefont {Bell}}, \bibinfo {author} {\bibfnamefont {L.}~\bibnamefont {Fu}}, \bibinfo {author} {\bibfnamefont {R.}~\bibnamefont {Comin}},\ and\ \bibinfo {author} {\bibfnamefont {J.~G.}\ \bibnamefont {Checkelsky}},\ }\href {https://doi.org/10.1038/nature25987} {\bibfield  {journal} {\bibinfo  {journal} {Nature}\ }\textbf {\bibinfo {volume} {555}},\ \bibinfo {pages} {638} (\bibinfo {year}
  {2018})}\BibitemShut {NoStop}%
\bibitem [{\citenamefont {Wang}\ \emph {et~al.}(2018)\citenamefont {Wang}, \citenamefont {Xu}, \citenamefont {Lou}, \citenamefont {Liu}, \citenamefont {Li}, \citenamefont {Huang}, \citenamefont {Shen}, \citenamefont {Weng}, \citenamefont {Wang},\ and\ \citenamefont {Lei}}]{Wang2018_sm}%
  \BibitemOpen
  \bibfield  {author} {\bibinfo {author} {\bibfnamefont {Q.}~\bibnamefont {Wang}}, \bibinfo {author} {\bibfnamefont {Y.}~\bibnamefont {Xu}}, \bibinfo {author} {\bibfnamefont {R.}~\bibnamefont {Lou}}, \bibinfo {author} {\bibfnamefont {Z.}~\bibnamefont {Liu}}, \bibinfo {author} {\bibfnamefont {M.}~\bibnamefont {Li}}, \bibinfo {author} {\bibfnamefont {Y.}~\bibnamefont {Huang}}, \bibinfo {author} {\bibfnamefont {D.}~\bibnamefont {Shen}}, \bibinfo {author} {\bibfnamefont {H.}~\bibnamefont {Weng}}, \bibinfo {author} {\bibfnamefont {S.}~\bibnamefont {Wang}},\ and\ \bibinfo {author} {\bibfnamefont {H.}~\bibnamefont {Lei}},\ }\href {https://doi.org/10.1038/s41467-018-06088-2} {\bibfield  {journal} {\bibinfo  {journal} {Nat. Commun.}\ }\textbf {\bibinfo {volume} {9}},\ \bibinfo {pages} {3681} (\bibinfo {year} {2018})}\BibitemShut {NoStop}%
\bibitem [{\citenamefont {Miyasato}\ \emph {et~al.}(2007)\citenamefont {Miyasato}, \citenamefont {Abe}, \citenamefont {Fujii}, \citenamefont {Asamitsu}, \citenamefont {Onoda}, \citenamefont {Onose}, \citenamefont {Nagaosa},\ and\ \citenamefont {Tokura}}]{Miyasato2007_sm}%
  \BibitemOpen
  \bibfield  {author} {\bibinfo {author} {\bibfnamefont {T.}~\bibnamefont {Miyasato}}, \bibinfo {author} {\bibfnamefont {N.}~\bibnamefont {Abe}}, \bibinfo {author} {\bibfnamefont {T.}~\bibnamefont {Fujii}}, \bibinfo {author} {\bibfnamefont {A.}~\bibnamefont {Asamitsu}}, \bibinfo {author} {\bibfnamefont {S.}~\bibnamefont {Onoda}}, \bibinfo {author} {\bibfnamefont {Y.}~\bibnamefont {Onose}}, \bibinfo {author} {\bibfnamefont {N.}~\bibnamefont {Nagaosa}},\ and\ \bibinfo {author} {\bibfnamefont {Y.}~\bibnamefont {Tokura}},\ }\href {https://doi.org/10.1103/PhysRevLett.99.086602} {\bibfield  {journal} {\bibinfo  {journal} {Phys. Rev. Lett.}\ }\textbf {\bibinfo {volume} {99}},\ \bibinfo {pages} {086602} (\bibinfo {year} {2007})}\BibitemShut {NoStop}%
\bibitem [{\citenamefont {Onoda}\ \emph {et~al.}(2006)\citenamefont {Onoda}, \citenamefont {Sugimoto},\ and\ \citenamefont {Nagaosa}}]{Onoda2006_sm}%
  \BibitemOpen
  \bibfield  {author} {\bibinfo {author} {\bibfnamefont {S.}~\bibnamefont {Onoda}}, \bibinfo {author} {\bibfnamefont {N.}~\bibnamefont {Sugimoto}},\ and\ \bibinfo {author} {\bibfnamefont {N.}~\bibnamefont {Nagaosa}},\ }\href {https://doi.org/10.1103/PhysRevLett.97.126602} {\bibfield  {journal} {\bibinfo  {journal} {Phys. Rev. Lett.}\ }\textbf {\bibinfo {volume} {97}},\ \bibinfo {pages} {126602} (\bibinfo {year} {2006})}\BibitemShut {NoStop}%
\bibitem [{\citenamefont {Guin}\ \emph {et~al.}(2019)\citenamefont {Guin}, \citenamefont {Manna}, \citenamefont {Noky}, \citenamefont {Watzman}, \citenamefont {Fu}, \citenamefont {Kumar}, \citenamefont {Schnelle}, \citenamefont {Shekhar}, \citenamefont {Sun}, \citenamefont {Gooth},\ and\ \citenamefont {Felser}}]{Guin2019_sm}%
  \BibitemOpen
  \bibfield  {author} {\bibinfo {author} {\bibfnamefont {S.~N.}\ \bibnamefont {Guin}}, \bibinfo {author} {\bibfnamefont {K.}~\bibnamefont {Manna}}, \bibinfo {author} {\bibfnamefont {J.}~\bibnamefont {Noky}}, \bibinfo {author} {\bibfnamefont {S.~J.}\ \bibnamefont {Watzman}}, \bibinfo {author} {\bibfnamefont {C.}~\bibnamefont {Fu}}, \bibinfo {author} {\bibfnamefont {N.}~\bibnamefont {Kumar}}, \bibinfo {author} {\bibfnamefont {W.}~\bibnamefont {Schnelle}}, \bibinfo {author} {\bibfnamefont {C.}~\bibnamefont {Shekhar}}, \bibinfo {author} {\bibfnamefont {Y.}~\bibnamefont {Sun}}, \bibinfo {author} {\bibfnamefont {J.}~\bibnamefont {Gooth}},\ and\ \bibinfo {author} {\bibfnamefont {C.}~\bibnamefont {Felser}},\ }\href {https://doi.org/10.1038/s41427-019-0116-z} {\bibfield  {journal} {\bibinfo  {journal} {NPG Asia Mater.}\ }\textbf {\bibinfo {volume} {11}},\ \bibinfo {pages} {16} (\bibinfo {year} {2019})}\BibitemShut {NoStop}%
\bibitem [{\citenamefont {Wang}\ \emph {et~al.}(2022)\citenamefont {Wang}, \citenamefont {Lau}, \citenamefont {Zhou}, \citenamefont {Seki}, \citenamefont {Sakuraba}, \citenamefont {Kubota}, \citenamefont {Ito},\ and\ \citenamefont {Takanashi}}]{Wang2022_sm}%
  \BibitemOpen
  \bibfield  {author} {\bibinfo {author} {\bibfnamefont {J.}~\bibnamefont {Wang}}, \bibinfo {author} {\bibfnamefont {Y.-C.}\ \bibnamefont {Lau}}, \bibinfo {author} {\bibfnamefont {W.}~\bibnamefont {Zhou}}, \bibinfo {author} {\bibfnamefont {T.}~\bibnamefont {Seki}}, \bibinfo {author} {\bibfnamefont {Y.}~\bibnamefont {Sakuraba}}, \bibinfo {author} {\bibfnamefont {T.}~\bibnamefont {Kubota}}, \bibinfo {author} {\bibfnamefont {K.}~\bibnamefont {Ito}},\ and\ \bibinfo {author} {\bibfnamefont {K.}~\bibnamefont {Takanashi}},\ }\href {https://doi.org/https://doi.org/10.1002/aelm.202101380} {\bibfield  {journal} {\bibinfo  {journal} {Adv. Electron. Mater.}\ }\textbf {\bibinfo {volume} {8}},\ \bibinfo {pages} {2101380} (\bibinfo {year} {2022})}\BibitemShut {NoStop}%
\bibitem [{\citenamefont {{Ashworth, T. and Loomer, J. E. and Kreitman, M. M.}}(1973)}]{Ashworth_sm}%
  \BibitemOpen
  \bibfield  {author} {\bibinfo {author} {\bibnamefont {{Ashworth, T. and Loomer, J. E. and Kreitman, M. M.}}},\ }in\ \href {https://doi.org/10.1007/978-1-4684-3111-7_31} {\emph {\bibinfo {booktitle} {{Advances in Cryogenic Engineering}}}},\ \bibinfo {editor} {edited by\ \bibinfo {editor} {\bibnamefont {{Timmerhaus, K. D.}}}}\ (\bibinfo  {publisher} {{Springer US}},\ \bibinfo {address} {{Boston, MA}},\ \bibinfo {year} {{1973}})\ pp.\ \bibinfo {pages} {{271--279}}\BibitemShut {NoStop}%
\bibitem [{\citenamefont {Sun}\ \emph {et~al.}(2015)\citenamefont {Sun}, \citenamefont {Ruzsinszky},\ and\ \citenamefont {Perdew}}]{Sun2015_sm}%
  \BibitemOpen
  \bibfield  {author} {\bibinfo {author} {\bibfnamefont {J.}~\bibnamefont {Sun}}, \bibinfo {author} {\bibfnamefont {A.}~\bibnamefont {Ruzsinszky}},\ and\ \bibinfo {author} {\bibfnamefont {J.~P.}\ \bibnamefont {Perdew}},\ }\href {https://doi.org/10.1103/PhysRevLett.115.036402} {\bibfield  {journal} {\bibinfo  {journal} {Phys. Rev. Lett.}\ }\textbf {\bibinfo {volume} {115}},\ \bibinfo {pages} {036402} (\bibinfo {year} {2015})}\BibitemShut {NoStop}%
\end{thebibliography}%


\begin{thebibliography}{75}%
\makeatletter
\providecommand \@ifxundefined [1]{%
 \@ifx{#1\undefined}
}%
\providecommand \@ifnum [1]{%
 \ifnum #1\expandafter \@firstoftwo
 \else \expandafter \@secondoftwo
 \fi
}%
\providecommand \@ifx [1]{%
 \ifx #1\expandafter \@firstoftwo
 \else \expandafter \@secondoftwo
 \fi
}%
\providecommand \natexlab [1]{#1}%
\providecommand \enquote  [1]{``#1''}%
\providecommand \bibnamefont  [1]{#1}%
\providecommand \bibfnamefont [1]{#1}%
\providecommand \citenamefont [1]{#1}%
\providecommand \href@noop [0]{\@secondoftwo}%
\providecommand \href [0]{\begingroup \@sanitize@url \@href}%
\providecommand \@href[1]{\@@startlink{#1}\@@href}%
\providecommand \@@href[1]{\endgroup#1\@@endlink}%
\providecommand \@sanitize@url [0]{\catcode `\\12\catcode `\$12\catcode `\&12\catcode `\#12\catcode `\^12\catcode `\_12\catcode `\%12\relax}%
\providecommand \@@startlink[1]{}%
\providecommand \@@endlink[0]{}%
\providecommand \url  [0]{\begingroup\@sanitize@url \@url }%
\providecommand \@url [1]{\endgroup\@href {#1}{\urlprefix }}%
\providecommand \urlprefix  [0]{URL }%
\providecommand \Eprint [0]{\href }%
\providecommand \doibase [0]{https://doi.org/}%
\providecommand \selectlanguage [0]{\@gobble}%
\providecommand \bibinfo  [0]{\@secondoftwo}%
\providecommand \bibfield  [0]{\@secondoftwo}%
\providecommand \translation [1]{[#1]}%
\providecommand \BibitemOpen [0]{}%
\providecommand \bibitemStop [0]{}%
\providecommand \bibitemNoStop [0]{.\EOS\space}%
\providecommand \EOS [0]{\spacefactor3000\relax}%
\providecommand \BibitemShut  [1]{\csname bibitem#1\endcsname}%
\let\auto@bib@innerbib\@empty
\bibitem [{\citenamefont {Hirohata}\ and\ \citenamefont {Takanashi}(2014)}]{Hirohata2014}%
  \BibitemOpen
  \bibfield  {author} {\bibinfo {author} {\bibfnamefont {A.}~\bibnamefont {Hirohata}}\ and\ \bibinfo {author} {\bibfnamefont {K.}~\bibnamefont {Takanashi}},\ }\bibfield  {title} {\bibinfo {title} {{Future perspectives for spintronic devices}},\ }\href {https://doi.org/10.1088/0022-3727/47/19/193001} {\bibfield  {journal} {\bibinfo  {journal} {J. Phys. D: Appl. Phys.}\ }\textbf {\bibinfo {volume} {47}},\ \bibinfo {pages} {193001} (\bibinfo {year} {2014})}\BibitemShut {NoStop}%
\bibitem [{\citenamefont {Ikhlas}\ \emph {et~al.}(2017)\citenamefont {Ikhlas}, \citenamefont {Tomita}, \citenamefont {Koretsune}, \citenamefont {Suzuki}, \citenamefont {Nishio-Hamane}, \citenamefont {Arita}, \citenamefont {Otani},\ and\ \citenamefont {Nakatsuji}}]{Ikhlas2017}%
  \BibitemOpen
  \bibfield  {author} {\bibinfo {author} {\bibfnamefont {M.}~\bibnamefont {Ikhlas}}, \bibinfo {author} {\bibfnamefont {T.}~\bibnamefont {Tomita}}, \bibinfo {author} {\bibfnamefont {T.}~\bibnamefont {Koretsune}}, \bibinfo {author} {\bibfnamefont {M.-T.}\ \bibnamefont {Suzuki}}, \bibinfo {author} {\bibfnamefont {D.}~\bibnamefont {Nishio-Hamane}}, \bibinfo {author} {\bibfnamefont {R.}~\bibnamefont {Arita}}, \bibinfo {author} {\bibfnamefont {Y.}~\bibnamefont {Otani}},\ and\ \bibinfo {author} {\bibfnamefont {S.}~\bibnamefont {Nakatsuji}},\ }\bibfield  {title} {\bibinfo {title} {{Large anomalous Nernst effect at room temperature in a chiral antiferromagnet}},\ }\href {https://doi.org/10.1038/nphys4181} {\bibfield  {journal} {\bibinfo  {journal} {Nat. Phys.}\ }\textbf {\bibinfo {volume} {13}},\ \bibinfo {pages} {1085} (\bibinfo {year} {2017})}\BibitemShut {NoStop}%
\bibitem [{\citenamefont {Pu}\ \emph {et~al.}(2008)\citenamefont {Pu}, \citenamefont {Chiba}, \citenamefont {Matsukura}, \citenamefont {Ohno},\ and\ \citenamefont {Shi}}]{Pu2008}%
  \BibitemOpen
  \bibfield  {author} {\bibinfo {author} {\bibfnamefont {Y.}~\bibnamefont {Pu}}, \bibinfo {author} {\bibfnamefont {D.}~\bibnamefont {Chiba}}, \bibinfo {author} {\bibfnamefont {F.}~\bibnamefont {Matsukura}}, \bibinfo {author} {\bibfnamefont {H.}~\bibnamefont {Ohno}},\ and\ \bibinfo {author} {\bibfnamefont {J.}~\bibnamefont {Shi}},\ }\bibfield  {title} {\bibinfo {title} {{Mott Relation for Anomalous Hall and Nernst Effects in ${\mathrm{Ga}}_{1\ensuremath{-}x}{\mathrm{Mn}}_{x}\mathrm{As}$ Ferromagnetic Semiconductors}},\ }\href {https://doi.org/10.1103/PhysRevLett.101.117208} {\bibfield  {journal} {\bibinfo  {journal} {Phys. Rev. Lett.}\ }\textbf {\bibinfo {volume} {101}},\ \bibinfo {pages} {117208} (\bibinfo {year} {2008})}\BibitemShut {NoStop}%
\bibitem [{\citenamefont {Armitage}\ \emph {et~al.}(2018)\citenamefont {Armitage}, \citenamefont {Mele},\ and\ \citenamefont {Vishwanath}}]{Armitage2018}%
  \BibitemOpen
  \bibfield  {author} {\bibinfo {author} {\bibfnamefont {N.~P.}\ \bibnamefont {Armitage}}, \bibinfo {author} {\bibfnamefont {E.~J.}\ \bibnamefont {Mele}},\ and\ \bibinfo {author} {\bibfnamefont {A.}~\bibnamefont {Vishwanath}},\ }\bibfield  {title} {\bibinfo {title} {{Weyl and Dirac semimetals in three-dimensional solids}},\ }\href {https://doi.org/10.1103/RevModPhys.90.015001} {\bibfield  {journal} {\bibinfo  {journal} {Rev. Mod. Phys.}\ }\textbf {\bibinfo {volume} {90}},\ \bibinfo {pages} {015001} (\bibinfo {year} {2018})}\BibitemShut {NoStop}%
\bibitem [{\citenamefont {Breidenbach}\ \emph {et~al.}(2022)\citenamefont {Breidenbach}, \citenamefont {Yu}, \citenamefont {Peterson}, \citenamefont {McFadden}, \citenamefont {Peria}, \citenamefont {Palmstr\o{}m},\ and\ \citenamefont {Crowell}}]{Breidenbach2022}%
  \BibitemOpen
  \bibfield  {author} {\bibinfo {author} {\bibfnamefont {A.~T.}\ \bibnamefont {Breidenbach}}, \bibinfo {author} {\bibfnamefont {H.}~\bibnamefont {Yu}}, \bibinfo {author} {\bibfnamefont {T.~A.}\ \bibnamefont {Peterson}}, \bibinfo {author} {\bibfnamefont {A.~P.}\ \bibnamefont {McFadden}}, \bibinfo {author} {\bibfnamefont {W.~K.}\ \bibnamefont {Peria}}, \bibinfo {author} {\bibfnamefont {C.~J.}\ \bibnamefont {Palmstr\o{}m}},\ and\ \bibinfo {author} {\bibfnamefont {P.~A.}\ \bibnamefont {Crowell}},\ }\bibfield  {title} {\bibinfo {title} {{Anomalous Nernst and Seebeck coefficients in epitaxial thin film ${\mathrm{Co}}_{2}{\mathrm{MnAl}}_{x}{\mathrm{Si}}_{1\ensuremath{-}x}$ and ${\mathrm{Co}}_{2}\mathrm{FeAl}$}},\ }\href {https://doi.org/10.1103/PhysRevB.105.144405} {\bibfield  {journal} {\bibinfo  {journal} {Phys. Rev. B}\ }\textbf {\bibinfo {volume} {105}},\ \bibinfo {pages} {144405} (\bibinfo {year} {2022})}\BibitemShut {NoStop}%
\bibitem [{\citenamefont {Sakuraba}\ \emph {et~al.}(2020)\citenamefont {Sakuraba}, \citenamefont {Hyodo}, \citenamefont {Sakuma},\ and\ \citenamefont {Mitani}}]{Sakuraba2020}%
  \BibitemOpen
  \bibfield  {author} {\bibinfo {author} {\bibfnamefont {Y.}~\bibnamefont {Sakuraba}}, \bibinfo {author} {\bibfnamefont {K.}~\bibnamefont {Hyodo}}, \bibinfo {author} {\bibfnamefont {A.}~\bibnamefont {Sakuma}},\ and\ \bibinfo {author} {\bibfnamefont {S.}~\bibnamefont {Mitani}},\ }\bibfield  {title} {\bibinfo {title} {{Giant anomalous Nernst effect in the $\mathrm{C}{\mathrm{o}}_{2}\mathrm{MnA}{\mathrm{l}}_{1\ensuremath{-}x}\mathrm{S}{\mathrm{i}}_{x}$ Heusler alloy induced by Fermi level tuning and atomic ordering}},\ }\href {https://doi.org/10.1103/PhysRevB.101.134407} {\bibfield  {journal} {\bibinfo  {journal} {Phys. Rev. B}\ }\textbf {\bibinfo {volume} {101}},\ \bibinfo {pages} {134407} (\bibinfo {year} {2020})}\BibitemShut {NoStop}%
\bibitem [{\citenamefont {Hu}\ \emph {et~al.}(2018)\citenamefont {Hu}, \citenamefont {Ernst}, \citenamefont {Tu}, \citenamefont {Kuve\ifmmode \check{z}\else \v{z}\fi{}di\ifmmode~\acute{c}\else \'{c}\fi{}}, \citenamefont {Hamzi\ifmmode~\acute{c}\else \'{c}\fi{}}, \citenamefont {Tafra}, \citenamefont {Basleti\ifmmode~\acute{c}\else \'{c}\fi{}}, \citenamefont {Zhang}, \citenamefont {Markou}, \citenamefont {Felser}, \citenamefont {Fert}, \citenamefont {Zhao}, \citenamefont {Ansermet},\ and\ \citenamefont {Yu}}]{Hu2018}%
  \BibitemOpen
  \bibfield  {author} {\bibinfo {author} {\bibfnamefont {J.}~\bibnamefont {Hu}}, \bibinfo {author} {\bibfnamefont {B.}~\bibnamefont {Ernst}}, \bibinfo {author} {\bibfnamefont {S.}~\bibnamefont {Tu}}, \bibinfo {author} {\bibfnamefont {M.}~\bibnamefont {Kuve\ifmmode \check{z}\else \v{z}\fi{}di\ifmmode~\acute{c}\else \'{c}\fi{}}}, \bibinfo {author} {\bibfnamefont {A.}~\bibnamefont {Hamzi\ifmmode~\acute{c}\else \'{c}\fi{}}}, \bibinfo {author} {\bibfnamefont {E.}~\bibnamefont {Tafra}}, \bibinfo {author} {\bibfnamefont {M.}~\bibnamefont {Basleti\ifmmode~\acute{c}\else \'{c}\fi{}}}, \bibinfo {author} {\bibfnamefont {Y.}~\bibnamefont {Zhang}}, \bibinfo {author} {\bibfnamefont {A.}~\bibnamefont {Markou}}, \bibinfo {author} {\bibfnamefont {C.}~\bibnamefont {Felser}}, \bibinfo {author} {\bibfnamefont {A.}~\bibnamefont {Fert}}, \bibinfo {author} {\bibfnamefont {W.}~\bibnamefont {Zhao}}, \bibinfo {author} {\bibfnamefont {J.-P.}\ \bibnamefont {Ansermet}},\ and\ \bibinfo {author} {\bibfnamefont {H.}~\bibnamefont {Yu}},\
  }\bibfield  {title} {\bibinfo {title} {{Anomalous Hall and Nernst Effects in ${\mathrm{Co}}_{2}\mathrm{Ti}\mathrm{Sn}$ and ${\mathrm{Co}}_{2}{\mathrm{Ti}}_{0.6}{\mathrm{V}}_{0.4}\mathrm{Sn}$ Heusler Thin Films}},\ }\href {https://doi.org/10.1103/PhysRevApplied.10.044037} {\bibfield  {journal} {\bibinfo  {journal} {Phys. Rev. Appl.}\ }\textbf {\bibinfo {volume} {10}},\ \bibinfo {pages} {044037} (\bibinfo {year} {2018})}\BibitemShut {NoStop}%
\bibitem [{\citenamefont {Hu}\ \emph {et~al.}(2020)\citenamefont {Hu}, \citenamefont {Butler}, \citenamefont {Cabero~Z.}, \citenamefont {Wang}, \citenamefont {Wei}, \citenamefont {Tu}, \citenamefont {Guo}, \citenamefont {Wan}, \citenamefont {Han}, \citenamefont {Liu}, \citenamefont {Zhao}, \citenamefont {Ansermet}, \citenamefont {Granville},\ and\ \citenamefont {Yu}}]{Hu2020}%
  \BibitemOpen
  \bibfield  {author} {\bibinfo {author} {\bibfnamefont {J.}~\bibnamefont {Hu}}, \bibinfo {author} {\bibfnamefont {T.}~\bibnamefont {Butler}}, \bibinfo {author} {\bibfnamefont {M.~A.}\ \bibnamefont {Cabero~Z.}}, \bibinfo {author} {\bibfnamefont {H.}~\bibnamefont {Wang}}, \bibinfo {author} {\bibfnamefont {B.}~\bibnamefont {Wei}}, \bibinfo {author} {\bibfnamefont {S.}~\bibnamefont {Tu}}, \bibinfo {author} {\bibfnamefont {C.}~\bibnamefont {Guo}}, \bibinfo {author} {\bibfnamefont {C.}~\bibnamefont {Wan}}, \bibinfo {author} {\bibfnamefont {X.}~\bibnamefont {Han}}, \bibinfo {author} {\bibfnamefont {S.}~\bibnamefont {Liu}}, \bibinfo {author} {\bibfnamefont {W.}~\bibnamefont {Zhao}}, \bibinfo {author} {\bibfnamefont {J.-P.}\ \bibnamefont {Ansermet}}, \bibinfo {author} {\bibfnamefont {S.}~\bibnamefont {Granville}},\ and\ \bibinfo {author} {\bibfnamefont {H.}~\bibnamefont {Yu}},\ }\bibfield  {title} {\bibinfo {title} {{Regulating the anomalous Hall and Nernst effects in Heusler-based trilayers}},\ }\href
  {https://doi.org/10.1063/5.0014879} {\bibfield  {journal} {\bibinfo  {journal} {Appl. Phys. Lett.}\ }\textbf {\bibinfo {volume} {117}},\ \bibinfo {pages} {062405} (\bibinfo {year} {2020})}\BibitemShut {NoStop}%
\bibitem [{\citenamefont {Mende}\ \emph {et~al.}(2021)\citenamefont {Mende}, \citenamefont {Noky}, \citenamefont {Guin}, \citenamefont {Fecher}, \citenamefont {Manna}, \citenamefont {Adler}, \citenamefont {Schnelle}, \citenamefont {Sun}, \citenamefont {Fu},\ and\ \citenamefont {Felser}}]{Mende2021}%
  \BibitemOpen
  \bibfield  {author} {\bibinfo {author} {\bibfnamefont {F.}~\bibnamefont {Mende}}, \bibinfo {author} {\bibfnamefont {J.}~\bibnamefont {Noky}}, \bibinfo {author} {\bibfnamefont {S.~N.}\ \bibnamefont {Guin}}, \bibinfo {author} {\bibfnamefont {G.~H.}\ \bibnamefont {Fecher}}, \bibinfo {author} {\bibfnamefont {K.}~\bibnamefont {Manna}}, \bibinfo {author} {\bibfnamefont {P.}~\bibnamefont {Adler}}, \bibinfo {author} {\bibfnamefont {W.}~\bibnamefont {Schnelle}}, \bibinfo {author} {\bibfnamefont {Y.}~\bibnamefont {Sun}}, \bibinfo {author} {\bibfnamefont {C.}~\bibnamefont {Fu}},\ and\ \bibinfo {author} {\bibfnamefont {C.}~\bibnamefont {Felser}},\ }\bibfield  {title} {\bibinfo {title} {{Large Anomalous Hall and Nernst Effects in High Curie-Temperature Iron-Based Heusler Compounds}},\ }\href {https://doi.org/https://doi.org/10.1002/advs.202100782} {\bibfield  {journal} {\bibinfo  {journal} {Adv. Sci.}\ }\textbf {\bibinfo {volume} {8}},\ \bibinfo {pages} {2100782} (\bibinfo {year} {2021})}\BibitemShut {NoStop}%
\bibitem [{\citenamefont {De}\ \emph {et~al.}(2021)\citenamefont {De}, \citenamefont {Singh}, \citenamefont {Singh},\ and\ \citenamefont {Nair}}]{De2021}%
  \BibitemOpen
  \bibfield  {author} {\bibinfo {author} {\bibfnamefont {A.}~\bibnamefont {De}}, \bibinfo {author} {\bibfnamefont {A.~K.}\ \bibnamefont {Singh}}, \bibinfo {author} {\bibfnamefont {S.}~\bibnamefont {Singh}},\ and\ \bibinfo {author} {\bibfnamefont {S.}~\bibnamefont {Nair}},\ }\bibfield  {title} {\bibinfo {title} {{Temperature dependence of the anomalous Nernst effect in ${\mathrm{Ni}}_{2}\mathrm{Mn}\mathrm{Ga}$ shape memory alloy}},\ }\href {https://doi.org/10.1103/PhysRevB.103.L020404} {\bibfield  {journal} {\bibinfo  {journal} {Phys. Rev. B}\ }\textbf {\bibinfo {volume} {103}},\ \bibinfo {pages} {L020404} (\bibinfo {year} {2021})}\BibitemShut {NoStop}%
\bibitem [{\citenamefont {Xiao}\ \emph {et~al.}(2006)\citenamefont {Xiao}, \citenamefont {Yao}, \citenamefont {Fang},\ and\ \citenamefont {Niu}}]{Xiao2006}%
  \BibitemOpen
  \bibfield  {author} {\bibinfo {author} {\bibfnamefont {D.}~\bibnamefont {Xiao}}, \bibinfo {author} {\bibfnamefont {Y.}~\bibnamefont {Yao}}, \bibinfo {author} {\bibfnamefont {Z.}~\bibnamefont {Fang}},\ and\ \bibinfo {author} {\bibfnamefont {Q.}~\bibnamefont {Niu}},\ }\bibfield  {title} {\bibinfo {title} {{Berry-Phase Effect in Anomalous Thermoelectric Transport}},\ }\href {https://doi.org/10.1103/PhysRevLett.97.026603} {\bibfield  {journal} {\bibinfo  {journal} {Phys. Rev. Lett.}\ }\textbf {\bibinfo {volume} {97}},\ \bibinfo {pages} {026603} (\bibinfo {year} {2006})}\BibitemShut {NoStop}%
\bibitem [{\citenamefont {Belopolski}\ \emph {et~al.}(2019)\citenamefont {Belopolski}, \citenamefont {Manna}, \citenamefont {Sanchez}, \citenamefont {Chang}, \citenamefont {Ernst}, \citenamefont {Yin}, \citenamefont {Zhang}, \citenamefont {Cochran}, \citenamefont {Shumiya}, \citenamefont {Zheng}, \citenamefont {Singh}, \citenamefont {Bian}, \citenamefont {Multer}, \citenamefont {Litskevich}, \citenamefont {Zhou}, \citenamefont {Huang}, \citenamefont {Wang}, \citenamefont {Chang}, \citenamefont {Xu}, \citenamefont {Bansil}, \citenamefont {Felser}, \citenamefont {Lin},\ and\ \citenamefont {Hasan}}]{p3}%
  \BibitemOpen
  \bibfield  {author} {\bibinfo {author} {\bibfnamefont {I.}~\bibnamefont {Belopolski}}, \bibinfo {author} {\bibfnamefont {K.}~\bibnamefont {Manna}}, \bibinfo {author} {\bibfnamefont {D.~S.}\ \bibnamefont {Sanchez}}, \bibinfo {author} {\bibfnamefont {G.}~\bibnamefont {Chang}}, \bibinfo {author} {\bibfnamefont {B.}~\bibnamefont {Ernst}}, \bibinfo {author} {\bibfnamefont {J.}~\bibnamefont {Yin}}, \bibinfo {author} {\bibfnamefont {S.~S.}\ \bibnamefont {Zhang}}, \bibinfo {author} {\bibfnamefont {T.}~\bibnamefont {Cochran}}, \bibinfo {author} {\bibfnamefont {N.}~\bibnamefont {Shumiya}}, \bibinfo {author} {\bibfnamefont {H.}~\bibnamefont {Zheng}}, \bibinfo {author} {\bibfnamefont {B.}~\bibnamefont {Singh}}, \bibinfo {author} {\bibfnamefont {G.}~\bibnamefont {Bian}}, \bibinfo {author} {\bibfnamefont {D.}~\bibnamefont {Multer}}, \bibinfo {author} {\bibfnamefont {M.}~\bibnamefont {Litskevich}}, \bibinfo {author} {\bibfnamefont {X.}~\bibnamefont {Zhou}}, \bibinfo {author} {\bibfnamefont {S.-M.}\ \bibnamefont {Huang}},
  \bibinfo {author} {\bibfnamefont {B.}~\bibnamefont {Wang}}, \bibinfo {author} {\bibfnamefont {T.-R.}\ \bibnamefont {Chang}}, \bibinfo {author} {\bibfnamefont {S.-Y.}\ \bibnamefont {Xu}}, \bibinfo {author} {\bibfnamefont {A.}~\bibnamefont {Bansil}}, \bibinfo {author} {\bibfnamefont {C.}~\bibnamefont {Felser}}, \bibinfo {author} {\bibfnamefont {H.}~\bibnamefont {Lin}},\ and\ \bibinfo {author} {\bibfnamefont {M.~Z.}\ \bibnamefont {Hasan}},\ }\bibfield  {title} {\bibinfo {title} {{Discovery of topological Weyl fermion lines and drumhead surface states in a room temperature magnet}},\ }\href {https://doi.org/10.1126/science.aav2327} {\bibfield  {journal} {\bibinfo  {journal} {Science}\ }\textbf {\bibinfo {volume} {365}},\ \bibinfo {pages} {1278} (\bibinfo {year} {2019})}\BibitemShut {NoStop}%
\bibitem [{\citenamefont {Li}\ \emph {et~al.}(2020)\citenamefont {Li}, \citenamefont {Koo}, \citenamefont {Ning}, \citenamefont {Li}, \citenamefont {Miao}, \citenamefont {Min}, \citenamefont {Zhu}, \citenamefont {Wang}, \citenamefont {Alem}, \citenamefont {Liu}, \citenamefont {Mao},\ and\ \citenamefont {Yan}}]{Li2020}%
  \BibitemOpen
  \bibfield  {author} {\bibinfo {author} {\bibfnamefont {P.}~\bibnamefont {Li}}, \bibinfo {author} {\bibfnamefont {J.}~\bibnamefont {Koo}}, \bibinfo {author} {\bibfnamefont {W.}~\bibnamefont {Ning}}, \bibinfo {author} {\bibfnamefont {J.}~\bibnamefont {Li}}, \bibinfo {author} {\bibfnamefont {L.}~\bibnamefont {Miao}}, \bibinfo {author} {\bibfnamefont {L.}~\bibnamefont {Min}}, \bibinfo {author} {\bibfnamefont {Y.}~\bibnamefont {Zhu}}, \bibinfo {author} {\bibfnamefont {Y.}~\bibnamefont {Wang}}, \bibinfo {author} {\bibfnamefont {N.}~\bibnamefont {Alem}}, \bibinfo {author} {\bibfnamefont {C.-X.}\ \bibnamefont {Liu}}, \bibinfo {author} {\bibfnamefont {Z.}~\bibnamefont {Mao}},\ and\ \bibinfo {author} {\bibfnamefont {B.}~\bibnamefont {Yan}},\ }\bibfield  {title} {\bibinfo {title} {{Giant room temperature anomalous Hall effect and tunable topology in a ferromagnetic topological semimetal Co$_2$MnAl}},\ }\href {https://doi.org/10.1038/s41467-020-17174-9} {\bibfield  {journal} {\bibinfo  {journal} {Nat. Commun.}\
  }\textbf {\bibinfo {volume} {11}},\ \bibinfo {pages} {3476} (\bibinfo {year} {2020})}\BibitemShut {NoStop}%
\bibitem [{\citenamefont {Chatterjee}\ \emph {et~al.}(2023)\citenamefont {Chatterjee}, \citenamefont {Sau}, \citenamefont {Samanta}, \citenamefont {Ghosh}, \citenamefont {Kumar}, \citenamefont {Kumar},\ and\ \citenamefont {Mandal}}]{p7}%
  \BibitemOpen
  \bibfield  {author} {\bibinfo {author} {\bibfnamefont {S.}~\bibnamefont {Chatterjee}}, \bibinfo {author} {\bibfnamefont {J.}~\bibnamefont {Sau}}, \bibinfo {author} {\bibfnamefont {S.}~\bibnamefont {Samanta}}, \bibinfo {author} {\bibfnamefont {B.}~\bibnamefont {Ghosh}}, \bibinfo {author} {\bibfnamefont {N.}~\bibnamefont {Kumar}}, \bibinfo {author} {\bibfnamefont {M.}~\bibnamefont {Kumar}},\ and\ \bibinfo {author} {\bibfnamefont {K.}~\bibnamefont {Mandal}},\ }\bibfield  {title} {\bibinfo {title} {{Nodal-line and triple point fermion induced anomalous Hall effect in the topological Heusler compound $\mathrm{Co}_{2}\mathrm{CrGa}$}},\ }\href {https://doi.org/10.1103/PhysRevB.107.125138} {\bibfield  {journal} {\bibinfo  {journal} {Phys. Rev. B}\ }\textbf {\bibinfo {volume} {107}},\ \bibinfo {pages} {125138} (\bibinfo {year} {2023})}\BibitemShut {NoStop}%
\bibitem [{\citenamefont {Kim}\ \emph {et~al.}(2018)\citenamefont {Kim}, \citenamefont {Seo}, \citenamefont {Lee}, \citenamefont {Ko}, \citenamefont {Kim}, \citenamefont {Jang}, \citenamefont {Ok}, \citenamefont {Lee}, \citenamefont {Jo}, \citenamefont {Kang} \emph {et~al.}}]{p10}%
  \BibitemOpen
  \bibfield  {author} {\bibinfo {author} {\bibfnamefont {K.}~\bibnamefont {Kim}}, \bibinfo {author} {\bibfnamefont {J.}~\bibnamefont {Seo}}, \bibinfo {author} {\bibfnamefont {E.}~\bibnamefont {Lee}}, \bibinfo {author} {\bibfnamefont {K.-T.}\ \bibnamefont {Ko}}, \bibinfo {author} {\bibfnamefont {B.}~\bibnamefont {Kim}}, \bibinfo {author} {\bibfnamefont {B.~G.}\ \bibnamefont {Jang}}, \bibinfo {author} {\bibfnamefont {J.~M.}\ \bibnamefont {Ok}}, \bibinfo {author} {\bibfnamefont {J.}~\bibnamefont {Lee}}, \bibinfo {author} {\bibfnamefont {Y.~J.}\ \bibnamefont {Jo}}, \bibinfo {author} {\bibfnamefont {W.}~\bibnamefont {Kang}}, \emph {et~al.},\ }\bibfield  {title} {\bibinfo {title} {{Large anomalous Hall current induced by topological nodal lines in a ferromagnetic van der Waals semimetal}},\ }\href {https://doi.org/10.1038/s41563-018-0132-3} {\bibfield  {journal} {\bibinfo  {journal} {Nat. Mater.}\ }\textbf {\bibinfo {volume} {17}},\ \bibinfo {pages} {794} (\bibinfo {year} {2018})}\BibitemShut {NoStop}%
\bibitem [{\citenamefont {Wang}\ \emph {et~al.}(2018)\citenamefont {Wang}, \citenamefont {Xu}, \citenamefont {Lou}, \citenamefont {Liu}, \citenamefont {Li}, \citenamefont {Huang}, \citenamefont {Shen}, \citenamefont {Weng}, \citenamefont {Wang},\ and\ \citenamefont {Lei}}]{p13}%
  \BibitemOpen
  \bibfield  {author} {\bibinfo {author} {\bibfnamefont {Q.}~\bibnamefont {Wang}}, \bibinfo {author} {\bibfnamefont {Y.}~\bibnamefont {Xu}}, \bibinfo {author} {\bibfnamefont {R.}~\bibnamefont {Lou}}, \bibinfo {author} {\bibfnamefont {Z.}~\bibnamefont {Liu}}, \bibinfo {author} {\bibfnamefont {M.}~\bibnamefont {Li}}, \bibinfo {author} {\bibfnamefont {Y.}~\bibnamefont {Huang}}, \bibinfo {author} {\bibfnamefont {D.}~\bibnamefont {Shen}}, \bibinfo {author} {\bibfnamefont {H.}~\bibnamefont {Weng}}, \bibinfo {author} {\bibfnamefont {S.}~\bibnamefont {Wang}},\ and\ \bibinfo {author} {\bibfnamefont {H.}~\bibnamefont {Lei}},\ }\bibfield  {title} {\bibinfo {title} {{Large intrinsic anomalous Hall effect in half-metallic ferromagnet Co$_3$Sn$_2$S$_2$ with magnetic Weyl fermions}},\ }\href {https://doi.org/10.1038/s41467-018-06088-2} {\bibfield  {journal} {\bibinfo  {journal} {Nat. Commun.}\ }\textbf {\bibinfo {volume} {9}},\ \bibinfo {pages} {3681} (\bibinfo {year} {2018})}\BibitemShut {NoStop}%
\bibitem [{\citenamefont {Onoda}\ \emph {et~al.}(2008)\citenamefont {Onoda}, \citenamefont {Sugimoto},\ and\ \citenamefont {Nagaosa}}]{p15}%
  \BibitemOpen
  \bibfield  {author} {\bibinfo {author} {\bibfnamefont {S.}~\bibnamefont {Onoda}}, \bibinfo {author} {\bibfnamefont {N.}~\bibnamefont {Sugimoto}},\ and\ \bibinfo {author} {\bibfnamefont {N.}~\bibnamefont {Nagaosa}},\ }\bibfield  {title} {\bibinfo {title} {{Quantum transport theory of anomalous electric, thermoelectric, and thermal Hall effects in ferromagnets}},\ }\href {https://doi.org/10.1103/PhysRevB.77.165103} {\bibfield  {journal} {\bibinfo  {journal} {Phys. Rev. B}\ }\textbf {\bibinfo {volume} {77}},\ \bibinfo {pages} {165103} (\bibinfo {year} {2008})}\BibitemShut {NoStop}%
\bibitem [{\citenamefont {Wang}\ \emph {et~al.}(2006)\citenamefont {Wang}, \citenamefont {Yates}, \citenamefont {Souza},\ and\ \citenamefont {Vanderbilt}}]{Wang2006}%
  \BibitemOpen
  \bibfield  {author} {\bibinfo {author} {\bibfnamefont {X.}~\bibnamefont {Wang}}, \bibinfo {author} {\bibfnamefont {J.~R.}\ \bibnamefont {Yates}}, \bibinfo {author} {\bibfnamefont {I.}~\bibnamefont {Souza}},\ and\ \bibinfo {author} {\bibfnamefont {D.}~\bibnamefont {Vanderbilt}},\ }\bibfield  {title} {\bibinfo {title} {{Ab initio calculation of the anomalous Hall conductivity by Wannier interpolation}},\ }\href {https://doi.org/10.1103/PhysRevB.74.195118} {\bibfield  {journal} {\bibinfo  {journal} {Phys. Rev. B}\ }\textbf {\bibinfo {volume} {74}},\ \bibinfo {pages} {195118} (\bibinfo {year} {2006})}\BibitemShut {NoStop}%
\bibitem [{\citenamefont {Sakai}\ \emph {et~al.}(2018)\citenamefont {Sakai}, \citenamefont {Mizuta}, \citenamefont {Nugroho}, \citenamefont {Sihombing}, \citenamefont {Koretsune}, \citenamefont {Suzuki}, \citenamefont {Takemori}, \citenamefont {Ishii}, \citenamefont {Nishio-Hamane}, \citenamefont {Arita}, \citenamefont {Goswami},\ and\ \citenamefont {Nakatsuji}}]{Sakai2018}%
  \BibitemOpen
  \bibfield  {author} {\bibinfo {author} {\bibfnamefont {A.}~\bibnamefont {Sakai}}, \bibinfo {author} {\bibfnamefont {Y.~P.}\ \bibnamefont {Mizuta}}, \bibinfo {author} {\bibfnamefont {A.~A.}\ \bibnamefont {Nugroho}}, \bibinfo {author} {\bibfnamefont {R.}~\bibnamefont {Sihombing}}, \bibinfo {author} {\bibfnamefont {T.}~\bibnamefont {Koretsune}}, \bibinfo {author} {\bibfnamefont {M.-T.}\ \bibnamefont {Suzuki}}, \bibinfo {author} {\bibfnamefont {N.}~\bibnamefont {Takemori}}, \bibinfo {author} {\bibfnamefont {R.}~\bibnamefont {Ishii}}, \bibinfo {author} {\bibfnamefont {D.}~\bibnamefont {Nishio-Hamane}}, \bibinfo {author} {\bibfnamefont {R.}~\bibnamefont {Arita}}, \bibinfo {author} {\bibfnamefont {P.}~\bibnamefont {Goswami}},\ and\ \bibinfo {author} {\bibfnamefont {S.}~\bibnamefont {Nakatsuji}},\ }\bibfield  {title} {\bibinfo {title} {{Giant anomalous Nernst effect and quantum-critical scaling in a ferromagnetic semimetal}},\ }\href {https://doi.org/10.1038/s41567-018-0225-6} {\bibfield  {journal} {\bibinfo
  {journal} {Nat. Phys.}\ }\textbf {\bibinfo {volume} {14}},\ \bibinfo {pages} {1119} (\bibinfo {year} {2018})}\BibitemShut {NoStop}%
\bibitem [{\citenamefont {Guin}\ \emph {et~al.}(2019)\citenamefont {Guin}, \citenamefont {Manna}, \citenamefont {Noky}, \citenamefont {Watzman}, \citenamefont {Fu}, \citenamefont {Kumar}, \citenamefont {Schnelle}, \citenamefont {Shekhar}, \citenamefont {Sun}, \citenamefont {Gooth},\ and\ \citenamefont {Felser}}]{Guin2019}%
  \BibitemOpen
  \bibfield  {author} {\bibinfo {author} {\bibfnamefont {S.~N.}\ \bibnamefont {Guin}}, \bibinfo {author} {\bibfnamefont {K.}~\bibnamefont {Manna}}, \bibinfo {author} {\bibfnamefont {J.}~\bibnamefont {Noky}}, \bibinfo {author} {\bibfnamefont {S.~J.}\ \bibnamefont {Watzman}}, \bibinfo {author} {\bibfnamefont {C.}~\bibnamefont {Fu}}, \bibinfo {author} {\bibfnamefont {N.}~\bibnamefont {Kumar}}, \bibinfo {author} {\bibfnamefont {W.}~\bibnamefont {Schnelle}}, \bibinfo {author} {\bibfnamefont {C.}~\bibnamefont {Shekhar}}, \bibinfo {author} {\bibfnamefont {Y.}~\bibnamefont {Sun}}, \bibinfo {author} {\bibfnamefont {J.}~\bibnamefont {Gooth}},\ and\ \bibinfo {author} {\bibfnamefont {C.}~\bibnamefont {Felser}},\ }\bibfield  {title} {\bibinfo {title} {{Anomalous Nernst effect beyond the magnetization scaling relation in the ferromagnetic Heusler compound Co$_2$MnGa}},\ }\href {https://doi.org/10.1038/s41427-019-0116-z} {\bibfield  {journal} {\bibinfo  {journal} {NPG Asia Mater.}\ }\textbf {\bibinfo {volume} {11}},\
  \bibinfo {pages} {16} (\bibinfo {year} {2019})}\BibitemShut {NoStop}%
\bibitem [{\citenamefont {Cox}\ \emph {et~al.}(2019)\citenamefont {Cox}, \citenamefont {Caruana}, \citenamefont {Cropper},\ and\ \citenamefont {Morrison}}]{Cox2019}%
  \BibitemOpen
  \bibfield  {author} {\bibinfo {author} {\bibfnamefont {C.}~\bibnamefont {Cox}}, \bibinfo {author} {\bibfnamefont {A.}~\bibnamefont {Caruana}}, \bibinfo {author} {\bibfnamefont {M.}~\bibnamefont {Cropper}},\ and\ \bibinfo {author} {\bibfnamefont {K.}~\bibnamefont {Morrison}},\ }\bibfield  {title} {\bibinfo {title} {{Anomalous Nernst effect in Co$_2$MnSi thin films}},\ }\href {https://doi.org/10.1088/1361-6463/ab4eeb} {\bibfield  {journal} {\bibinfo  {journal} {J. Phys. D: Appl. Phys.}\ }\textbf {\bibinfo {volume} {53}},\ \bibinfo {pages} {035005} (\bibinfo {year} {2019})}\BibitemShut {NoStop}%
\bibitem [{\citenamefont {Park}\ \emph {et~al.}(2020)\citenamefont {Park}, \citenamefont {Reichlova}, \citenamefont {Schlitz}, \citenamefont {Lammel}, \citenamefont {Markou}, \citenamefont {Swekis}, \citenamefont {Ritzinger}, \citenamefont {Kriegner}, \citenamefont {Noky}, \citenamefont {Gayles}, \citenamefont {Sun}, \citenamefont {Felser}, \citenamefont {Nielsch}, \citenamefont {Goennenwein},\ and\ \citenamefont {Thomas}}]{Park2020}%
  \BibitemOpen
  \bibfield  {author} {\bibinfo {author} {\bibfnamefont {G.-H.}\ \bibnamefont {Park}}, \bibinfo {author} {\bibfnamefont {H.}~\bibnamefont {Reichlova}}, \bibinfo {author} {\bibfnamefont {R.}~\bibnamefont {Schlitz}}, \bibinfo {author} {\bibfnamefont {M.}~\bibnamefont {Lammel}}, \bibinfo {author} {\bibfnamefont {A.}~\bibnamefont {Markou}}, \bibinfo {author} {\bibfnamefont {P.}~\bibnamefont {Swekis}}, \bibinfo {author} {\bibfnamefont {P.}~\bibnamefont {Ritzinger}}, \bibinfo {author} {\bibfnamefont {D.}~\bibnamefont {Kriegner}}, \bibinfo {author} {\bibfnamefont {J.}~\bibnamefont {Noky}}, \bibinfo {author} {\bibfnamefont {J.}~\bibnamefont {Gayles}}, \bibinfo {author} {\bibfnamefont {Y.}~\bibnamefont {Sun}}, \bibinfo {author} {\bibfnamefont {C.}~\bibnamefont {Felser}}, \bibinfo {author} {\bibfnamefont {K.}~\bibnamefont {Nielsch}}, \bibinfo {author} {\bibfnamefont {S.~T.~B.}\ \bibnamefont {Goennenwein}},\ and\ \bibinfo {author} {\bibfnamefont {A.}~\bibnamefont {Thomas}},\ }\bibfield  {title} {\bibinfo {title}
  {{Thickness dependence of the anomalous Nernst effect and the Mott relation of Weyl semimetal ${\mathrm{Co}}_{2}\mathrm{MnGa}$ thin films}},\ }\href {https://doi.org/10.1103/PhysRevB.101.060406} {\bibfield  {journal} {\bibinfo  {journal} {Phys. Rev. B}\ }\textbf {\bibinfo {volume} {101}},\ \bibinfo {pages} {060406} (\bibinfo {year} {2020})}\BibitemShut {NoStop}%
\bibitem [{\citenamefont {Uesugi}\ \emph {et~al.}(2023)\citenamefont {Uesugi}, \citenamefont {Higo},\ and\ \citenamefont {Nakatsuji}}]{Uesugi2023}%
  \BibitemOpen
  \bibfield  {author} {\bibinfo {author} {\bibfnamefont {R.}~\bibnamefont {Uesugi}}, \bibinfo {author} {\bibfnamefont {T.}~\bibnamefont {Higo}},\ and\ \bibinfo {author} {\bibfnamefont {S.}~\bibnamefont {Nakatsuji}},\ }\bibfield  {title} {\bibinfo {title} {{Giant anomalous Nernst effect in polycrystalline thin films of the Weyl ferromagnet Co$_2$MnGa}},\ }\href {https://doi.org/10.1063/5.0174663} {\bibfield  {journal} {\bibinfo  {journal} {Appl. Phys. Lett.}\ }\textbf {\bibinfo {volume} {123}},\ \bibinfo {pages} {252401} (\bibinfo {year} {2023})}\BibitemShut {NoStop}%
\bibitem [{\citenamefont {Zhou}\ \emph {et~al.}(2023)\citenamefont {Zhou}, \citenamefont {Miura}, \citenamefont {Hirai}, \citenamefont {Sakuraba},\ and\ \citenamefont {Uchida}}]{Zhou2023}%
  \BibitemOpen
  \bibfield  {author} {\bibinfo {author} {\bibfnamefont {W.}~\bibnamefont {Zhou}}, \bibinfo {author} {\bibfnamefont {A.}~\bibnamefont {Miura}}, \bibinfo {author} {\bibfnamefont {T.}~\bibnamefont {Hirai}}, \bibinfo {author} {\bibfnamefont {Y.}~\bibnamefont {Sakuraba}},\ and\ \bibinfo {author} {\bibfnamefont {K.-i.}\ \bibnamefont {Uchida}},\ }\bibfield  {title} {\bibinfo {title} {{Seebeck-driven transverse thermoelectric generation in magnetic hybrid bulk materials}},\ }\href {https://doi.org/10.1063/5.0126870} {\bibfield  {journal} {\bibinfo  {journal} {Appl. Phys. Lett.}\ }\textbf {\bibinfo {volume} {122}},\ \bibinfo {pages} {062402} (\bibinfo {year} {2023})}\BibitemShut {NoStop}%
\bibitem [{\citenamefont {Fujiwara}\ \emph {et~al.}(2023)\citenamefont {Fujiwara}, \citenamefont {Kato}, \citenamefont {Abe}, \citenamefont {Noguchi}, \citenamefont {Shiogai}, \citenamefont {Niwa}, \citenamefont {Kumigashira}, \citenamefont {Motome},\ and\ \citenamefont {Tsukazaki}}]{p20}%
  \BibitemOpen
  \bibfield  {author} {\bibinfo {author} {\bibfnamefont {K.}~\bibnamefont {Fujiwara}}, \bibinfo {author} {\bibfnamefont {Y.}~\bibnamefont {Kato}}, \bibinfo {author} {\bibfnamefont {H.}~\bibnamefont {Abe}}, \bibinfo {author} {\bibfnamefont {S.}~\bibnamefont {Noguchi}}, \bibinfo {author} {\bibfnamefont {J.}~\bibnamefont {Shiogai}}, \bibinfo {author} {\bibfnamefont {Y.}~\bibnamefont {Niwa}}, \bibinfo {author} {\bibfnamefont {H.}~\bibnamefont {Kumigashira}}, \bibinfo {author} {\bibfnamefont {Y.}~\bibnamefont {Motome}},\ and\ \bibinfo {author} {\bibfnamefont {A.}~\bibnamefont {Tsukazaki}},\ }\bibfield  {title} {\bibinfo {title} {{Berry curvature contributions of kagome-lattice fragments in amorphous {Fe--Sn} thin films}},\ }\href {https://doi.org/10.1038/s41467-023-39112-1} {\bibfield  {journal} {\bibinfo  {journal} {Nat. Commun.}\ }\textbf {\bibinfo {volume} {14}},\ \bibinfo {pages} {3399} (\bibinfo {year} {2023})}\BibitemShut {NoStop}%
\bibitem [{\citenamefont {Ye}\ \emph {et~al.}(2018)\citenamefont {Ye}, \citenamefont {Kang}, \citenamefont {Liu}, \citenamefont {Von~Cube}, \citenamefont {Wicker}, \citenamefont {Suzuki}, \citenamefont {Jozwiak}, \citenamefont {Bostwick}, \citenamefont {Rotenberg}, \citenamefont {Bell}, \citenamefont {Fu}, \citenamefont {Comin},\ and\ \citenamefont {Checkelsky}}]{p21}%
  \BibitemOpen
  \bibfield  {author} {\bibinfo {author} {\bibfnamefont {L.}~\bibnamefont {Ye}}, \bibinfo {author} {\bibfnamefont {M.}~\bibnamefont {Kang}}, \bibinfo {author} {\bibfnamefont {J.}~\bibnamefont {Liu}}, \bibinfo {author} {\bibfnamefont {F.}~\bibnamefont {Von~Cube}}, \bibinfo {author} {\bibfnamefont {C.~R.}\ \bibnamefont {Wicker}}, \bibinfo {author} {\bibfnamefont {T.}~\bibnamefont {Suzuki}}, \bibinfo {author} {\bibfnamefont {C.}~\bibnamefont {Jozwiak}}, \bibinfo {author} {\bibfnamefont {A.}~\bibnamefont {Bostwick}}, \bibinfo {author} {\bibfnamefont {E.}~\bibnamefont {Rotenberg}}, \bibinfo {author} {\bibfnamefont {D.~C.}\ \bibnamefont {Bell}}, \bibinfo {author} {\bibfnamefont {L.}~\bibnamefont {Fu}}, \bibinfo {author} {\bibfnamefont {R.}~\bibnamefont {Comin}},\ and\ \bibinfo {author} {\bibfnamefont {J.~G.}\ \bibnamefont {Checkelsky}},\ }\bibfield  {title} {\bibinfo {title} {{Massive Dirac fermions in a ferromagnetic kagome metal}},\ }\href {https://doi.org/10.1038/nature25987} {\bibfield  {journal} {\bibinfo
  {journal} {Nature}\ }\textbf {\bibinfo {volume} {555}},\ \bibinfo {pages} {638} (\bibinfo {year} {2018})}\BibitemShut {NoStop}%
\bibitem [{\citenamefont {Chen}\ \emph {et~al.}(2022)\citenamefont {Chen}, \citenamefont {Minami}, \citenamefont {Sakai}, \citenamefont {Wang}, \citenamefont {Feng}, \citenamefont {Nomoto}, \citenamefont {Hirayama}, \citenamefont {Ishii}, \citenamefont {Koretsune}, \citenamefont {Arita},\ and\ \citenamefont {Nakatsuji}}]{p22}%
  \BibitemOpen
  \bibfield  {author} {\bibinfo {author} {\bibfnamefont {T.}~\bibnamefont {Chen}}, \bibinfo {author} {\bibfnamefont {S.}~\bibnamefont {Minami}}, \bibinfo {author} {\bibfnamefont {A.}~\bibnamefont {Sakai}}, \bibinfo {author} {\bibfnamefont {Y.}~\bibnamefont {Wang}}, \bibinfo {author} {\bibfnamefont {Z.}~\bibnamefont {Feng}}, \bibinfo {author} {\bibfnamefont {T.}~\bibnamefont {Nomoto}}, \bibinfo {author} {\bibfnamefont {M.}~\bibnamefont {Hirayama}}, \bibinfo {author} {\bibfnamefont {R.}~\bibnamefont {Ishii}}, \bibinfo {author} {\bibfnamefont {T.}~\bibnamefont {Koretsune}}, \bibinfo {author} {\bibfnamefont {R.}~\bibnamefont {Arita}},\ and\ \bibinfo {author} {\bibfnamefont {S.}~\bibnamefont {Nakatsuji}},\ }\bibfield  {title} {\bibinfo {title} {{Large anomalous Nernst effect and nodal plane in an iron-based kagome ferromagnet}},\ }\href {https://doi.org/10.1126/sciadv.abk1480} {\bibfield  {journal} {\bibinfo  {journal} {Sci. Adv.}\ }\textbf {\bibinfo {volume} {8}},\ \bibinfo {pages} {eabk1480} (\bibinfo {year}
  {2022})}\BibitemShut {NoStop}%
\bibitem [{\citenamefont {K\"ubler}\ and\ \citenamefont {Felser}(2012)}]{p26}%
  \BibitemOpen
  \bibfield  {author} {\bibinfo {author} {\bibfnamefont {J.}~\bibnamefont {K\"ubler}}\ and\ \bibinfo {author} {\bibfnamefont {C.}~\bibnamefont {Felser}},\ }\bibfield  {title} {\bibinfo {title} {{Berry curvature and the anomalous Hall effect in Heusler compounds}},\ }\href {https://doi.org/10.1103/PhysRevB.85.012405} {\bibfield  {journal} {\bibinfo  {journal} {Phys. Rev. B}\ }\textbf {\bibinfo {volume} {85}},\ \bibinfo {pages} {012405} (\bibinfo {year} {2012})}\BibitemShut {NoStop}%
\bibitem [{\citenamefont {Chang}\ \emph {et~al.}(2017)\citenamefont {Chang}, \citenamefont {Xu}, \citenamefont {Zhou}, \citenamefont {Huang}, \citenamefont {Singh}, \citenamefont {Wang}, \citenamefont {Belopolski}, \citenamefont {Yin}, \citenamefont {Zhang}, \citenamefont {Bansil}, \citenamefont {Lin},\ and\ \citenamefont {Hasan}}]{p27}%
  \BibitemOpen
  \bibfield  {author} {\bibinfo {author} {\bibfnamefont {G.}~\bibnamefont {Chang}}, \bibinfo {author} {\bibfnamefont {S.-Y.}\ \bibnamefont {Xu}}, \bibinfo {author} {\bibfnamefont {X.}~\bibnamefont {Zhou}}, \bibinfo {author} {\bibfnamefont {S.-M.}\ \bibnamefont {Huang}}, \bibinfo {author} {\bibfnamefont {B.}~\bibnamefont {Singh}}, \bibinfo {author} {\bibfnamefont {B.}~\bibnamefont {Wang}}, \bibinfo {author} {\bibfnamefont {I.}~\bibnamefont {Belopolski}}, \bibinfo {author} {\bibfnamefont {J.}~\bibnamefont {Yin}}, \bibinfo {author} {\bibfnamefont {S.}~\bibnamefont {Zhang}}, \bibinfo {author} {\bibfnamefont {A.}~\bibnamefont {Bansil}}, \bibinfo {author} {\bibfnamefont {H.}~\bibnamefont {Lin}},\ and\ \bibinfo {author} {\bibfnamefont {M.~Z.}\ \bibnamefont {Hasan}},\ }\bibfield  {title} {\bibinfo {title} {{Topological Hopf and Chain Link Semimetal States and Their Application to ${\mathrm{Co}}_{2}\mathrm{Mn}\text{G}\text{a}$}},\ }\href {https://doi.org/10.1103/PhysRevLett.119.156401} {\bibfield  {journal} {\bibinfo
  {journal} {Phys. Rev. Lett.}\ }\textbf {\bibinfo {volume} {119}},\ \bibinfo {pages} {156401} (\bibinfo {year} {2017})}\BibitemShut {NoStop}%
\bibitem [{\citenamefont {Manna}\ \emph {et~al.}(2018)\citenamefont {Manna}, \citenamefont {Muechler}, \citenamefont {Kao}, \citenamefont {Stinshoff}, \citenamefont {Zhang}, \citenamefont {Gooth}, \citenamefont {Kumar}, \citenamefont {Kreiner}, \citenamefont {Koepernik}, \citenamefont {Car}, \citenamefont {K\"ubler}, \citenamefont {Fecher}, \citenamefont {Shekhar}, \citenamefont {Sun},\ and\ \citenamefont {Felser}}]{p28}%
  \BibitemOpen
  \bibfield  {author} {\bibinfo {author} {\bibfnamefont {K.}~\bibnamefont {Manna}}, \bibinfo {author} {\bibfnamefont {L.}~\bibnamefont {Muechler}}, \bibinfo {author} {\bibfnamefont {T.-H.}\ \bibnamefont {Kao}}, \bibinfo {author} {\bibfnamefont {R.}~\bibnamefont {Stinshoff}}, \bibinfo {author} {\bibfnamefont {Y.}~\bibnamefont {Zhang}}, \bibinfo {author} {\bibfnamefont {J.}~\bibnamefont {Gooth}}, \bibinfo {author} {\bibfnamefont {N.}~\bibnamefont {Kumar}}, \bibinfo {author} {\bibfnamefont {G.}~\bibnamefont {Kreiner}}, \bibinfo {author} {\bibfnamefont {K.}~\bibnamefont {Koepernik}}, \bibinfo {author} {\bibfnamefont {R.}~\bibnamefont {Car}}, \bibinfo {author} {\bibfnamefont {J.}~\bibnamefont {K\"ubler}}, \bibinfo {author} {\bibfnamefont {G.~H.}\ \bibnamefont {Fecher}}, \bibinfo {author} {\bibfnamefont {C.}~\bibnamefont {Shekhar}}, \bibinfo {author} {\bibfnamefont {Y.}~\bibnamefont {Sun}},\ and\ \bibinfo {author} {\bibfnamefont {C.}~\bibnamefont {Felser}},\ }\bibfield  {title} {\bibinfo {title} {{From Colossal to
  Zero: Controlling the Anomalous Hall Effect in Magnetic Heusler Compounds via Berry Curvature Design}},\ }\href {https://doi.org/10.1103/PhysRevX.8.041045} {\bibfield  {journal} {\bibinfo  {journal} {Phys. Rev. X}\ }\textbf {\bibinfo {volume} {8}},\ \bibinfo {pages} {041045} (\bibinfo {year} {2018})}\BibitemShut {NoStop}%
\bibitem [{\citenamefont {Graf}\ \emph {et~al.}(2010)\citenamefont {Graf}, \citenamefont {Barth}, \citenamefont {Blum}, \citenamefont {Balke}, \citenamefont {Felser}, \citenamefont {Klaer},\ and\ \citenamefont {Elmers}}]{p30}%
  \BibitemOpen
  \bibfield  {author} {\bibinfo {author} {\bibfnamefont {T.}~\bibnamefont {Graf}}, \bibinfo {author} {\bibfnamefont {J.}~\bibnamefont {Barth}}, \bibinfo {author} {\bibfnamefont {C.~G.~F.}\ \bibnamefont {Blum}}, \bibinfo {author} {\bibfnamefont {B.}~\bibnamefont {Balke}}, \bibinfo {author} {\bibfnamefont {C.}~\bibnamefont {Felser}}, \bibinfo {author} {\bibfnamefont {P.}~\bibnamefont {Klaer}},\ and\ \bibinfo {author} {\bibfnamefont {H.-J.}\ \bibnamefont {Elmers}},\ }\bibfield  {title} {\bibinfo {title} {{Phase-separation-induced changes in the magnetic and transport properties of the quaternary Heusler alloy $\text{Co}_{2}\text{Mn}_{1\ensuremath{-}x}\text{Ti}_{x}\text{Sn}$}},\ }\href {https://doi.org/10.1103/PhysRevB.82.104420} {\bibfield  {journal} {\bibinfo  {journal} {Phys. Rev. B}\ }\textbf {\bibinfo {volume} {82}},\ \bibinfo {pages} {104420} (\bibinfo {year} {2010})}\BibitemShut {NoStop}%
\bibitem [{\citenamefont {Chanda}\ \emph {et~al.}(2022{\natexlab{a}})\citenamefont {Chanda}, \citenamefont {Rani}, \citenamefont {Nag}, \citenamefont {Alam}, \citenamefont {Suresh}, \citenamefont {Phan},\ and\ \citenamefont {Srikanth}}]{Chanda2022_1}%
  \BibitemOpen
  \bibfield  {author} {\bibinfo {author} {\bibfnamefont {A.}~\bibnamefont {Chanda}}, \bibinfo {author} {\bibfnamefont {D.}~\bibnamefont {Rani}}, \bibinfo {author} {\bibfnamefont {J.}~\bibnamefont {Nag}}, \bibinfo {author} {\bibfnamefont {A.}~\bibnamefont {Alam}}, \bibinfo {author} {\bibfnamefont {K.~G.}\ \bibnamefont {Suresh}}, \bibinfo {author} {\bibfnamefont {M.~H.}\ \bibnamefont {Phan}},\ and\ \bibinfo {author} {\bibfnamefont {H.}~\bibnamefont {Srikanth}},\ }\bibfield  {title} {\bibinfo {title} {{Emergence of asymmetric skew-scattering dominated anomalous Nernst effect in the spin gapless semiconductors ${\mathrm{Co}}_{1+x}{\mathrm{Fe}}_{1\ensuremath{-}x}\mathrm{CrGa}$}},\ }\href {https://doi.org/10.1103/PhysRevB.106.134416} {\bibfield  {journal} {\bibinfo  {journal} {Phys. Rev. B}\ }\textbf {\bibinfo {volume} {106}},\ \bibinfo {pages} {134416} (\bibinfo {year} {2022}{\natexlab{a}})}\BibitemShut {NoStop}%
\bibitem [{\citenamefont {Chanda}\ \emph {et~al.}(2022{\natexlab{b}})\citenamefont {Chanda}, \citenamefont {Holzmann}, \citenamefont {Schulz}, \citenamefont {Seyd}, \citenamefont {Albrecht}, \citenamefont {Phan},\ and\ \citenamefont {Srikanth}}]{Chanda2022_2}%
  \BibitemOpen
  \bibfield  {author} {\bibinfo {author} {\bibfnamefont {A.}~\bibnamefont {Chanda}}, \bibinfo {author} {\bibfnamefont {C.}~\bibnamefont {Holzmann}}, \bibinfo {author} {\bibfnamefont {N.}~\bibnamefont {Schulz}}, \bibinfo {author} {\bibfnamefont {J.}~\bibnamefont {Seyd}}, \bibinfo {author} {\bibfnamefont {M.}~\bibnamefont {Albrecht}}, \bibinfo {author} {\bibfnamefont {M.-H.}\ \bibnamefont {Phan}},\ and\ \bibinfo {author} {\bibfnamefont {H.}~\bibnamefont {Srikanth}},\ }\bibfield  {title} {\bibinfo {title} {{Scaling of the Thermally Induced Sign Inversion of Longitudinal Spin Seebeck Effect in a Compensated Ferrimagnet: Role of Magnetic Anisotropy}},\ }\href {https://doi.org/https://doi.org/10.1002/adfm.202109170} {\bibfield  {journal} {\bibinfo  {journal} {Adv. Funct. Mater.}\ }\textbf {\bibinfo {volume} {32}},\ \bibinfo {pages} {2109170} (\bibinfo {year} {2022}{\natexlab{b}})}\BibitemShut {NoStop}%
\bibitem [{\citenamefont {Chanda}\ \emph {et~al.}(2023{\natexlab{a}})\citenamefont {Chanda}, \citenamefont {Nag}, \citenamefont {Alam}, \citenamefont {Suresh}, \citenamefont {Phan},\ and\ \citenamefont {Srikanth}}]{Chanda2023_1}%
  \BibitemOpen
  \bibfield  {author} {\bibinfo {author} {\bibfnamefont {A.}~\bibnamefont {Chanda}}, \bibinfo {author} {\bibfnamefont {J.}~\bibnamefont {Nag}}, \bibinfo {author} {\bibfnamefont {A.}~\bibnamefont {Alam}}, \bibinfo {author} {\bibfnamefont {K.~G.}\ \bibnamefont {Suresh}}, \bibinfo {author} {\bibfnamefont {M.-H.}\ \bibnamefont {Phan}},\ and\ \bibinfo {author} {\bibfnamefont {H.}~\bibnamefont {Srikanth}},\ }\bibfield  {title} {\bibinfo {title} {{Intrinsic Berry curvature driven anomalous Nernst thermopower in the semimetallic Heusler alloy CoFeVSb}},\ }\href {https://doi.org/10.1103/PhysRevB.107.L220403} {\bibfield  {journal} {\bibinfo  {journal} {Phys. Rev. B}\ }\textbf {\bibinfo {volume} {107}},\ \bibinfo {pages} {L220403} (\bibinfo {year} {2023}{\natexlab{a}})}\BibitemShut {NoStop}%
\bibitem [{\citenamefont {Chanda}\ \emph {et~al.}(2023{\natexlab{b}})\citenamefont {Chanda}, \citenamefont {Rani}, \citenamefont {DeTellem}, \citenamefont {Alzahrani}, \citenamefont {Arena}, \citenamefont {Witanachchi}, \citenamefont {Chatterjee}, \citenamefont {Phan},\ and\ \citenamefont {Srikanth}}]{Chanda2023_2}%
  \BibitemOpen
  \bibfield  {author} {\bibinfo {author} {\bibfnamefont {A.}~\bibnamefont {Chanda}}, \bibinfo {author} {\bibfnamefont {D.}~\bibnamefont {Rani}}, \bibinfo {author} {\bibfnamefont {D.}~\bibnamefont {DeTellem}}, \bibinfo {author} {\bibfnamefont {N.}~\bibnamefont {Alzahrani}}, \bibinfo {author} {\bibfnamefont {D.~A.}\ \bibnamefont {Arena}}, \bibinfo {author} {\bibfnamefont {S.}~\bibnamefont {Witanachchi}}, \bibinfo {author} {\bibfnamefont {R.}~\bibnamefont {Chatterjee}}, \bibinfo {author} {\bibfnamefont {M.-H.}\ \bibnamefont {Phan}},\ and\ \bibinfo {author} {\bibfnamefont {H.}~\bibnamefont {Srikanth}},\ }\bibfield  {title} {\bibinfo {title} {{Large Thermo-Spin Effects in Heusler Alloy-Based Spin Gapless Semiconductor Thin Films}},\ }\href {https://doi.org/10.1021/acsami.3c12342} {\bibfield  {journal} {\bibinfo  {journal} {ACS Appl. Mater. Interfaces}\ }\textbf {\bibinfo {volume} {15}},\ \bibinfo {pages} {53697} (\bibinfo {year} {2023}{\natexlab{b}})}\BibitemShut {NoStop}%
\bibitem [{\citenamefont {Chanda}\ \emph {et~al.}(2024)\citenamefont {Chanda}, \citenamefont {Nag}, \citenamefont {Schulz}, \citenamefont {Alam}, \citenamefont {Suresh}, \citenamefont {Phan},\ and\ \citenamefont {Srikanth}}]{Chanda2024}%
  \BibitemOpen
  \bibfield  {author} {\bibinfo {author} {\bibfnamefont {A.}~\bibnamefont {Chanda}}, \bibinfo {author} {\bibfnamefont {J.}~\bibnamefont {Nag}}, \bibinfo {author} {\bibfnamefont {N.}~\bibnamefont {Schulz}}, \bibinfo {author} {\bibfnamefont {A.}~\bibnamefont {Alam}}, \bibinfo {author} {\bibfnamefont {K.~G.}\ \bibnamefont {Suresh}}, \bibinfo {author} {\bibfnamefont {M.-H.}\ \bibnamefont {Phan}},\ and\ \bibinfo {author} {\bibfnamefont {H.}~\bibnamefont {Srikanth}},\ }\bibfield  {title} {\bibinfo {title} {{Large anomalous Nernst effect and its bipolarity in the quaternary equiatomic Heusler alloys $\mathrm{CrRu}X\mathrm{Ge}$ ($X=\mathrm{Co}$ and Mn)}},\ }\href {https://doi.org/10.1103/PhysRevB.109.224415} {\bibfield  {journal} {\bibinfo  {journal} {Phys. Rev. B}\ }\textbf {\bibinfo {volume} {109}},\ \bibinfo {pages} {224415} (\bibinfo {year} {2024})}\BibitemShut {NoStop}%
\bibitem [{\citenamefont {Hohenberg}\ and\ \citenamefont {Kohn}(1964)}]{Hohenberg_Kohn1964}%
  \BibitemOpen
  \bibfield  {author} {\bibinfo {author} {\bibfnamefont {P.}~\bibnamefont {Hohenberg}}\ and\ \bibinfo {author} {\bibfnamefont {W.}~\bibnamefont {Kohn}},\ }\bibfield  {title} {\bibinfo {title} {{Inhomogeneous Electron Gas}},\ }\href {https://doi.org/10.1103/PhysRev.136.B864} {\bibfield  {journal} {\bibinfo  {journal} {Phys. Rev.}\ }\textbf {\bibinfo {volume} {136}},\ \bibinfo {pages} {B864} (\bibinfo {year} {1964})}\BibitemShut {NoStop}%
\bibitem [{\citenamefont {Kohn}\ and\ \citenamefont {Sham}(1965)}]{Kohn_Sham1965}%
  \BibitemOpen
  \bibfield  {author} {\bibinfo {author} {\bibfnamefont {W.}~\bibnamefont {Kohn}}\ and\ \bibinfo {author} {\bibfnamefont {L.~J.}\ \bibnamefont {Sham}},\ }\bibfield  {title} {\bibinfo {title} {{Self-Consistent Equations Including Exchange and Correlation Effects}},\ }\href {https://doi.org/10.1103/PhysRev.140.A1133} {\bibfield  {journal} {\bibinfo  {journal} {Phys. Rev.}\ }\textbf {\bibinfo {volume} {140}},\ \bibinfo {pages} {A1133} (\bibinfo {year} {1965})}\BibitemShut {NoStop}%
\bibitem [{\citenamefont {Kresse}\ and\ \citenamefont {Hafner}(1993)}]{Kresse1993}%
  \BibitemOpen
  \bibfield  {author} {\bibinfo {author} {\bibfnamefont {G.}~\bibnamefont {Kresse}}\ and\ \bibinfo {author} {\bibfnamefont {J.}~\bibnamefont {Hafner}},\ }\bibfield  {title} {\bibinfo {title} {{Ab initio molecular dynamics for liquid metals}},\ }\href {https://doi.org/10.1103/PhysRevB.47.558} {\bibfield  {journal} {\bibinfo  {journal} {Phys. Rev. B}\ }\textbf {\bibinfo {volume} {47}},\ \bibinfo {pages} {558} (\bibinfo {year} {1993})}\BibitemShut {NoStop}%
\bibitem [{\citenamefont {Kresse}\ and\ \citenamefont {Furthm\"uller}(1996)}]{Kresse1996_1}%
  \BibitemOpen
  \bibfield  {author} {\bibinfo {author} {\bibfnamefont {G.}~\bibnamefont {Kresse}}\ and\ \bibinfo {author} {\bibfnamefont {J.}~\bibnamefont {Furthm\"uller}},\ }\bibfield  {title} {\bibinfo {title} {{Efficient iterative schemes for ab initio total-energy calculations using a plane-wave basis set}},\ }\href {https://doi.org/10.1103/PhysRevB.54.11169} {\bibfield  {journal} {\bibinfo  {journal} {Phys. Rev. B}\ }\textbf {\bibinfo {volume} {54}},\ \bibinfo {pages} {11169} (\bibinfo {year} {1996})}\BibitemShut {NoStop}%
\bibitem [{\citenamefont {Kresse}\ and\ \citenamefont {Furthm{\"u}ller}(1996)}]{Kresse1996_2}%
  \BibitemOpen
  \bibfield  {author} {\bibinfo {author} {\bibfnamefont {G.}~\bibnamefont {Kresse}}\ and\ \bibinfo {author} {\bibfnamefont {J.}~\bibnamefont {Furthm{\"u}ller}},\ }\bibfield  {title} {\bibinfo {title} {{Efficiency of ab-initio total energy calculations for metals and semiconductors using a plane-wave basis set}},\ }\href {https://doi.org/10.1016/0927-0256(96)00008-0} {\bibfield  {journal} {\bibinfo  {journal} {Comput. Mater. Sci.}\ }\textbf {\bibinfo {volume} {6}},\ \bibinfo {pages} {15} (\bibinfo {year} {1996})}\BibitemShut {NoStop}%
\bibitem [{\citenamefont {Kresse}\ and\ \citenamefont {Joubert}(1999)}]{Kresse1999}%
  \BibitemOpen
  \bibfield  {author} {\bibinfo {author} {\bibfnamefont {G.}~\bibnamefont {Kresse}}\ and\ \bibinfo {author} {\bibfnamefont {D.}~\bibnamefont {Joubert}},\ }\bibfield  {title} {\bibinfo {title} {{From ultrasoft pseudopotentials to the projector augmented-wave method}},\ }\href {https://doi.org/10.1103/PhysRevB.59.1758} {\bibfield  {journal} {\bibinfo  {journal} {Phys. Rev. B}\ }\textbf {\bibinfo {volume} {59}},\ \bibinfo {pages} {1758} (\bibinfo {year} {1999})}\BibitemShut {NoStop}%
\bibitem [{\citenamefont {Bl\"ochl}(1994)}]{Blochl1994_paw}%
  \BibitemOpen
  \bibfield  {author} {\bibinfo {author} {\bibfnamefont {P.~E.}\ \bibnamefont {Bl\"ochl}},\ }\bibfield  {title} {\bibinfo {title} {{Projector augmented-wave method}},\ }\href {https://doi.org/10.1103/PhysRevB.50.17953} {\bibfield  {journal} {\bibinfo  {journal} {Phys. Rev. B}\ }\textbf {\bibinfo {volume} {50}},\ \bibinfo {pages} {17953} (\bibinfo {year} {1994})}\BibitemShut {NoStop}%
\bibitem [{\citenamefont {Sun}\ \emph {et~al.}(2015)\citenamefont {Sun}, \citenamefont {Ruzsinszky},\ and\ \citenamefont {Perdew}}]{Sun2015}%
  \BibitemOpen
  \bibfield  {author} {\bibinfo {author} {\bibfnamefont {J.}~\bibnamefont {Sun}}, \bibinfo {author} {\bibfnamefont {A.}~\bibnamefont {Ruzsinszky}},\ and\ \bibinfo {author} {\bibfnamefont {J.~P.}\ \bibnamefont {Perdew}},\ }\bibfield  {title} {\bibinfo {title} {{Strongly Constrained and Appropriately Normed Semilocal Density Functional}},\ }\href {https://doi.org/10.1103/PhysRevLett.115.036402} {\bibfield  {journal} {\bibinfo  {journal} {Phys. Rev. Lett.}\ }\textbf {\bibinfo {volume} {115}},\ \bibinfo {pages} {036402} (\bibinfo {year} {2015})}\BibitemShut {NoStop}%
\bibitem [{\citenamefont {Bl\"ochl}\ \emph {et~al.}(1994)\citenamefont {Bl\"ochl}, \citenamefont {Jepsen},\ and\ \citenamefont {Andersen}}]{Blochl1994_tet}%
  \BibitemOpen
  \bibfield  {author} {\bibinfo {author} {\bibfnamefont {P.~E.}\ \bibnamefont {Bl\"ochl}}, \bibinfo {author} {\bibfnamefont {O.}~\bibnamefont {Jepsen}},\ and\ \bibinfo {author} {\bibfnamefont {O.~K.}\ \bibnamefont {Andersen}},\ }\bibfield  {title} {\bibinfo {title} {{Improved tetrahedron method for Brillouin-zone integrations}},\ }\href {https://doi.org/10.1103/PhysRevB.49.16223} {\bibfield  {journal} {\bibinfo  {journal} {Phys. Rev. B}\ }\textbf {\bibinfo {volume} {49}},\ \bibinfo {pages} {16223} (\bibinfo {year} {1994})}\BibitemShut {NoStop}%
\bibitem [{\citenamefont {Marzari}\ and\ \citenamefont {Vanderbilt}(1997)}]{Marzari1997}%
  \BibitemOpen
  \bibfield  {author} {\bibinfo {author} {\bibfnamefont {N.}~\bibnamefont {Marzari}}\ and\ \bibinfo {author} {\bibfnamefont {D.}~\bibnamefont {Vanderbilt}},\ }\bibfield  {title} {\bibinfo {title} {{Maximally localized generalized Wannier functions for composite energy bands}},\ }\href {https://doi.org/10.1103/PhysRevB.56.12847} {\bibfield  {journal} {\bibinfo  {journal} {Phys. Rev. B}\ }\textbf {\bibinfo {volume} {56}},\ \bibinfo {pages} {12847} (\bibinfo {year} {1997})}\BibitemShut {NoStop}%
\bibitem [{\citenamefont {Souza}\ \emph {et~al.}(2001)\citenamefont {Souza}, \citenamefont {Marzari},\ and\ \citenamefont {Vanderbilt}}]{Souza2001}%
  \BibitemOpen
  \bibfield  {author} {\bibinfo {author} {\bibfnamefont {I.}~\bibnamefont {Souza}}, \bibinfo {author} {\bibfnamefont {N.}~\bibnamefont {Marzari}},\ and\ \bibinfo {author} {\bibfnamefont {D.}~\bibnamefont {Vanderbilt}},\ }\bibfield  {title} {\bibinfo {title} {{Maximally localized Wannier functions for entangled energy bands}},\ }\href {https://doi.org/10.1103/PhysRevB.65.035109} {\bibfield  {journal} {\bibinfo  {journal} {Phys. Rev. B}\ }\textbf {\bibinfo {volume} {65}},\ \bibinfo {pages} {035109} (\bibinfo {year} {2001})}\BibitemShut {NoStop}%
\bibitem [{\citenamefont {Marzari}\ \emph {et~al.}(2012)\citenamefont {Marzari}, \citenamefont {Mostofi}, \citenamefont {Yates}, \citenamefont {Souza},\ and\ \citenamefont {Vanderbilt}}]{Marzari2012}%
  \BibitemOpen
  \bibfield  {author} {\bibinfo {author} {\bibfnamefont {N.}~\bibnamefont {Marzari}}, \bibinfo {author} {\bibfnamefont {A.~A.}\ \bibnamefont {Mostofi}}, \bibinfo {author} {\bibfnamefont {J.~R.}\ \bibnamefont {Yates}}, \bibinfo {author} {\bibfnamefont {I.}~\bibnamefont {Souza}},\ and\ \bibinfo {author} {\bibfnamefont {D.}~\bibnamefont {Vanderbilt}},\ }\bibfield  {title} {\bibinfo {title} {{Maximally localized Wannier functions: Theory and applications}},\ }\href {https://doi.org/10.1103/RevModPhys.84.1419} {\bibfield  {journal} {\bibinfo  {journal} {Rev. Mod. Phys.}\ }\textbf {\bibinfo {volume} {84}},\ \bibinfo {pages} {1419} (\bibinfo {year} {2012})}\BibitemShut {NoStop}%
\bibitem [{\citenamefont {Mostofi}\ \emph {et~al.}(2008)\citenamefont {Mostofi}, \citenamefont {Yates}, \citenamefont {Lee}, \citenamefont {Souza}, \citenamefont {Vanderbilt},\ and\ \citenamefont {Marzari}}]{wannier90v1}%
  \BibitemOpen
  \bibfield  {author} {\bibinfo {author} {\bibfnamefont {A.~A.}\ \bibnamefont {Mostofi}}, \bibinfo {author} {\bibfnamefont {J.~R.}\ \bibnamefont {Yates}}, \bibinfo {author} {\bibfnamefont {Y.-S.}\ \bibnamefont {Lee}}, \bibinfo {author} {\bibfnamefont {I.}~\bibnamefont {Souza}}, \bibinfo {author} {\bibfnamefont {D.}~\bibnamefont {Vanderbilt}},\ and\ \bibinfo {author} {\bibfnamefont {N.}~\bibnamefont {Marzari}},\ }\bibfield  {title} {\bibinfo {title} {{wannier90: A tool for obtaining maximally-localised Wannier functions}},\ }\href {https://doi.org/10.1016/j.cpc.2007.11.016} {\bibfield  {journal} {\bibinfo  {journal} {Comput. Phys. Commun.}\ }\textbf {\bibinfo {volume} {178}},\ \bibinfo {pages} {685} (\bibinfo {year} {2008})}\BibitemShut {NoStop}%
\bibitem [{\citenamefont {Mostofi}\ \emph {et~al.}(2014)\citenamefont {Mostofi}, \citenamefont {Yates}, \citenamefont {Pizzi}, \citenamefont {Lee}, \citenamefont {Souza}, \citenamefont {Vanderbilt},\ and\ \citenamefont {Marzari}}]{wannier90v2}%
  \BibitemOpen
  \bibfield  {author} {\bibinfo {author} {\bibfnamefont {A.~A.}\ \bibnamefont {Mostofi}}, \bibinfo {author} {\bibfnamefont {J.~R.}\ \bibnamefont {Yates}}, \bibinfo {author} {\bibfnamefont {G.}~\bibnamefont {Pizzi}}, \bibinfo {author} {\bibfnamefont {Y.-S.}\ \bibnamefont {Lee}}, \bibinfo {author} {\bibfnamefont {I.}~\bibnamefont {Souza}}, \bibinfo {author} {\bibfnamefont {D.}~\bibnamefont {Vanderbilt}},\ and\ \bibinfo {author} {\bibfnamefont {N.}~\bibnamefont {Marzari}},\ }\bibfield  {title} {\bibinfo {title} {{An updated version of wannier90: A tool for obtaining maximally-localised Wannier functions}},\ }\href {https://doi.org/https://doi.org/10.1016/j.cpc.2014.05.003} {\bibfield  {journal} {\bibinfo  {journal} {Comput. Phys. Commun.}\ }\textbf {\bibinfo {volume} {185}},\ \bibinfo {pages} {2309} (\bibinfo {year} {2014})}\BibitemShut {NoStop}%
\bibitem [{\citenamefont {Pizzi}\ \emph {et~al.}(2020)\citenamefont {Pizzi}, \citenamefont {Vitale}, \citenamefont {Arita}, \citenamefont {Blügel}, \citenamefont {Freimuth}, \citenamefont {Géranton}, \citenamefont {Gibertini}, \citenamefont {Gresch}, \citenamefont {Johnson}, \citenamefont {Koretsune}, \citenamefont {Ibañez-Azpiroz}, \citenamefont {Lee}, \citenamefont {Lihm}, \citenamefont {Marchand}, \citenamefont {Marrazzo}, \citenamefont {Mokrousov}, \citenamefont {Mustafa}, \citenamefont {Nohara}, \citenamefont {Nomura}, \citenamefont {Paulatto}, \citenamefont {Poncé}, \citenamefont {Ponweiser}, \citenamefont {Qiao}, \citenamefont {Thöle}, \citenamefont {Tsirkin}, \citenamefont {Wierzbowska}, \citenamefont {Marzari}, \citenamefont {Vanderbilt}, \citenamefont {Souza}, \citenamefont {Mostofi},\ and\ \citenamefont {Yates}}]{wannier90v3}%
  \BibitemOpen
  \bibfield  {author} {\bibinfo {author} {\bibfnamefont {G.}~\bibnamefont {Pizzi}}, \bibinfo {author} {\bibfnamefont {V.}~\bibnamefont {Vitale}}, \bibinfo {author} {\bibfnamefont {R.}~\bibnamefont {Arita}}, \bibinfo {author} {\bibfnamefont {S.}~\bibnamefont {Blügel}}, \bibinfo {author} {\bibfnamefont {F.}~\bibnamefont {Freimuth}}, \bibinfo {author} {\bibfnamefont {G.}~\bibnamefont {Géranton}}, \bibinfo {author} {\bibfnamefont {M.}~\bibnamefont {Gibertini}}, \bibinfo {author} {\bibfnamefont {D.}~\bibnamefont {Gresch}}, \bibinfo {author} {\bibfnamefont {C.}~\bibnamefont {Johnson}}, \bibinfo {author} {\bibfnamefont {T.}~\bibnamefont {Koretsune}}, \bibinfo {author} {\bibfnamefont {J.}~\bibnamefont {Ibañez-Azpiroz}}, \bibinfo {author} {\bibfnamefont {H.}~\bibnamefont {Lee}}, \bibinfo {author} {\bibfnamefont {J.-M.}\ \bibnamefont {Lihm}}, \bibinfo {author} {\bibfnamefont {D.}~\bibnamefont {Marchand}}, \bibinfo {author} {\bibfnamefont {A.}~\bibnamefont {Marrazzo}}, \bibinfo {author} {\bibfnamefont
  {Y.}~\bibnamefont {Mokrousov}}, \bibinfo {author} {\bibfnamefont {J.~I.}\ \bibnamefont {Mustafa}}, \bibinfo {author} {\bibfnamefont {Y.}~\bibnamefont {Nohara}}, \bibinfo {author} {\bibfnamefont {Y.}~\bibnamefont {Nomura}}, \bibinfo {author} {\bibfnamefont {L.}~\bibnamefont {Paulatto}}, \bibinfo {author} {\bibfnamefont {S.}~\bibnamefont {Poncé}}, \bibinfo {author} {\bibfnamefont {T.}~\bibnamefont {Ponweiser}}, \bibinfo {author} {\bibfnamefont {J.}~\bibnamefont {Qiao}}, \bibinfo {author} {\bibfnamefont {F.}~\bibnamefont {Thöle}}, \bibinfo {author} {\bibfnamefont {S.~S.}\ \bibnamefont {Tsirkin}}, \bibinfo {author} {\bibfnamefont {M.}~\bibnamefont {Wierzbowska}}, \bibinfo {author} {\bibfnamefont {N.}~\bibnamefont {Marzari}}, \bibinfo {author} {\bibfnamefont {D.}~\bibnamefont {Vanderbilt}}, \bibinfo {author} {\bibfnamefont {I.}~\bibnamefont {Souza}}, \bibinfo {author} {\bibfnamefont {A.~A.}\ \bibnamefont {Mostofi}},\ and\ \bibinfo {author} {\bibfnamefont {J.~R.}\ \bibnamefont {Yates}},\ }\bibfield  {title}
  {\bibinfo {title} {{Wannier90 as a community code: new features and applications}},\ }\href {https://doi.org/10.1088/1361-648X/ab51ff} {\bibfield  {journal} {\bibinfo  {journal} {J. Phys.: Condens. Matter}\ }\textbf {\bibinfo {volume} {32}},\ \bibinfo {pages} {165902} (\bibinfo {year} {2020})}\BibitemShut {NoStop}%
\bibitem [{sup()}]{supp}%
  \BibitemOpen
  \href@noop {} {}\bibinfo {note} {{See Supplementary Materials for further auxiliary informations about the crystal structure, magnetoresistance, anomalous Hall and Nernst effects, {\it ab-initio} spin polarized electronic structure and temperature dependence of anomalous Hall conductivity.}}\BibitemShut {Stop}%
\bibitem [{\citenamefont {Nagaosa}\ \emph {et~al.}(2010)\citenamefont {Nagaosa}, \citenamefont {Sinova}, \citenamefont {Onoda}, \citenamefont {MacDonald},\ and\ \citenamefont {Ong}}]{Nagaosa2010}%
  \BibitemOpen
  \bibfield  {author} {\bibinfo {author} {\bibfnamefont {N.}~\bibnamefont {Nagaosa}}, \bibinfo {author} {\bibfnamefont {J.}~\bibnamefont {Sinova}}, \bibinfo {author} {\bibfnamefont {S.}~\bibnamefont {Onoda}}, \bibinfo {author} {\bibfnamefont {A.~H.}\ \bibnamefont {MacDonald}},\ and\ \bibinfo {author} {\bibfnamefont {N.~P.}\ \bibnamefont {Ong}},\ }\bibfield  {title} {\bibinfo {title} {{Anomalous Hall effect}},\ }\href {https://doi.org/10.1103/RevModPhys.82.1539} {\bibfield  {journal} {\bibinfo  {journal} {Rev. Mod. Phys.}\ }\textbf {\bibinfo {volume} {82}},\ \bibinfo {pages} {1539} (\bibinfo {year} {2010})}\BibitemShut {NoStop}%
\bibitem [{\citenamefont {Liu}\ \emph {et~al.}(2021{\natexlab{a}})\citenamefont {Liu}, \citenamefont {Wang}, \citenamefont {Fu}, \citenamefont {Ge}, \citenamefont {Li}, \citenamefont {Xi}, \citenamefont {Zhang}, \citenamefont {Yan}, \citenamefont {Mandrus}, \citenamefont {Yan},\ and\ \citenamefont {Wang}}]{p8}%
  \BibitemOpen
  \bibfield  {author} {\bibinfo {author} {\bibfnamefont {Y.}~\bibnamefont {Liu}}, \bibinfo {author} {\bibfnamefont {H.}~\bibnamefont {Wang}}, \bibinfo {author} {\bibfnamefont {H.}~\bibnamefont {Fu}}, \bibinfo {author} {\bibfnamefont {J.}~\bibnamefont {Ge}}, \bibinfo {author} {\bibfnamefont {Y.}~\bibnamefont {Li}}, \bibinfo {author} {\bibfnamefont {C.}~\bibnamefont {Xi}}, \bibinfo {author} {\bibfnamefont {J.}~\bibnamefont {Zhang}}, \bibinfo {author} {\bibfnamefont {J.}~\bibnamefont {Yan}}, \bibinfo {author} {\bibfnamefont {D.}~\bibnamefont {Mandrus}}, \bibinfo {author} {\bibfnamefont {B.}~\bibnamefont {Yan}},\ and\ \bibinfo {author} {\bibfnamefont {J.}~\bibnamefont {Wang}},\ }\bibfield  {title} {\bibinfo {title} {{Induced anomalous Hall effect of massive Dirac fermions in $\mathrm{Zr}\mathrm{Te}_{5}$ and $\mathrm{Hf}\mathrm{Te}_{5}$ thin flakes}},\ }\href {https://doi.org/10.1103/PhysRevB.103.L201110} {\bibfield  {journal} {\bibinfo  {journal} {Phys. Rev. B}\ }\textbf {\bibinfo {volume} {103}},\ \bibinfo
  {pages} {L201110} (\bibinfo {year} {2021}{\natexlab{a}})}\BibitemShut {NoStop}%
\bibitem [{\citenamefont {Tian}\ \emph {et~al.}(2009)\citenamefont {Tian}, \citenamefont {Ye},\ and\ \citenamefont {Jin}}]{p18}%
  \BibitemOpen
  \bibfield  {author} {\bibinfo {author} {\bibfnamefont {Y.}~\bibnamefont {Tian}}, \bibinfo {author} {\bibfnamefont {L.}~\bibnamefont {Ye}},\ and\ \bibinfo {author} {\bibfnamefont {X.}~\bibnamefont {Jin}},\ }\bibfield  {title} {\bibinfo {title} {{Proper Scaling of the Anomalous Hall Effect}},\ }\href {https://doi.org/10.1103/PhysRevLett.103.087206} {\bibfield  {journal} {\bibinfo  {journal} {Phys. Rev. Lett.}\ }\textbf {\bibinfo {volume} {103}},\ \bibinfo {pages} {087206} (\bibinfo {year} {2009})}\BibitemShut {NoStop}%
\bibitem [{\citenamefont {Chakraborty}\ \emph {et~al.}(2022)\citenamefont {Chakraborty}, \citenamefont {Samanta}, \citenamefont {Guin}, \citenamefont {Noky}, \citenamefont {Robredo}, \citenamefont {Prasad}, \citenamefont {Kuebler}, \citenamefont {Shekhar}, \citenamefont {Vergniory},\ and\ \citenamefont {Felser}}]{p19}%
  \BibitemOpen
  \bibfield  {author} {\bibinfo {author} {\bibfnamefont {T.}~\bibnamefont {Chakraborty}}, \bibinfo {author} {\bibfnamefont {K.}~\bibnamefont {Samanta}}, \bibinfo {author} {\bibfnamefont {S.~N.}\ \bibnamefont {Guin}}, \bibinfo {author} {\bibfnamefont {J.}~\bibnamefont {Noky}}, \bibinfo {author} {\bibfnamefont {I.~n.}\ \bibnamefont {Robredo}}, \bibinfo {author} {\bibfnamefont {S.}~\bibnamefont {Prasad}}, \bibinfo {author} {\bibfnamefont {J.}~\bibnamefont {Kuebler}}, \bibinfo {author} {\bibfnamefont {C.}~\bibnamefont {Shekhar}}, \bibinfo {author} {\bibfnamefont {M.~G.}\ \bibnamefont {Vergniory}},\ and\ \bibinfo {author} {\bibfnamefont {C.}~\bibnamefont {Felser}},\ }\bibfield  {title} {\bibinfo {title} {{Berry curvature induced anomalous Hall conductivity in the magnetic topological oxide double perovskite $\mathrm{Sr}_{2}\mathrm{FeMoO}_{6}$}},\ }\href {https://doi.org/10.1103/PhysRevB.106.155141} {\bibfield  {journal} {\bibinfo  {journal} {Phys. Rev. B}\ }\textbf {\bibinfo {volume} {106}},\ \bibinfo {pages}
  {155141} (\bibinfo {year} {2022})}\BibitemShut {NoStop}%
\bibitem [{\citenamefont {Shukla}\ \emph {et~al.}(2021)\citenamefont {Shukla}, \citenamefont {Sau}, \citenamefont {Shahi}, \citenamefont {Singh}, \citenamefont {Kumar},\ and\ \citenamefont {Singh}}]{p12}%
  \BibitemOpen
  \bibfield  {author} {\bibinfo {author} {\bibfnamefont {G.~K.}\ \bibnamefont {Shukla}}, \bibinfo {author} {\bibfnamefont {J.}~\bibnamefont {Sau}}, \bibinfo {author} {\bibfnamefont {N.}~\bibnamefont {Shahi}}, \bibinfo {author} {\bibfnamefont {A.~K.}\ \bibnamefont {Singh}}, \bibinfo {author} {\bibfnamefont {M.}~\bibnamefont {Kumar}},\ and\ \bibinfo {author} {\bibfnamefont {S.}~\bibnamefont {Singh}},\ }\bibfield  {title} {\bibinfo {title} {{Anomalous Hall effect from gapped nodal line in the $\mathrm{Co}_{2}\mathrm{FeGe}$ Heusler compound}},\ }\href {https://doi.org/10.1103/PhysRevB.104.195108} {\bibfield  {journal} {\bibinfo  {journal} {Phys. Rev. B}\ }\textbf {\bibinfo {volume} {104}},\ \bibinfo {pages} {195108} (\bibinfo {year} {2021})}\BibitemShut {NoStop}%
\bibitem [{\citenamefont {Liu}\ \emph {et~al.}(2021{\natexlab{b}})\citenamefont {Liu}, \citenamefont {Tan}, \citenamefont {Hu}, \citenamefont {Yan},\ and\ \citenamefont {Petrovic}}]{p14}%
  \BibitemOpen
  \bibfield  {author} {\bibinfo {author} {\bibfnamefont {Y.}~\bibnamefont {Liu}}, \bibinfo {author} {\bibfnamefont {H.}~\bibnamefont {Tan}}, \bibinfo {author} {\bibfnamefont {Z.}~\bibnamefont {Hu}}, \bibinfo {author} {\bibfnamefont {B.}~\bibnamefont {Yan}},\ and\ \bibinfo {author} {\bibfnamefont {C.}~\bibnamefont {Petrovic}},\ }\bibfield  {title} {\bibinfo {title} {{Anomalous Hall effect in the weak-itinerant ferrimagnet $\mathrm{FeCr}_{2}\mathrm{Te}_{4}$}},\ }\href {https://doi.org/10.1103/PhysRevB.103.045106} {\bibfield  {journal} {\bibinfo  {journal} {Phys. Rev. B}\ }\textbf {\bibinfo {volume} {103}},\ \bibinfo {pages} {045106} (\bibinfo {year} {2021}{\natexlab{b}})}\BibitemShut {NoStop}%
\bibitem [{\citenamefont {Ramos}\ \emph {et~al.}(2014)\citenamefont {Ramos}, \citenamefont {Aguirre}, \citenamefont {Anad\'on}, \citenamefont {Blasco}, \citenamefont {Lucas}, \citenamefont {Uchida}, \citenamefont {Algarabel}, \citenamefont {Morell\'on}, \citenamefont {Saitoh},\ and\ \citenamefont {Ibarra}}]{Ramos2014}%
  \BibitemOpen
  \bibfield  {author} {\bibinfo {author} {\bibfnamefont {R.}~\bibnamefont {Ramos}}, \bibinfo {author} {\bibfnamefont {M.~H.}\ \bibnamefont {Aguirre}}, \bibinfo {author} {\bibfnamefont {A.}~\bibnamefont {Anad\'on}}, \bibinfo {author} {\bibfnamefont {J.}~\bibnamefont {Blasco}}, \bibinfo {author} {\bibfnamefont {I.}~\bibnamefont {Lucas}}, \bibinfo {author} {\bibfnamefont {K.}~\bibnamefont {Uchida}}, \bibinfo {author} {\bibfnamefont {P.~A.}\ \bibnamefont {Algarabel}}, \bibinfo {author} {\bibfnamefont {L.}~\bibnamefont {Morell\'on}}, \bibinfo {author} {\bibfnamefont {E.}~\bibnamefont {Saitoh}},\ and\ \bibinfo {author} {\bibfnamefont {M.~R.}\ \bibnamefont {Ibarra}},\ }\bibfield  {title} {\bibinfo {title} {{Anomalous Nernst effect of ${\mathrm{Fe}}_{3}{\mathrm{O}}_{4}$ single crystal}},\ }\href {https://doi.org/10.1103/PhysRevB.90.054422} {\bibfield  {journal} {\bibinfo  {journal} {Phys. Rev. B}\ }\textbf {\bibinfo {volume} {90}},\ \bibinfo {pages} {054422} (\bibinfo {year} {2014})}\BibitemShut {NoStop}%
\bibitem [{\citenamefont {Ghosh}\ \emph {et~al.}(2021)\citenamefont {Ghosh}, \citenamefont {Chanda},\ and\ \citenamefont {Mahendiran}}]{Ghosh2021}%
  \BibitemOpen
  \bibfield  {author} {\bibinfo {author} {\bibfnamefont {A.}~\bibnamefont {Ghosh}}, \bibinfo {author} {\bibfnamefont {A.}~\bibnamefont {Chanda}},\ and\ \bibinfo {author} {\bibfnamefont {R.}~\bibnamefont {Mahendiran}},\ }\bibfield  {title} {\bibinfo {title} {{Anomalous Nernst effect in Pr$_{0.5}$Sr$_{0.5}$CoO$_3$}},\ }\href {https://doi.org/10.1063/5.0039709} {\bibfield  {journal} {\bibinfo  {journal} {AIP Adv.}\ }\textbf {\bibinfo {volume} {11}},\ \bibinfo {pages} {035031} (\bibinfo {year} {2021})}\BibitemShut {NoStop}%
\bibitem [{\citenamefont {De}\ \emph {et~al.}(2020)\citenamefont {De}, \citenamefont {Ghosh}, \citenamefont {Mandal}, \citenamefont {Ogale},\ and\ \citenamefont {Nair}}]{De2020}%
  \BibitemOpen
  \bibfield  {author} {\bibinfo {author} {\bibfnamefont {A.}~\bibnamefont {De}}, \bibinfo {author} {\bibfnamefont {A.}~\bibnamefont {Ghosh}}, \bibinfo {author} {\bibfnamefont {R.}~\bibnamefont {Mandal}}, \bibinfo {author} {\bibfnamefont {S.}~\bibnamefont {Ogale}},\ and\ \bibinfo {author} {\bibfnamefont {S.}~\bibnamefont {Nair}},\ }\bibfield  {title} {\bibinfo {title} {{Temperature Dependence of the Spin Seebeck Effect in a Mixed Valent Manganite}},\ }\href {https://doi.org/10.1103/PhysRevLett.124.017203} {\bibfield  {journal} {\bibinfo  {journal} {Phys. Rev. Lett.}\ }\textbf {\bibinfo {volume} {124}},\ \bibinfo {pages} {017203} (\bibinfo {year} {2020})}\BibitemShut {NoStop}%
\bibitem [{\citenamefont {{Ashworth, T. and Loomer, J. E. and Kreitman, M. M.}}(1973)}]{Ashworth}%
  \BibitemOpen
  \bibfield  {author} {\bibinfo {author} {\bibnamefont {{Ashworth, T. and Loomer, J. E. and Kreitman, M. M.}}},\ }\bibfield  {title} {\bibinfo {title} {{Thermal Conductivity of Nylons and Apiezon Greases}},\ }in\ \href {https://doi.org/10.1007/978-1-4684-3111-7_31} {\emph {\bibinfo {booktitle} {{Advances in Cryogenic Engineering}}}},\ \bibinfo {editor} {edited by\ \bibinfo {editor} {\bibnamefont {{Timmerhaus, K. D.}}}}\ (\bibinfo  {publisher} {{Springer US}},\ \bibinfo {address} {{Boston, MA}},\ \bibinfo {year} {{1973}})\ pp.\ \bibinfo {pages} {{271--279}}\BibitemShut {NoStop}%
\bibitem [{\citenamefont {Wang}\ \emph {et~al.}(2024)\citenamefont {Wang}, \citenamefont {Sakai}, \citenamefont {Minami}, \citenamefont {Gu}, \citenamefont {Chen}, \citenamefont {Feng}, \citenamefont {Nishio-Hamane},\ and\ \citenamefont {Nakatsuji}}]{Wang2024}%
  \BibitemOpen
  \bibfield  {author} {\bibinfo {author} {\bibfnamefont {Y.}~\bibnamefont {Wang}}, \bibinfo {author} {\bibfnamefont {A.}~\bibnamefont {Sakai}}, \bibinfo {author} {\bibfnamefont {S.}~\bibnamefont {Minami}}, \bibinfo {author} {\bibfnamefont {H.}~\bibnamefont {Gu}}, \bibinfo {author} {\bibfnamefont {T.}~\bibnamefont {Chen}}, \bibinfo {author} {\bibfnamefont {Z.}~\bibnamefont {Feng}}, \bibinfo {author} {\bibfnamefont {D.}~\bibnamefont {Nishio-Hamane}},\ and\ \bibinfo {author} {\bibfnamefont {S.}~\bibnamefont {Nakatsuji}},\ }\bibfield  {title} {\bibinfo {title} {{Robust giant anomalous Nernst effect in polycrystalline nodal web ferromagnets}},\ }\href {https://doi.org/10.1063/5.0219416} {\bibfield  {journal} {\bibinfo  {journal} {Appl. Phys. Lett.}\ }\textbf {\bibinfo {volume} {125}},\ \bibinfo {pages} {081901} (\bibinfo {year} {2024})}\BibitemShut {NoStop}%
\bibitem [{\citenamefont {Ghosh}\ \emph {et~al.}(2019)\citenamefont {Ghosh}, \citenamefont {Das},\ and\ \citenamefont {Mahendiran}}]{Ghosh2019}%
  \BibitemOpen
  \bibfield  {author} {\bibinfo {author} {\bibfnamefont {A.}~\bibnamefont {Ghosh}}, \bibinfo {author} {\bibfnamefont {R.}~\bibnamefont {Das}},\ and\ \bibinfo {author} {\bibfnamefont {R.}~\bibnamefont {Mahendiran}},\ }\bibfield  {title} {\bibinfo {title} {{Skew scattering dominated anomalous Nernst effect in La$_{1-x}$Na$_x$MnO$_3$}},\ }\href {https://doi.org/10.1063/1.5081063} {\bibfield  {journal} {\bibinfo  {journal} {J. Appl. Phys.}\ }\textbf {\bibinfo {volume} {125}},\ \bibinfo {pages} {153902} (\bibinfo {year} {2019})}\BibitemShut {NoStop}%
\bibitem [{\citenamefont {Li}\ and\ \citenamefont {Spitzer}(2005)}]{Li2005}%
  \BibitemOpen
  \bibfield  {author} {\bibinfo {author} {\bibfnamefont {Y.}~\bibnamefont {Li}}\ and\ \bibinfo {author} {\bibfnamefont {K.}~\bibnamefont {Spitzer}},\ }\bibfield  {title} {\bibinfo {title} {{Finite element resistivity modelling for three-dimensional structures with arbitrary anisotropy}},\ }\href {https://doi.org/https://doi.org/10.1016/j.pepi.2004.08.014} {\bibfield  {journal} {\bibinfo  {journal} {Phys. Earth Planet. Inter.}\ }\textbf {\bibinfo {volume} {150}},\ \bibinfo {pages} {15} (\bibinfo {year} {2005})},\ \bibinfo {note} {electromagnetic Induction in the Earth}\BibitemShut {NoStop}%
\bibitem [{\citenamefont {Xu}\ \emph {et~al.}(2019)\citenamefont {Xu}, \citenamefont {Phelan},\ and\ \citenamefont {Chien}}]{Xu2019}%
  \BibitemOpen
  \bibfield  {author} {\bibinfo {author} {\bibfnamefont {J.}~\bibnamefont {Xu}}, \bibinfo {author} {\bibfnamefont {W.~A.}\ \bibnamefont {Phelan}},\ and\ \bibinfo {author} {\bibfnamefont {C.-L.}\ \bibnamefont {Chien}},\ }\bibfield  {title} {\bibinfo {title} {{Large Anomalous Nernst Effect in a van der Waals Ferromagnet Fe$_3$GeTe$_2$}},\ }\href {https://doi.org/10.1021/acs.nanolett.9b03739} {\bibfield  {journal} {\bibinfo  {journal} {Nano Lett.}\ }\textbf {\bibinfo {volume} {19}},\ \bibinfo {pages} {8250} (\bibinfo {year} {2019})},\ \bibinfo {note} {pMID: 31658813}\BibitemShut {NoStop}%
\bibitem [{\citenamefont {Mott}\ \emph {et~al.}(1958)\citenamefont {Mott}, \citenamefont {Jones},\ and\ \citenamefont {Jones}}]{Mott}%
  \BibitemOpen
  \bibfield  {author} {\bibinfo {author} {\bibfnamefont {N.~F.}\ \bibnamefont {Mott}}, \bibinfo {author} {\bibfnamefont {H.}~\bibnamefont {Jones}},\ and\ \bibinfo {author} {\bibfnamefont {H.}~\bibnamefont {Jones}},\ }\href@noop {} {\emph {\bibinfo {title} {{The Theory of the Properties of Metals and Alloys}}}}\ (\bibinfo  {publisher} {Courier Dover Publications},\ \bibinfo {year} {1958})\BibitemShut {NoStop}%
\bibitem [{\citenamefont {Ding}\ \emph {et~al.}(2019)\citenamefont {Ding}, \citenamefont {Koo}, \citenamefont {Xu}, \citenamefont {Li}, \citenamefont {Lu}, \citenamefont {Zhao}, \citenamefont {Wang}, \citenamefont {Yin}, \citenamefont {Lei}, \citenamefont {Yan}, \citenamefont {Zhu},\ and\ \citenamefont {Behnia}}]{Ding2019}%
  \BibitemOpen
  \bibfield  {author} {\bibinfo {author} {\bibfnamefont {L.}~\bibnamefont {Ding}}, \bibinfo {author} {\bibfnamefont {J.}~\bibnamefont {Koo}}, \bibinfo {author} {\bibfnamefont {L.}~\bibnamefont {Xu}}, \bibinfo {author} {\bibfnamefont {X.}~\bibnamefont {Li}}, \bibinfo {author} {\bibfnamefont {X.}~\bibnamefont {Lu}}, \bibinfo {author} {\bibfnamefont {L.}~\bibnamefont {Zhao}}, \bibinfo {author} {\bibfnamefont {Q.}~\bibnamefont {Wang}}, \bibinfo {author} {\bibfnamefont {Q.}~\bibnamefont {Yin}}, \bibinfo {author} {\bibfnamefont {H.}~\bibnamefont {Lei}}, \bibinfo {author} {\bibfnamefont {B.}~\bibnamefont {Yan}}, \bibinfo {author} {\bibfnamefont {Z.}~\bibnamefont {Zhu}},\ and\ \bibinfo {author} {\bibfnamefont {K.}~\bibnamefont {Behnia}},\ }\bibfield  {title} {\bibinfo {title} {{Intrinsic Anomalous Nernst Effect Amplified by Disorder in a Half-Metallic Semimetal}},\ }\href {https://doi.org/10.1103/PhysRevX.9.041061} {\bibfield  {journal} {\bibinfo  {journal} {Phys. Rev. X}\ }\textbf {\bibinfo {volume} {9}},\ \bibinfo
  {pages} {041061} (\bibinfo {year} {2019})}\BibitemShut {NoStop}%
\bibitem [{\citenamefont {Sumida}\ \emph {et~al.}(2020)\citenamefont {Sumida}, \citenamefont {Sakuraba}, \citenamefont {Masuda}, \citenamefont {Kono}, \citenamefont {Kakoki}, \citenamefont {Goto}, \citenamefont {Zhou}, \citenamefont {Miyamoto}, \citenamefont {Miura}, \citenamefont {Okuda},\ and\ \citenamefont {Kimura}}]{Sumida2020}%
  \BibitemOpen
  \bibfield  {author} {\bibinfo {author} {\bibfnamefont {K.}~\bibnamefont {Sumida}}, \bibinfo {author} {\bibfnamefont {Y.}~\bibnamefont {Sakuraba}}, \bibinfo {author} {\bibfnamefont {K.}~\bibnamefont {Masuda}}, \bibinfo {author} {\bibfnamefont {T.}~\bibnamefont {Kono}}, \bibinfo {author} {\bibfnamefont {M.}~\bibnamefont {Kakoki}}, \bibinfo {author} {\bibfnamefont {K.}~\bibnamefont {Goto}}, \bibinfo {author} {\bibfnamefont {W.}~\bibnamefont {Zhou}}, \bibinfo {author} {\bibfnamefont {K.}~\bibnamefont {Miyamoto}}, \bibinfo {author} {\bibfnamefont {Y.}~\bibnamefont {Miura}}, \bibinfo {author} {\bibfnamefont {T.}~\bibnamefont {Okuda}},\ and\ \bibinfo {author} {\bibfnamefont {A.}~\bibnamefont {Kimura}},\ }\bibfield  {title} {\bibinfo {title} {{Spin-polarized Weyl cones and giant anomalous Nernst effect in ferromagnetic Heusler films}},\ }\href {https://doi.org/10.1038/s43246-020-00088-w} {\bibfield  {journal} {\bibinfo  {journal} {Commun. Mater.}\ }\textbf {\bibinfo {volume} {1}},\ \bibinfo {pages} {89} (\bibinfo
  {year} {2020})}\BibitemShut {NoStop}%
\bibitem [{\citenamefont {Lee}\ \emph {et~al.}(2007)\citenamefont {Lee}, \citenamefont {Onose}, \citenamefont {Tokura},\ and\ \citenamefont {Ong}}]{p23}%
  \BibitemOpen
  \bibfield  {author} {\bibinfo {author} {\bibfnamefont {M.}~\bibnamefont {Lee}}, \bibinfo {author} {\bibfnamefont {Y.}~\bibnamefont {Onose}}, \bibinfo {author} {\bibfnamefont {Y.}~\bibnamefont {Tokura}},\ and\ \bibinfo {author} {\bibfnamefont {N.~P.}\ \bibnamefont {Ong}},\ }\bibfield  {title} {\bibinfo {title} {{Hidden constant in the anomalous Hall effect of high-purity magnet MnSi}},\ }\href {https://doi.org/10.1103/PhysRevB.75.172403} {\bibfield  {journal} {\bibinfo  {journal} {Phys. Rev. B}\ }\textbf {\bibinfo {volume} {75}},\ \bibinfo {pages} {172403} (\bibinfo {year} {2007})}\BibitemShut {NoStop}%
\bibitem [{\citenamefont {Bhattacharya}\ \emph {et~al.}(2024)\citenamefont {Bhattacharya}, \citenamefont {Habib}, \citenamefont {Ahmed}, \citenamefont {Satpati}, \citenamefont {DuttaGupta}, \citenamefont {Dasgupta},\ and\ \citenamefont {Das}}]{p24}%
  \BibitemOpen
  \bibfield  {author} {\bibinfo {author} {\bibfnamefont {A.}~\bibnamefont {Bhattacharya}}, \bibinfo {author} {\bibfnamefont {M.~R.}\ \bibnamefont {Habib}}, \bibinfo {author} {\bibfnamefont {A.}~\bibnamefont {Ahmed}}, \bibinfo {author} {\bibfnamefont {B.}~\bibnamefont {Satpati}}, \bibinfo {author} {\bibfnamefont {S.}~\bibnamefont {DuttaGupta}}, \bibinfo {author} {\bibfnamefont {I.}~\bibnamefont {Dasgupta}},\ and\ \bibinfo {author} {\bibfnamefont {I.}~\bibnamefont {Das}},\ }\bibfield  {title} {\bibinfo {title} {{Spin-valve-like magnetoresistance and anomalous Hall effect in magnetic Weyl metal $\mathrm{Mn}_{2}\mathrm{PdSn}$}},\ }\href {https://doi.org/10.1103/PhysRevB.110.014417} {\bibfield  {journal} {\bibinfo  {journal} {Phys. Rev. B}\ }\textbf {\bibinfo {volume} {110}},\ \bibinfo {pages} {014417} (\bibinfo {year} {2024})}\BibitemShut {NoStop}%
\bibitem [{\citenamefont {Miyasato}\ \emph {et~al.}(2007)\citenamefont {Miyasato}, \citenamefont {Abe}, \citenamefont {Fujii}, \citenamefont {Asamitsu}, \citenamefont {Onoda}, \citenamefont {Onose}, \citenamefont {Nagaosa},\ and\ \citenamefont {Tokura}}]{p25}%
  \BibitemOpen
  \bibfield  {author} {\bibinfo {author} {\bibfnamefont {T.}~\bibnamefont {Miyasato}}, \bibinfo {author} {\bibfnamefont {N.}~\bibnamefont {Abe}}, \bibinfo {author} {\bibfnamefont {T.}~\bibnamefont {Fujii}}, \bibinfo {author} {\bibfnamefont {A.}~\bibnamefont {Asamitsu}}, \bibinfo {author} {\bibfnamefont {S.}~\bibnamefont {Onoda}}, \bibinfo {author} {\bibfnamefont {Y.}~\bibnamefont {Onose}}, \bibinfo {author} {\bibfnamefont {N.}~\bibnamefont {Nagaosa}},\ and\ \bibinfo {author} {\bibfnamefont {Y.}~\bibnamefont {Tokura}},\ }\bibfield  {title} {\bibinfo {title} {{Crossover Behavior of the Anomalous Hall Effect and Anomalous Nernst Effect in Itinerant Ferromagnets}},\ }\href {https://doi.org/10.1103/PhysRevLett.99.086602} {\bibfield  {journal} {\bibinfo  {journal} {Phys. Rev. Lett.}\ }\textbf {\bibinfo {volume} {99}},\ \bibinfo {pages} {086602} (\bibinfo {year} {2007})}\BibitemShut {NoStop}%
\bibitem [{\citenamefont {Sakai}\ \emph {et~al.}(2020)\citenamefont {Sakai}, \citenamefont {Minami}, \citenamefont {Koretsune}, \citenamefont {Chen}, \citenamefont {Higo}, \citenamefont {Wang}, \citenamefont {Nomoto}, \citenamefont {Hirayama}, \citenamefont {Miwa}, \citenamefont {Nishio-Hamane} \emph {et~al.}}]{sakai2020iron}%
  \BibitemOpen
  \bibfield  {author} {\bibinfo {author} {\bibfnamefont {A.}~\bibnamefont {Sakai}}, \bibinfo {author} {\bibfnamefont {S.}~\bibnamefont {Minami}}, \bibinfo {author} {\bibfnamefont {T.}~\bibnamefont {Koretsune}}, \bibinfo {author} {\bibfnamefont {T.}~\bibnamefont {Chen}}, \bibinfo {author} {\bibfnamefont {T.}~\bibnamefont {Higo}}, \bibinfo {author} {\bibfnamefont {Y.}~\bibnamefont {Wang}}, \bibinfo {author} {\bibfnamefont {T.}~\bibnamefont {Nomoto}}, \bibinfo {author} {\bibfnamefont {M.}~\bibnamefont {Hirayama}}, \bibinfo {author} {\bibfnamefont {S.}~\bibnamefont {Miwa}}, \bibinfo {author} {\bibfnamefont {D.}~\bibnamefont {Nishio-Hamane}}, \emph {et~al.},\ }\bibfield  {title} {\bibinfo {title} {{Iron-based binary ferromagnets for transverse thermoelectric conversion}},\ }\href {https://doi.org/10.1038/s41586-020-2230-z} {\bibfield  {journal} {\bibinfo  {journal} {Nature}\ }\textbf {\bibinfo {volume} {581}},\ \bibinfo {pages} {53} (\bibinfo {year} {2020})}\BibitemShut {NoStop}%
\bibitem [{\citenamefont {Asaba}\ \emph {et~al.}(2021)\citenamefont {Asaba}, \citenamefont {Ivanov}, \citenamefont {Thomas}, \citenamefont {Savrasov}, \citenamefont {Thompson}, \citenamefont {Bauer},\ and\ \citenamefont {Ronning}}]{asaba2021colossal}%
  \BibitemOpen
  \bibfield  {author} {\bibinfo {author} {\bibfnamefont {T.}~\bibnamefont {Asaba}}, \bibinfo {author} {\bibfnamefont {V.}~\bibnamefont {Ivanov}}, \bibinfo {author} {\bibfnamefont {S.~M.}\ \bibnamefont {Thomas}}, \bibinfo {author} {\bibfnamefont {S.}~\bibnamefont {Savrasov}}, \bibinfo {author} {\bibfnamefont {J.~D.}\ \bibnamefont {Thompson}}, \bibinfo {author} {\bibfnamefont {E.~D.}\ \bibnamefont {Bauer}},\ and\ \bibinfo {author} {\bibfnamefont {F.}~\bibnamefont {Ronning}},\ }\bibfield  {title} {\bibinfo {title} {{Colossal anomalous Nernst effect in a correlated noncentrosymmetric kagome ferromagnet}},\ }\href {https://doi.org/10.1126/sciadv.abf1467} {\bibfield  {journal} {\bibinfo  {journal} {Sci. Adv.}\ }\textbf {\bibinfo {volume} {7}},\ \bibinfo {pages} {eabf1467} (\bibinfo {year} {2021})}\BibitemShut {NoStop}%
\bibitem [{\citenamefont {Badura}\ \emph {et~al.}(2025)\citenamefont {Badura}, \citenamefont {Campos}, \citenamefont {Bharadwaj}, \citenamefont {Kounta}, \citenamefont {Michez}, \citenamefont {Petit}, \citenamefont {Rial}, \citenamefont {Leivisk{\"a}}, \citenamefont {Baltz}, \citenamefont {Krizek} \emph {et~al.}}]{badura2025observation}%
  \BibitemOpen
  \bibfield  {author} {\bibinfo {author} {\bibfnamefont {A.}~\bibnamefont {Badura}}, \bibinfo {author} {\bibfnamefont {W.~H.}\ \bibnamefont {Campos}}, \bibinfo {author} {\bibfnamefont {V.~K.}\ \bibnamefont {Bharadwaj}}, \bibinfo {author} {\bibfnamefont {I.}~\bibnamefont {Kounta}}, \bibinfo {author} {\bibfnamefont {L.}~\bibnamefont {Michez}}, \bibinfo {author} {\bibfnamefont {M.}~\bibnamefont {Petit}}, \bibinfo {author} {\bibfnamefont {J.}~\bibnamefont {Rial}}, \bibinfo {author} {\bibfnamefont {M.}~\bibnamefont {Leivisk{\"a}}}, \bibinfo {author} {\bibfnamefont {V.}~\bibnamefont {Baltz}}, \bibinfo {author} {\bibfnamefont {F.}~\bibnamefont {Krizek}}, \emph {et~al.},\ }\bibfield  {title} {\bibinfo {title} {{Observation of the anomalous Nernst effect in altermagnetic candidate Mn$_5$Si$_3$}},\ }\href {https://doi.org/10.1038/s41467-025-62331-7} {\bibfield  {journal} {\bibinfo  {journal} {Nat. Commun.}\ }\textbf {\bibinfo {volume} {16}},\ \bibinfo {pages} {7111} (\bibinfo {year} {2025})}\BibitemShut {NoStop}%
\end{thebibliography}%
\clearpage

\resumetoc
\title{Supplementary Material for `` Intrinsic Berry Curvature Driven Anomalous Hall and Nernst Effect in Co$_2$MnSn "}

\maketitle
\onecolumngrid
\tableofcontents
\clearpage

\setcounter{section}{0}
\setcounter{equation}{0}
\setcounter{figure}{0}
\setcounter{table}{0}
\makeatletter
\renewcommand{\bibnumfmt}[1]{[S#1]}
\renewcommand{\citenumfont}[1]{S#1}
\renewcommand{\thesection}{S\arabic{section}}
\renewcommand{\thefigure}{S\arabic{figure}}
\renewcommand{\thetable}{S\arabic{table}}

\begin{bibunit}
    \noindent Here, we present further auxiliary information about the crystal structure, magnetoresistance, anomalous Hall and Nernst effects, \textit{ab-initio} spin-polarized electronic structure and temperature dependence of anomalous Hall conductivity.

    \section{Experimental Results}
    
    \subsection{Crystal Structure}
    \begin{figure}[h]
        \centering
        \includegraphics[width=0.5\linewidth]{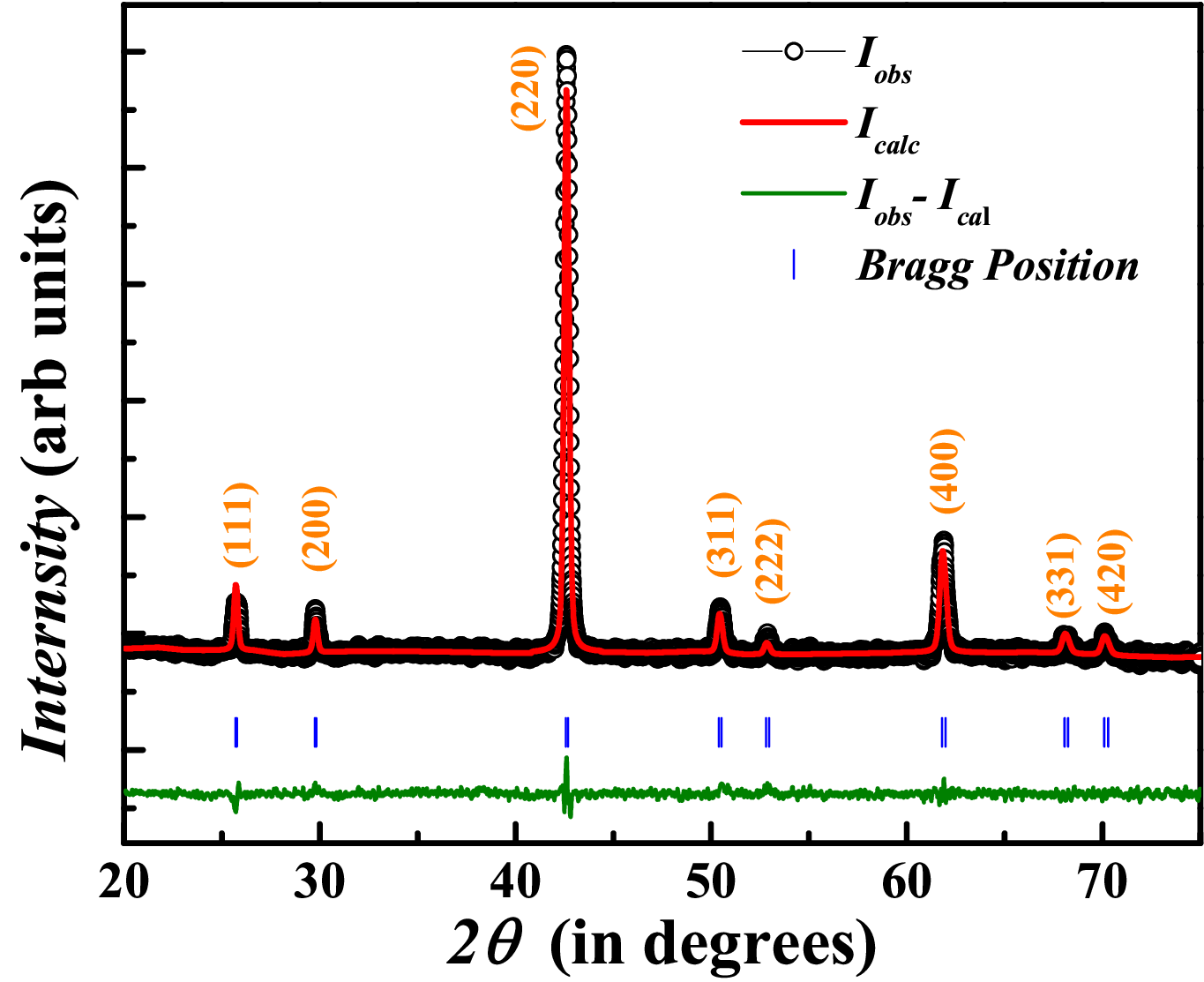}
        \caption{(Colour online) Rietveld refinement of room temperature X-Ray diffraction pattern of Co$_2$MnSn.}
        \label{XRD}
    \end{figure}
        
    Figure \ref{XRD} depicts the Rietveld refinement of the X-ray diffraction pattern. The presence of equal intensity superlattice (111) and (200) peaks attests to the L2$_1$ ordering of Co$_2$MnSn. The lattice and Rietveld refinement parameters are listed in Table \ref{XRDt}.
        
    \begin{table}[h]
        \begin{center}
            \caption{(Colour online) Crystallographic parameters of full Rietveld analysis of the compound Co$_2$MnSn, which crystallizes in cubic Fm$\bar{3}$m space group (Group No. 225). Here $a=b=c$ are the lattice parameters \label{XRDt}}
            \begin{tabular}{cccccc}
                \hline\hline
                \noalign{\smallskip}
                \noalign{\smallskip}
                \multicolumn{1}{l}{Composition} & \multicolumn{1}{l}{\hspace{1cm} Co$_2$MnSn} \\
                \multicolumn{1}{l}{Space Group} & \multicolumn{1}{l}{\hspace{1cm} Fm$\bar{3}$m(Group No. 225)} \\
                \multicolumn{1}{l}{$a=$} & \multicolumn{1}{l}{\hspace{1cm} 5.99(6)~\AA} \\
                \multicolumn{1}{l}{$R_{f}=$} & \multicolumn{1}{l}{\hspace{1cm} 3.191} \\
                \multicolumn{1}{l}{$R_{Bragg}=$} & \multicolumn{1}{l}{\hspace{1cm} 4.379} \\
                \multicolumn{1}{l}{$\chi^2=$} & \multicolumn{1}{l}{\hspace{1cm} 2.1} \\
                \noalign{\smallskip}
                \noalign{\smallskip}
                \hline\hline
                \noalign{\smallskip}
                \noalign{\smallskip}
                \multicolumn{6}{c}{Atomic Coordinates} \\
                \noalign{\smallskip}
                \noalign{\smallskip}
                \hline\hline
                Atom & Wyckoff Position &  x   &   y  &   z & occu. \\
                \hline
                \noalign{\smallskip}
                Co & 8c & 0.25 & 0.25 & 0.25 & 1\\
                Mn & 4b & 0.5 & 0.5 & 0.5 & 1\\
                Sn & 4a & 0 & 0 & 0 & 1\\
                \noalign{\smallskip}
                \hline\hline
            \end{tabular}
        \end{center}
    \end{table} 
    
    \subsection{Magnetoresistance}
\begin{figure*}[h]
        \centering
        \includegraphics[width=0.55\linewidth]{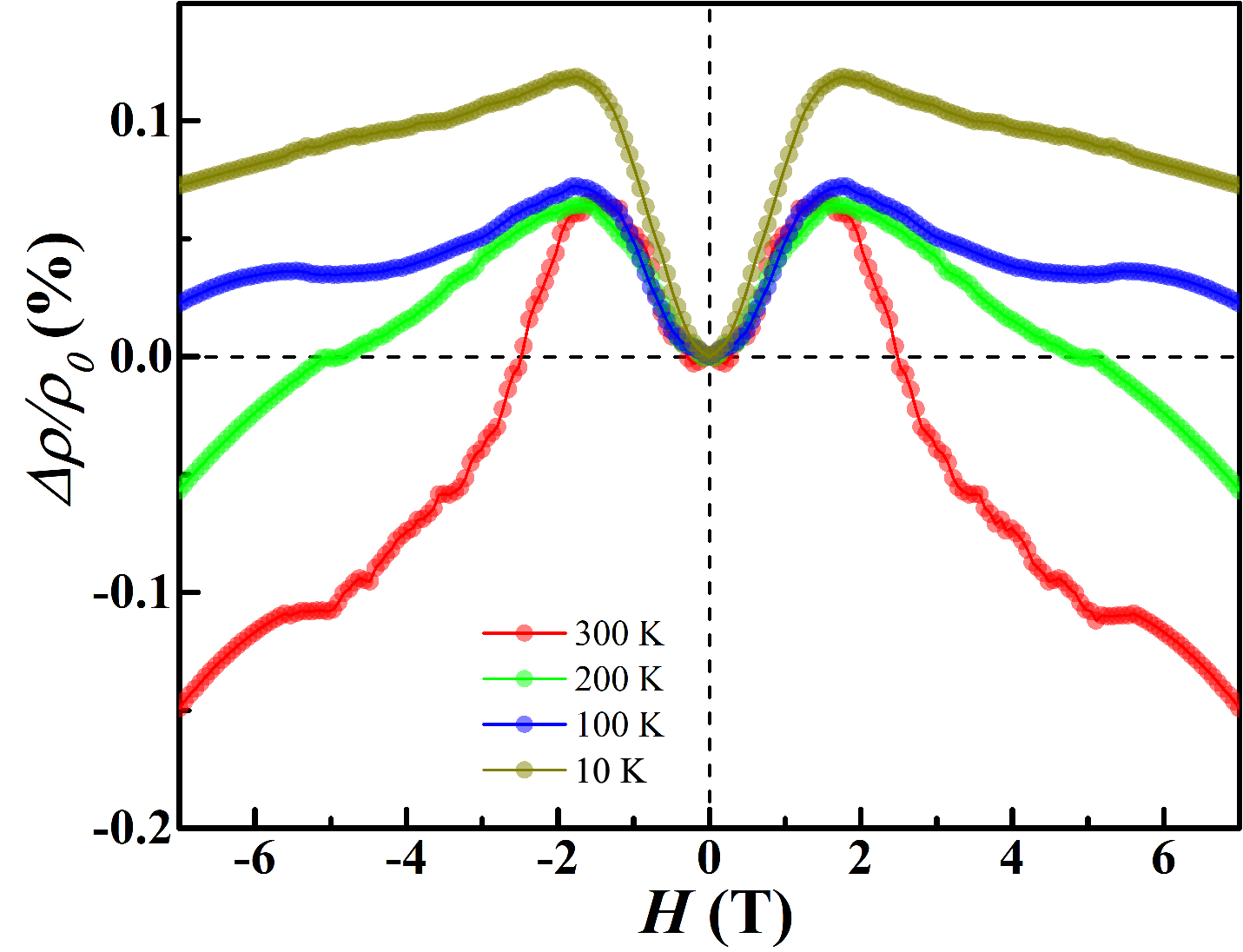}
        \caption{(Colour online) Field dependence of magnetoresistance across the entire measured temperature range. }
        \label{MR}
    \end{figure*}

    The observed negligible magnetoresistance ($\approx$ 0.1 $\%$) over the entire measured range of temperature, i.e., 5 K to 300 K, in Fig.\,\ref{MR}, suggest that the transverse resistivity ($\rho_{xy}$) is dominated by the anomalous Hall resistivity ($\rho_{xy}^A$) with a negligible contributions from longitudinal resistivity ($\rho_{xx}$).  
        
    \begin{figure*}[b!]
        \centering
        \includegraphics[width=0.55\linewidth]{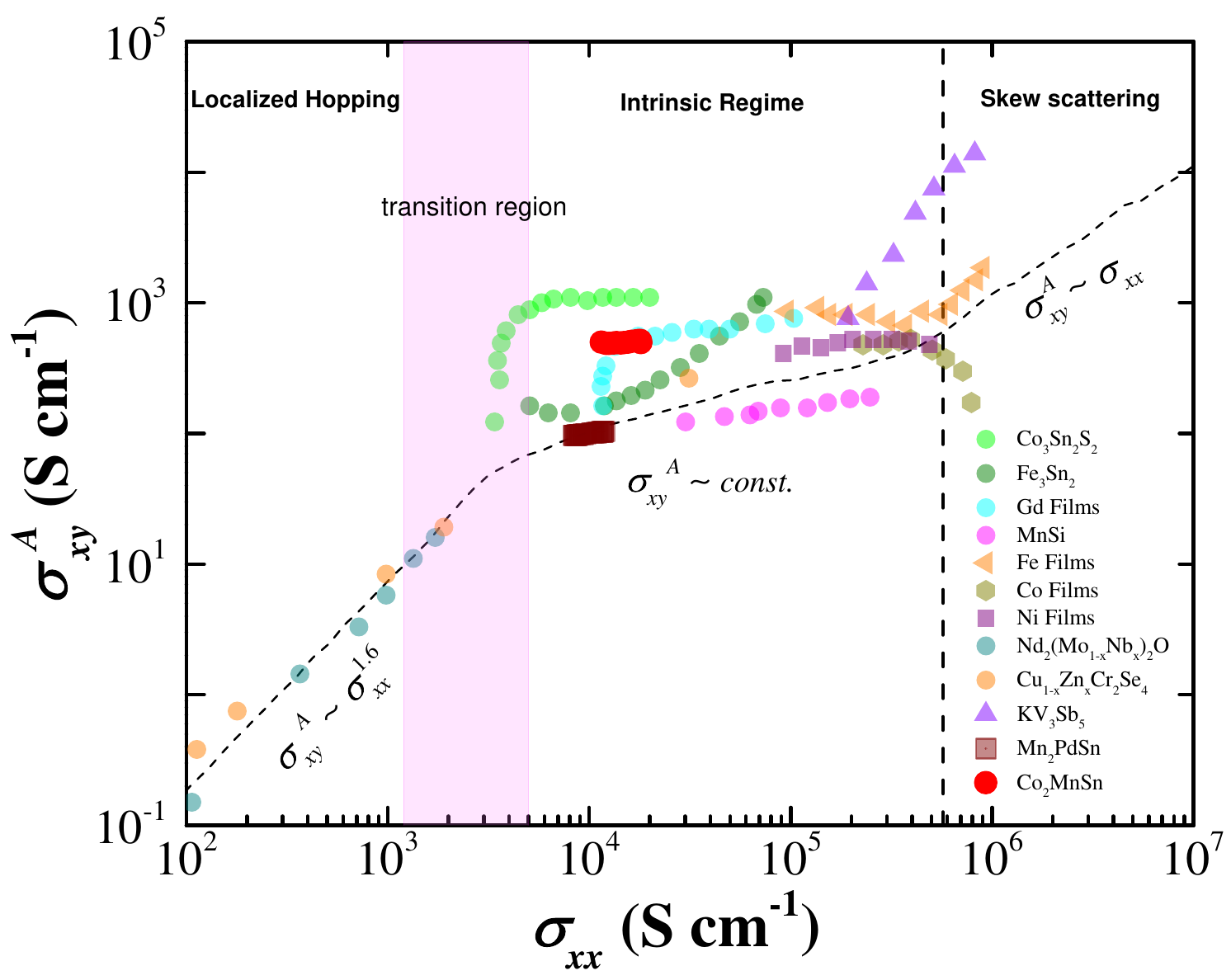}
        \caption{(Colour online) Universal plot of anomalous Hall conductivity ($\sigma_{xy}^A$) as a function of longitudinal conductivity ($\sigma_{xx}$). The red curve depicts our experimental result for Co$_2$MnSn, along with other materials such as MnSi \cite{Lee2007_sm}, Mn$_2$PdSn \cite{Bhattacharya2024_sm}, Fe$_3$Sn$_2$ \cite{Ye2018_sm}, Co$_3$Sn$_2$S$_2$ \cite{Wang2018_sm} and several other ferromagnets with trivial bulk band crossings \cite{Miyasato2007_sm}. The highlighted region shows the transition from the hopping regime to the intrinsic Berry-curvature-driven mechanism of AHC for Co$_2$MnSn.  }
        \label{master}
    \end{figure*}
    
    \subsection{Anomalous Hall effect}
    Figure~\ref{master} showcases the longitudinal conductivity ($\sigma_{xx}$) dependence of anomalous Hall conductivity ($\sigma_{xy}^A$) over the entire measured range of temperature, which lays well within the intrinsic conductivity regime for Co$_2$MnSn and constant over the entire range of temperature. This attests to the fact that $\sigma_{xy}^A$ remains impervious to the change of the dominant scattering mechanism observed in longitudinal resistivity.  The highlighted region represents the transition region from hopping-mechanism-driven AHC (with $\sigma_{xy}^A \propto \sigma_{xx}^{1.6}$) to intrinsic Berry-curvature-mediated AHC, where $\sigma_{xy}^A$ remains constant with $\sigma_{xx}$. The broken line marks the boundary from intrinsic regime to $sk$-mechanism dominated AHC in ultraclean limit or large longitudinal conductivity limit. The dotted line represents the theoretically predicted variation of $\sigma_{xy}^A$ with $\sigma_{xx}$ under these three mechanisms, by Onoda \textit{et al.} \cite{Onoda2006_sm}.    
    
    \subsection{Thermal and thermoelectric transport}
    \begin{figure*}[ht!]
        \centering
        \includegraphics[width=0.8\linewidth]{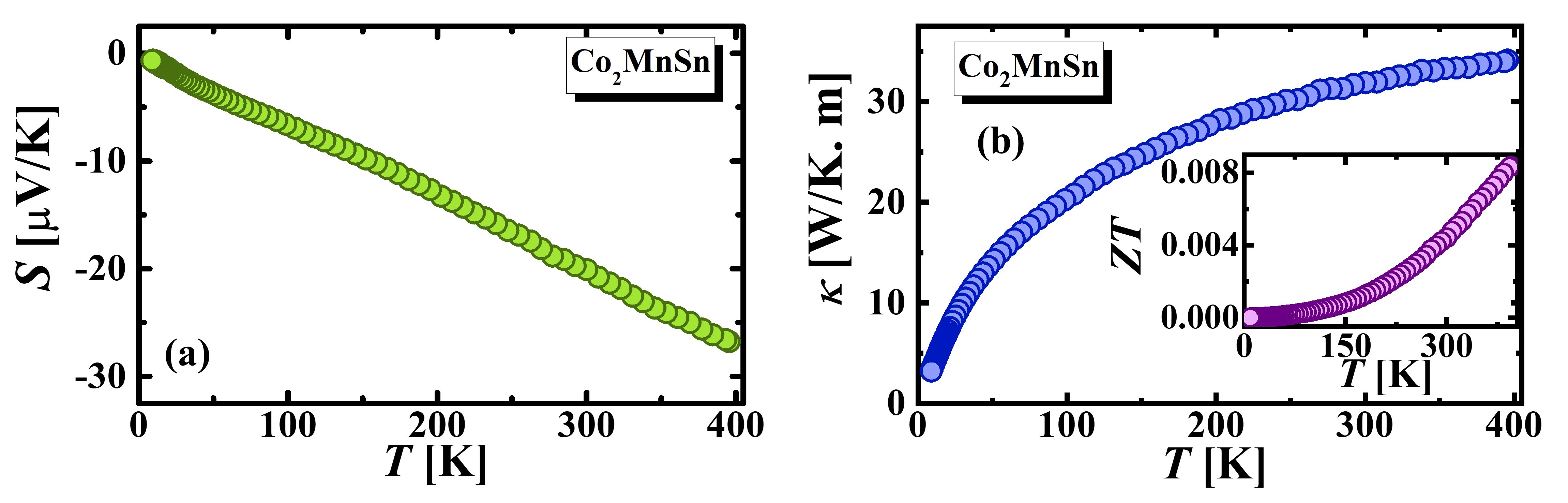}
        \caption{(Colour online) Temperature ($T$) variation of (a) longitudinal Seebeck coefficient ($S$) and (b) thermal conductivity ($\kappa$) of Co$_2$MnSn. The inset of (b) shows the $T$ dependence of thermoelectric figure of merit $ZT$.}
        \label{figS4}
    \end{figure*}

    \begin{figure*}[ht!]
        \centering
        \includegraphics[width=0.6\linewidth]{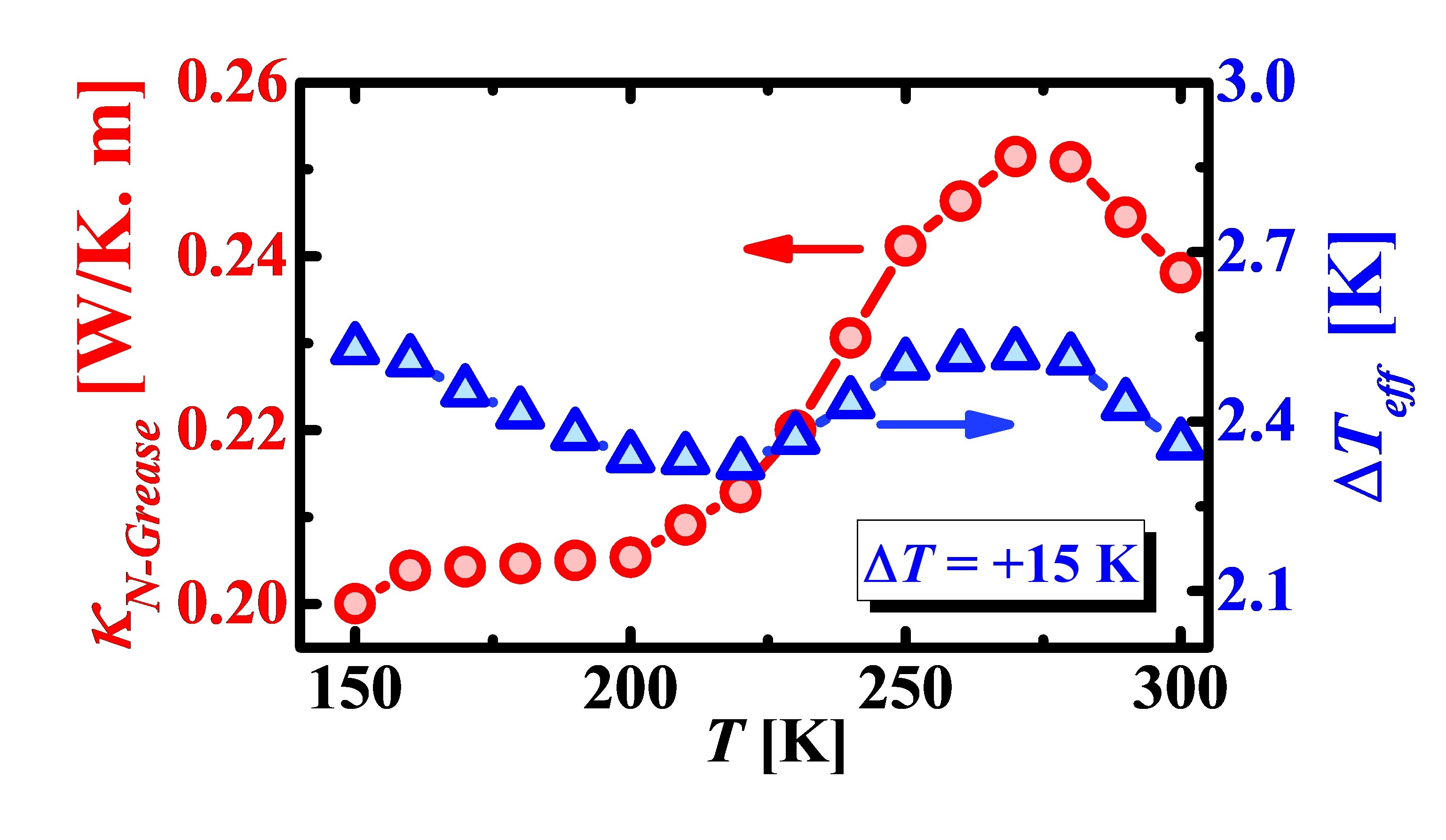}
        \caption{(Colour online) Temperature ($T$) variation of thermal conductivity of N-grease layers ($\kappa_{\text{N-Grease}}$) and effective temperature difference ($\Delta T_{eff}$) between the opposite isothermal surfaces as illustrated in the experimental configuration shown in Fig. 2(a) of main manuscript.}
        \label{figS5}
    \end{figure*}
    
    \noindent The temperature dependence of Seebeck coefficient $S$ for our Co$_2$MnSn alloy presented in Fig.~\ref{figS4}(a) shows a negative sign throughout the measured temperature range, indicating that electrons are the majority carriers for thermoelectric transport. The absolute value of $S$ increases nearly linearly with increasing temperature and reaches approximately $\approx -20\,\mu$V/K at 300 K, which is close to that of single crystal of Co$_2$MnGa ($S \approx -26\,\mu$V/K at 300 K) \cite{Guin2019_sm}. The temperature dependence of thermal conductivity $\kappa$ for our Co$_2$MnSn alloy (see main panel of Fig.~\ref{figS4}(b)) is $\approx$ 32 W/(K.m) at 300 K, which is higher than that of single crystal of Co$_2$MnGa ($\kappa \approx 22$ W/(K.m) at 300 K) \cite{Wang2022_sm}. The inset of Fig.~\ref{figS4}(b) shows the temperature evolution of the thermoelectric figure of merit $ZT$ for our Co$_2$MnSn alloy. The value of $ZT$ is $\approx$ 0.008 at 400 K.
    
    \noindent Left $y$-scale of Fig.~\ref{figS5} represents the thermal conductivity of N-grease ($\kappa_{\text{N-Grease}}(T)$) obtained by interpolating the reported values of the temperature dependence of $\kappa_{\text{N-Grease}}$ \cite{Ashworth_sm}. The right $y$-scale of Fig.~\ref{figS5} shows the effective temperature difference ($\Delta T_{eff}$) between the opposite isothermal surfaces of the sample during the Nernst measurements as illustrated in the experimental configuration shown in Fig.~\ref{F2}(a) of main manuscript and estimated by using the expression (Eq.~\eqref{eq5} of main manuscript): 
    
    \begin{equation*}
    \Delta T_{eff} = \frac{\Delta T}{\left[ 1 + \left(\frac{2L_{\text{N-Grease}}}{\kappa_{\text{N-Grease}}}\right)\cdot\left(\frac{\kappa}{L_S}\right) \right]}
    \end{equation*}
    
    where $L_S (= L_y)$ and $\kappa$ are the thickness and the thermal conductivity of Co$_2$MnSn sample, and $L_{\text{N-Grease}}$ and $\kappa_{\text{N-Grease}}$ are the thickness and thermal conductivity of the N-grease layers.

    \section{Ab-initio Results}
    
    \subsection{Spin-polarized electronic structure}
    \begin{figure*}[htb]
        \centering
        \includegraphics[width=0.55\linewidth]{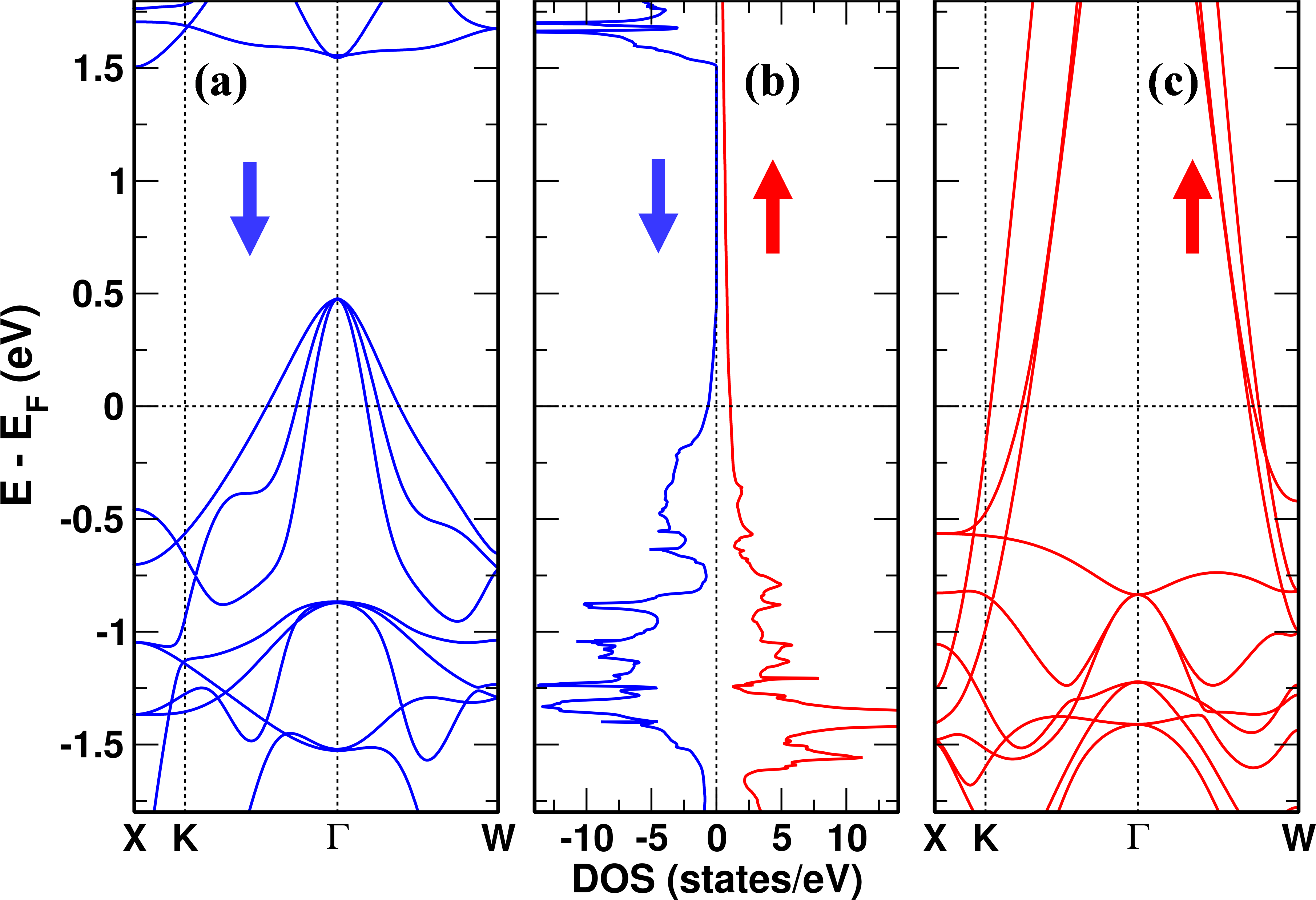}
        \caption{(Colour online) Spin-polarized electronic structure of Co$_2$MnSn. The minority-spin (blue down arrow) and majority-spin (red up arrow) channel resolved band structures are shown in panels (a) and (c) respectively. The density of states (DOS) for each individual spin channels are shown in panel (b).}
        \label{figS6}
    \end{figure*}
    
    The spin-polarized electronic structure and density of states (DOS) of ferromagnet Co$_2$MnSn, simulated using SCAN functional \cite{Sun2015_sm}, are shown in Fig.~\ref{figS6}. Both the majority and minority spin channels exhibits finite electronic states at the Fermi level ($E_F$), ruling out the half-metallic behavior of Co$_2$MnSn. However, the DOS in the minority spin channel remains relatively low and vanishes in the energy range between 0.48 and 1.51 eV, in contrast to finite DOS observed in the majority spin channel over the same energy window. This suggests a possibility of half-metallic behaviour in Co$_2$MnSn by tuning the chemical potential.
    
    Our simulations including spin-orbit coupling indicate negligible magnetocrystalline anisotropy ($\Delta E_{MAE} < 1\ \mu$eV) for different orientations of the magnetic moments. Consequently, the magnetization was aligned along the [001] direction (Cartesian $z$-axis) for all subsequent calculations.
    
    \subsection{Temperature dependence of anomalous Hall conductivity}
    \begin{figure*}[!h]
        \centering
        \includegraphics[width=0.5\linewidth]{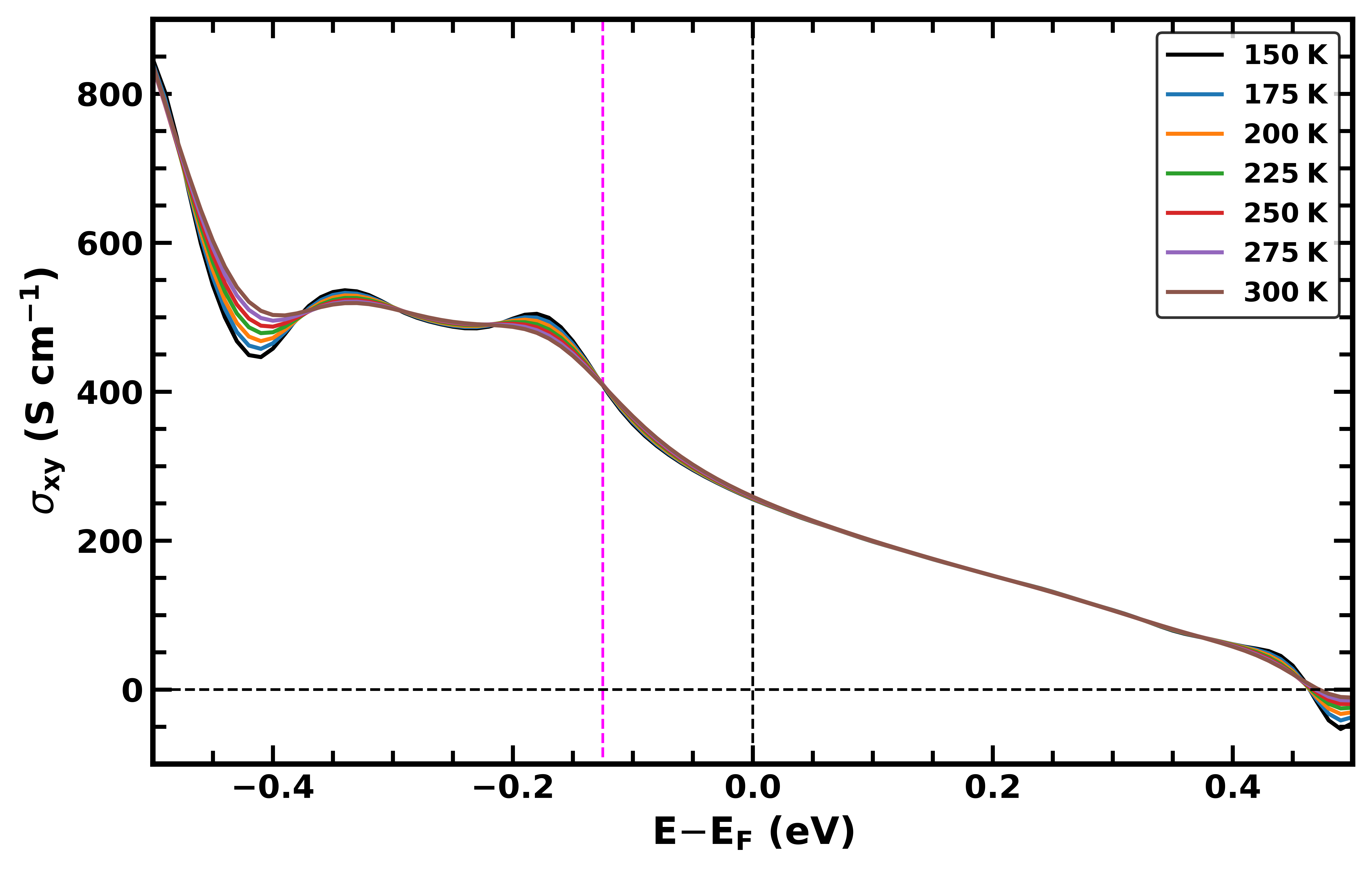}
        \caption{(Colour online) Variation of anomalous Hall conductivity ($\sigma_{xy}$) with chemical potential $\mu = E - E_F$ at different temperatures. The vertical dashed magenta line is at $\mu = -0.125$ eV.}
        \label{figS7}
    \end{figure*}
        
    \noindent As discussed in the main manuscript, the anomalous Hall conductivity ($\sigma_{xy}$) exhibits weak dependence on temperature ($T$). The variation of simulated $\sigma_{xy}$ with chemical potential $\mu = E - E_F$ at several temperatures are shown in Fig.~\ref{figS7}. This  clearly shows that $\sigma_{xy}$ remains nearly unchanged across the temperature range considered, reinforcing the conclusion that the intrinsic contribution to the anomalous Hall effect in Co$_2$MnSn is robust against thermal fluctuations.

    \putbib[sm]
    
\end{bibunit}

\end{document}